\documentclass[journal, twoside, final, 10pt, comsoc]{IEEEtran}
\pdfoutput=1

\usepackage[cmex10]{mathtools}   
\usepackage{amsthm, thmtools, amssymb}

\usepackage{graphicx}
\usepackage[caption = false]{subfig} 

\usepackage{tikz}
\usepackage{pgfplots}
\usetikzlibrary{arrows, arrows.meta, calc, intersections, positioning, fit}
\usetikzlibrary{decorations.markings, decorations.pathreplacing, shapes.misc}

\usepackage{stackrel} 
\usepackage{accents}  

\usepackage{etoolbox}

\usepackage[ruled,vlined]{algorithm2e}

\usepackage{url}

\usepackage{microtype}

\usepackage{booktabs}
\usepackage{multirow}

\usepackage{cleveref}

\theoremstyle{plain} 
\newtheorem{theorem}{Theorem}
\newtheorem{lemma}[theorem]{Lemma} 
\newtheorem{proposition}[theorem]{Proposition} 
\newtheorem{corollary}[theorem]{Corollary} 

\theoremstyle{definition} 
\newtheorem{definition}{Definition}
\newtheorem{example}{Example}

\theoremstyle{remark} 
\newtheorem{remark}{Remark}

\renewcommand\thmcontinues[1]{Continued}

\newcommand{\Prob}[1]{\ensuremath{\mathbb{P} \left\{#1 \right\}}} 
\newcommand{\bv}[1]{\mathbf{#1}} 
\newcommand{\kindi}[1]{\left[ #1 \right]}

\newcommand{\aconj}[1]{#1^{(\mathsf{a})}}
\newcommand{\bconj}[1]{#1^{(\mathsf{b})}}
\newcommand{\abconj}[1]{#1^{(\mathsf{ab})}}
\newcommand{\ba}{r}
\newcommand{\omegaa}{y_a,u_a}
\newcommand{\symm}[1]{\accentset{\circ}{#1}}
\newcommand{\properup}{\overset{p}{\succcurlyeq}}

\newcommand{\upgradec}[1]{\check{#1}}
\newcommand{\IMJP}{\textrm{\textup{IMJP}}}
\newcommand{\IML}{\textrm{\textup{IML}}}
\newcommand{\ML}{\textrm{\textup{ML}}}
\newcommand{\SC}{\textrm{\textup{SC}}}
\newcommand{\Dyauadb}{D_{y_a,u_a}^{d_b}}
\newcommand{\Dda}{D_{d_a}}
\newcommand{\Dd}{D_{d}}
\newcommand{\rhoyauadb}{\rho_{\omegaa}^{d_b}}
\newcommand{\muyauazb}{\mu_{\omegaa}^{z_b}}
\newcommand{\muyauazbkbar}{\mu_{\omegaa}^{\bar{z}_{bk}}}
\newcommand{\muyauazbibar}{\mu_{\omegaa}^{\bar{z}_{bi}}}
\newcommand{\muyauazbk}{\mu_{\omegaa}^{z_{bk}}}
\newcommand{\muyauazbi}{\mu_{\omegaa}^{z_{bi}}}
\newcommand{\bchanpi}[3][\pi]{{#1}_{#2}^{#3}}
\newcommand{\bchanPi}[2]{\Pi_{#1}^{#2}}
\newcommand{\achanpi}[2][\pi]{{#1}^{#2}}

\DeclareMathOperator*{\argmax}{arg\,max}

\newtoggle{twocol}
\toggletrue{twocol}

\begin{document}
\title{A Lower Bound on the Probability of Error of Polar Codes over BMS Channels}
\date{}
\author{\IEEEauthorblockN{Boaz~Shuval, Ido~Tal\\
Department of Electrical Engineering,\\
Technion, Haifa 32000, Israel.\\
Email: \{\texttt{bshuval@campus}, \texttt{idotal@ee}\}\texttt{.technion.ac.il}}
\thanks{An abbreviated version of this article, with the proofs omitted, has appeared in ISIT 2017.}}

\maketitle

\begin{abstract}
Polar codes are a family of capacity-achieving codes that have explicit and low-complexity construction, encoding, and decoding algorithms. Decoding of polar codes is based on the successive-cancellation decoder, which decodes in a bit-wise manner. A decoding error occurs when at least one bit is erroneously decoded. The various codeword bits are correlated, yet performance analysis of polar codes ignores this dependence: the upper bound is based on the union bound, and the lower bound is based on the worst-performing bit. 
Improvement of the lower bound is afforded by considering error probabilities of two bits simultaneously. These are difficult to compute explicitly due to the large alphabet size inherent to polar codes. In this research we propose a method to lower-bound the error probabilities of bit pairs. We develop several transformations on pairs of synthetic channels that make the resultant synthetic channels amenable to alphabet reduction. 
	Our method yields  lower bounds that significantly improve upon currently known lower bounds for polar codes under successive-cancellation decoding.  	
\end{abstract}

\section{Introduction}
\IEEEPARstart{P}{olar} codes~\cite{Arikan_2009} are a family of codes that achieve capacity on
binary, memoryless, symmetric (BMS) channels and have low-complexity construction, encoding, and
decoding algorithms. This is the setting we consider. Polar codes have since been extended to
a variety of settings including source-coding~\cite{arikan_2010_source, korada}, non-binary
channels~\cite{sasoglu_fnt}, asymmetric channels~\cite{Honda_Yamamoto_2013}, settings with
memory~\cite{sasoglu_2011_mem,sasoglu_2016,Shuval_Tal_Memory_2017}, and more. 	The probability of error of polar codes is
given by a union of correlated error events. The union bound, which ignores this correlation, is
used to upper-bound the error probability. In this work, we exploit the correlation between error
events to develop a general method for lower-bounding the probability of error of polar codes.

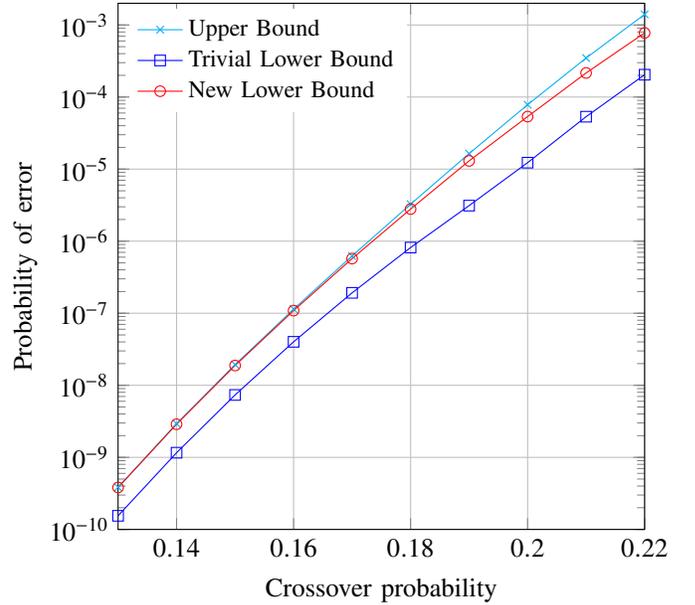
\begin{figure}[t]
\centering
\definecolor{mycolor1}{rgb}{1.00000,0.00000,1.00000}%
\begin{tikzpicture}

\begin{axis}[%
width=7cm,
height=7cm,
scale only axis,
grid = major,
xmin=0.13,
xmax=0.22,
xlabel={Crossover probability},
ymode=log,
ymin=1e-10,
ymax= 0.002,
ylabel={Probability of error},
legend style={at={(0.02,0.99)},anchor=north west,legend cell align=left,align=left,fill=white, draw=none}
]
\addplot [color=cyan,solid,mark=x,mark options={solid}]
  table[row sep=crcr]{%
0.13    3.850869663323411e-10\\
0.14    2.941405887386531e-09\\  
0.15	1.935488070643073e-08\\ 
0.16	1.139631130132584e-07\\ 
0.17	6.238867069515004e-07\\
0.18	3.261014180666720e-06\\
0.19	1.640412664303528e-05\\ 
0.2	    7.835253524825373e-05\\ 
0.21	3.477550110270991e-04\\ 
0.22	0.001403696167960\\ 
};
\addlegendentry{\small Upper Bound};

\addplot [color=blue,solid,mark=square,mark options={solid}]
  table[row sep=crcr]{%
0.13    1.545255000000000e-10\\
0.14    1.159648000000000e-09\\
0.15	7.346006e-09\\
0.16	4.011035e-08\\
0.17	1.918491e-07\\
0.18	8.154281e-07\\
0.19	3.115159e-06\\
0.2	1.228296e-05\\
0.21	5.337406e-05\\
0.22	0.0002041444\\
};
\addlegendentry{\small Trivial Lower Bound};  

\addplot [color=red,solid,mark=o,mark options={solid}]
  table[row sep=crcr]{%
0.13    3.821036930057008e-10\\
0.14    2.882313752550001e-09\\
0.15	1.881575980820003e-08\\
0.16    1.087182732840000e-07\\
0.17	5.730457175799999e-07\\
0.18	2.791437222699995e-06\\
0.19    1.299938258659999e-05\\
0.2     5.354927595200003e-05\\
0.21    2.158925888600008e-04 \\
0.22	7.756549415500035e-04\\
};
\addlegendentry{\small New Lower Bound};  

\end{axis}
\end{tikzpicture}%
\caption{Bounds on the probability of error of a rate $0.1$ polar code of length $2^{10}$ designed
for a BSC with crossover probability $0.2$. The code was used over BSCs with a range of crossover
probabilities. The upper bound is based on~\cite{Tal_2013}. The trivial lower bound is 
a lower bound on the probability of error of worst synthetic channel in the non-frozen set. 
The new lower bound was computed using the techniques of this paper.}
\label{fig_bounds}
\end{figure}
	
\Cref{fig_bounds} shows a numerical example of the lower bound developed in this paper. We designed
a polar code of length $N=2^{10}=1024$ and rate $R=0.1$ for a Binary Symmetric Channel (BSC) with
crossover probability $0.2$. We plot upper and lower bounds on the probability of error of this
code under successive cancellation decoding, when used over BSCs of varying crossover probabilities.
Our lower bound significantly improves upon the existing (trivial) lower bound, and is tight over
a large range of crossover probabilities. 

Our method is based on lower-bounding the probability of correlated error events. It consists of
several operations and transformations that we detail throughout this article. A high-level description of the
key steps appears at the end of the introduction, once we establish some notation.

Polar codes are based on an iterative construction that transforms $N=2^n$ identical and independent channel uses into low-entropy and high-entropy channels. The low-entropy channels are almost noiseless, whereas the high-entropy channels are almost pure noise. Ar\i{}kan showed~\cite{Arikan_2009} that for every $\epsilon>0$, as $N \to \infty$ the proportion of channels with capacity greater than $1-\epsilon$ tends to the channel capacity $C$ and the proportion of channels with capacity less than $\epsilon$ tends to $1-C$.

The polar construction begins with two identical and independent copies of a BMS channel $W$ and transforms them into two new channels,
\begin{align}
	W^-(y_1,y_2|u_1) &= \frac{1}{2} \sum_{u_2} W(y_1|	u_1\oplus u_2)W(y_2|u_2),\nonumber\\
	W^+(y_1,y_2,u_1|u_2) &= \frac{1}{2}  W(y_1|u_1\oplus u_2)W(y_2|u_2).\label{eq_plus transform}
\end{align}
Channel $W^+$ is a better channel than $W$ whereas channel $W^-$ is worse than $W$.\footnote{By this we mean that channel $W^+$ can be stochastically degraded to channel $W$, which in turn can be stochastically degraded to $W^-$.} This construction can be repeated multiple times; each time we take two identical copies of a channel, say $W^+$ and $W^+$, and polarize them, e.g., to $W^{+-}$ and $W^{++}$. We call the operation $W \mapsto W^{-}$ a `$-$'-transform, and the operation $W \mapsto W^{+}$ a `$+$'-transform. 

 There are $N=2^n$ possible combinations of $n$ `$-$'- and `$+$'-transforms; we define channel $W_a$
 as follows. Let $\langle \alpha_1,\alpha_2,\ldots, \alpha_n \rangle$ be the binary expansion of
 $a-1$, where $\alpha_1$
 is the most significant bit (MSB). Then, channel $W_a$ is obtained by $n$ transforms of $W$
 according to the sequence $\alpha_1,\alpha_2,\ldots, \alpha_n$, starting with the MSB: if $\alpha_j
 = 0$ we do a `$-$'-transform and if $\alpha_j = 1$ we do a `$+$'-transform. For example, if $n=3$, channel $W_5$ is $W^{+--}$, i.e., it first undergoes a `$+$'-transform and then two `$-$'-transforms. 

Overall, we obtain $N$ channels $W_1, \ldots, W_N$; channel $W_a$ has input $u_a$ and output  $y_1,\ldots, y_N,  u_1,\ldots, u_{a-1}$. That is, channel $W_a$ has binary input $u_a$, output that consists of the output and input of channel $W_{a-1}$, and assumes that the input bits of future channels $u_{a+1},\ldots,u_{N}$ are uniform. We call these \emph{synthetic channels}. One then determines which synthetic channels are low-entropy and which are high-entropy, and transmits information over the low-entropy synthetic channels and predetermined values over the high-entropy synthetic channels. Since the values transmitted over the latter are predetermined, we call the high-entropy synthetic channels \emph{frozen}. 

Decoding is accomplished via the successive-cancellation (SC) decoder. It decodes the synthetic channels in succession, using previous bit decisions as part of the output. The bit decision for a synthetic channel is either based on its likelihood or, if it is frozen, on its predetermined value. That is, denoting the set of non-frozen synthetic channels by $\mathcal{A}$, 
\[ \hat{U}_a(y_1^N, \hat{u}_1^{a-1}) = \begin{dcases} \argmax_{u_a} W_a(y_1^N,\hat{u}_1^{a-1}|u_a), & a \in \mathcal{A} \\ 
 u_a, & a \in \mathcal{A}^c,	
 \end{dcases}
\]
where we denoted $y_1^N = y_1,\ldots,y_N$ and similarly for the previous bit decisions $\hat{u}_1^{a-1}$. 
As non-frozen synthetic channels are almost noiseless, previous bit decisions are assumed to be
correct. Thus, when $N$ is sufficiently large, this scheme can be shown to achieve
capacity~\cite{Arikan_2009},  as the proportion of almost noiseless channels is $C$.  

To analyze the performance of polar codes, let $\mathcal{B}_a$ denote the event that channel $W_a$ errs under SC decoding while channels $1,2,\ldots,a-1$ do not. That is, 
\[ \mathcal{B}_a = \left\{u_1^N, y_1^N\,|\,\hat{u}_1^{a-1} = u_1^{a-1}, \, \hat{U}_a(y_1^N,\hat{u}_1^{a-1}) \neq u_a\right\}.\]
The probability of error of polar codes under SC decoding is given by $\Prob{\bigcup_{a \in \mathcal{A}} \mathcal{B}_a}$. Let $\mathcal{E}_a$ denote the event that channel $W_a$ errs given that a genie had revealed to it the true previous bits, i.e. 
\[ \mathcal{E}_a = \left\{u_1^N, y_1^N\,|\,\hat{U}_a(y_1^N,u_1^{a-1}) \neq u_a\right\}.\] We call an SC decoder with access to genie-provided previous bits a \emph{genie-aided decoder}. Some thought reveals that $\bigcup_{a \in \mathcal{A}} \mathcal{B}_a = \bigcup_{a \in \mathcal{A}} \mathcal{E}_a$ (see~\cite[Proposition 2.1]{sasoglu_fnt} or~\cite[Lemma 1]{Mori_Tanaka_2009}). Thus, the probability of error of polar codes under SC decoding is equivalently given by $P_e^{\SC}(W) = \Prob{\bigcup_{a \in \mathcal{A}} \mathcal{E}_a}$. In the sequel we assume a genie-aided decoder. 

The events $\{\mathcal{B}_a\}$ are disjoint but difficult to analyze. The events $\mathcal{E}_a$ are easier to analyze, but are no longer disjoint. A straightforward upper bound for $\Prob{\bigcup_{a \in \mathcal{A}} \mathcal{E}_a}$ is  the union bound: 
\begin{equation} \Prob{\bigcup_{a \in \mathcal{A}} \mathcal{E}_a} \leq \sum_{a \in \mathcal{A}} \Prob{\mathcal{E}_a}.\label{eq_union bound} \end{equation}
This bound facilitated the analysis of~\cite{Arikan_2009}. An important question is how tight this
upper bound is. To this end, one approach is to develop a lower bound to  $\Prob{\bigcup_{a\in
\mathcal{A}} \mathcal{E}_a}$, which is what we pursue in this work.  

A trivial lower bound on a union is 
\begin{equation} \Prob{\bigcup_{a \in \mathcal{A}} \mathcal{E}_a} \geq \max_{a \in \mathcal{A}} \Prob{\mathcal{E}_a}. \label{eq_trivial lower bound} \end{equation} Better lower bounds may be obtained by considering pairs of error events: 
\begin{equation*} \Prob{\bigcup_{a \in \mathcal{A}} \mathcal{E}_a} \geq \max_{a,b \in \mathcal{A}} \Prob{\mathcal{E}_a \cup \mathcal{E}_b}.\label{eq_lower bound union of two events} \end{equation*}
Via the inclusion-exclusion principle, one can combine lower bounds on multiple pairs of error events to obtain a better lower bound~\cite{Hoppe_1985}
\begin{equation} \Prob{\bigcup_{a \in \mathcal{A}} \mathcal{E}_a} \geq \sum_{a \in \mathcal{A}} \Prob{\mathcal{E}_a} - \sum_{\substack{a,b \in \mathcal{A}, \\a<b}} \Prob{\mathcal{E}_a \cap \mathcal{E}_b}.
\label{eq_inclusion exclusion lower bound}
\end{equation} 
This can also be cast in terms of unions of error events using $\Prob{\mathcal{E}_a\cap\mathcal{E}_b} =\Prob{\mathcal{E}_a}+\Prob{\mathcal{E}_b}- \Prob{\mathcal{E}_a\cup\mathcal{E}_b}$. 

To our knowledge, to date there have been two attempts to compute a lower bound on the performance of the SC decoder, both based on~\eqref{eq_inclusion exclusion lower bound}. The first attempt was in~\cite{Mori_Tanaka_2009}, using a density evolution approach, and the second attempt in~\cite{Parizi_2013} applies only to the Binary Erasure Channel (BEC). We briefly introduce these below, but first we explain where the difficulty lies. 

The probability $\Prob{\mathcal{E}_a}$ is given by an appropriate functional of the probability distribution of synthetic channel $W_a$. However, the output alphabet of $W_a$ is very large. If the output alphabet of $W$ is $\mathcal{Y}$ then the output alphabet of $W_a$ has size $|\mathcal{Y}|^N 2^{a-1}$. This quickly grows unwieldy, recalling that $N = 2^n$. It is infeasible to store this probability distribution and it must be approximated. Such approximations are the subject of~\cite{Tal_2013}; they enable one to compute upper and lower bounds on various functionals of the synthetic channel $W_a$. 

To compute probabilities of unions of events, one must know the joint distribution of two synthetic channels. The size of the joint channel's output alphabet is the product of each synthetic channel's alphabet size,  rendering the joint distribution infeasible to store. 

The authors of~\cite{Mori_Tanaka_2009} suggested to approximate the joint distribution of pairs of synthetic channels using a density evolution approach. This provides an iterative method to compute the joint channel, but does not address the problem of the amount of memory required to store it. Practical implementation of density evolution must involve quantization~\cite[Appendix B]{mct}. The probability of error derived from quantized joint channels approximates, but does not generally bound, the real probability of error. For the special case of the BEC, as noted and analyzed in~\cite{Mori_Tanaka_2009}, no quantization is needed, as the polar transform of a BEC is a BEC. Thus, they were able to precisely compute the probabilities of unions of error events of descendants of a BEC using density evolution. 

The same bounds for the BEC were developed in~\cite{Parizi_2013} using a different approach, again relying on the property that the polar transform of a BEC is a BEC. The authors were able to track the joint probability of erasure during the polarization process. Furthermore, they were able to show that the union bound is asymptotically tight for the BEC.

In this work, we develop an algorithm to compute lower bounds on the joint probability of error of two synthetic channels $\Prob{\mathcal{E}_a \cup \mathcal{E}_b}$. Our technique is general, and applies to synthetic channels that are polar descendants of any BMS channel. We use these bounds in~\eqref{eq_inclusion exclusion lower bound} to lower-bound the probability of error of polar codes. For the special case of the BEC, we recover the results of~\cite{Mori_Tanaka_2009} and~\cite{Parizi_2013} using our bounds.

Concretely, consider two synthetic channels, $W_a(y_a|u_a)$ and $W_b(y_b|u_b)$, which we call the
a-channel and the b-channel, respectively. 
Their joint channel is $W_{a,b}(y_a,y_b|u_a,u_b)$. Our algorithm lower-bound the probability that
a successive cancellation decoder errs on either channel. It is based on the following key steps:  
\begin{enumerate}
    \item Replace successive cancellation with a different decoding criterion
        (\Cref{sec_decoding two dependent channels}).
    \item Bring the joint channel to a different form that makes the b-channel decoding immediately
        apparent from the received symbol (\Cref{subsec_d value representation}).
    \item Apply the \emph{symmetrizing} transform, after which the output of the a-channel is independent
        of the input of the b-channel (\Cref{sec_Symmetrized Joint Bit-Channels}). 
    \item Apply the \emph{upgrade-couple} transform, which splits each a-channel output to multiple
        symbols.
        However, each such new symbol is constrained to appear with only a small subset of b-channel
        outputs (\Cref{subsec_Upgrading $W_a$}).
    \item Reduce each channel's alphabet size. This is done by stochastically upgrading one channel
        while keeping the other channel constant. Each channel has a different upgrading procedure; 
        the a-channel upgrading procedure is detailed in \Cref{subsec_Upgrading $W_a$}, and the
        b-channel upgrading procedure is detailed in \Cref{subsec_upgrading $W_b$}. 
\end{enumerate}

%

\section{Overview of Our Method} 
In this section we provide a brief overview of our method, and lay out the groundwork for the sections that follow. We aim to produce a lower bound on the probability of error of two synthetic channels. Since we cannot know the precise joint distribution, we must approximate it. The approximation is rooted in stochastic degradation. 

 Degradation is a partial ordering of channels. Let $W(y|u)$ and $Q(z|u)$ be two channels. We say that $W$ is (stochastically) degraded with respect to $Q$, denoted $W \preccurlyeq Q$, when there exists some channel $P(y|z)$ such that
\begin{equation}\label{eq_definition of degradation} W(y|u) = \sum_{z} P(y|z)Q(z|u).\end{equation} If $W$ is degraded with respect to $Q$ then $Q$ is upgraded with respect to $W$. Degradation  implies an ordering on the probability of error of the channels~\cite[Chapter 4]{mct}: if $ W \preccurlyeq Q$ then $P_e^{\star}(W) \geq P_e^{\star}(Q)$, where $P_e^{\star}$ denotes the probability of error of the optimal decoder (defined in \Cref{sec_decoding general case}).  

The notion of degradation readily applies to joint channels. If $W_{a,b}(y_a,y_b|u_a,u_b)$ and
$Q_{a,b}(z_a,z_b|u_a,u_b)$ are two joint channels, we say that $Q_{a,b}(z_a,z_b|u_a,u_b)
\succcurlyeq W_{a,b}(y_a,y_b|u_a,u_b)$ via some degrading channel $P(y_a,y_b|z_a,z_b)$ if 
\begin{equation} W_{a,b}(y_a,y_b|u_a,u_b) = \sum_{\mathclap{z_a,z_b}} P(y_a,y_b|z_a,z_b)Q_{a,b}(z_a,z_b|u_a,u_b). \label{eq_definition of joint degradation}\end{equation}
As for the single channel case, if $Q_{a,b} \succcurlyeq W_{a,b}$ then $P_e^{\star}(W_{a,b}) \geq
P_e^{\star}(Q_{a,b})$, where $P_e^{\star}$ is the probability of error of the \emph{optimal} decoder
for the joint channel. Indeed our approach will be to approximate the joint synthetic channel with
an upgraded joint channel with smaller output alphabet. 
There is a  snag, however: this ordering of error probabilities does not hold, in general, for
suboptimal decoders.

The SC decoder, used for polar codes, is suboptimal. In the genie-aided case, which we consider
here, it is equivalent to performing a maximum likelihood decision on each marginal separately. We
shall demonstrate the suboptimality of the SC decoder in \Cref{sec_decoding two dependent channels}.
Then, we will develop a different decoding criterion whose performance lower-bounds the SC decoder
performance and is ordered by degradation. While in general finding this decoder requires an
exhaustive search, for the special case of polar codes this decoder is easily found. It does,
however, imply a special structure for the degrading channel, which we use to our advantage. 

We investigate the joint distribution of two synthetic channels in \Cref{sec_properties of joint bit
channels}. We first bring it to a more convenient form that will be used in the sequel. Then, we
explain how to polarize a joint synthetic channel distribution and explore some consequences of
symmetry. Further consequences of symmetry are the subject of \Cref{sec_Symmetrized Joint
Bit-Channels}, in which we transform the channel to another  form that greatly simplifies the steps
that follow. This form exposes the inherent structure of the joint channel. 

How to actually upgrade joint channels is the subject of \Cref{sec_upgrading procedures for joint
bit channels}.  We upgrade the joint channel in two ways; each upgrades one marginal without
changing the other.  We cannot simply upgrade the marginals, as we must consider the joint channel
as a whole.  This is where the above-mentioned symmetrizing and upgrade-couple transforms come into play.   
	
We present our algorithm for lower-bounding the probability of error of polar codes in
\Cref{sec_lower bound procedures}. This algorithm is based on the building blocks presented in the
previous sections. Details of our implementation appears in \Cref{sec_implementation}. 
We demonstrate our algorithm with some numerical results in \Cref{sec_numerical
results}, and conclude with a short discussion in \Cref{sec_discussion}. 

\subsection{Notation}
We denote by $y_j^k = y_j,y_{j+1},\ldots,y_k$ for $j<k$. We use an Iverson-style notation (see~\cite{knuth_notation}) for indicator (characteristic) functions. That is, for a logical expression $\mathtt{expr}$, $\kindi{\mathtt{expr}}$ is $0$ whenever $\mathtt{expr}$ is not true and is $1$ otherwise. We assume that the indicator function takes precedence whenever it appears, e.g., $n^{-1}\kindi{n>0}$ is $0$ for $n=0$.   

\section{Decoding of Two Dependent Channels}\label{sec_decoding two dependent channels}
In this section, we tackle decoding of two dependent channels. We explain how this differs from the case of decoding a single channel, and dispel some misconceptions that may arise. We then specialize the discussion to polar codes. We explain the difficulty with combining the SC decoder with degradation procedures, and develop a different decoding criterion instead. Finally, we develop a special structure for the degrading channel that, combined with the decoding criterion, implies ordering of probability of error by degradation. 

\subsection{General Case}\label{sec_decoding general case}
A decoder for channel $W:\mathcal{U} \to \mathcal{Y}$ is a mapping $\phi$ that maps every output symbol $y \in \mathcal{Y}$ to some $u \in \mathcal{U}$. The average probability of error of the decoder for equiprobable inputs is given by
\[ P_e(W) = \sum_{u} \sum_{y} \frac{W(y|u)}{|\mathcal{U}|} \Prob{\phi(y) \neq u}. \]
The decoder is deterministic for symbols $y$ for which $\Prob{\phi(y) \neq u}$ assumes only the values $0$ and $1$. For some symbols, however, we allow the decoder to make a random decision. If $W(y|u) = W(y|u')$ for some $u,u' \in \mathcal{U}$, then $P_e(W)$ is the same whether $\phi(y) = u$ or $\phi(y) = u'$. Thus, the probability of error is insensitive to the resolution of ties. We denote the error event of a decoder by
$\mathcal{E} = \left\{ (u,y): \phi(y) \neq u \right\}.$ It is dependent on the decoder, i.e., $\mathcal{E} = \mathcal{E}(\phi)$;  we suppress this to avoid cumbersome notation. Clearly, $P_e(W) = \Prob{\mathcal{E}}$. 

The maximum-likelihood (ML) decoder, well known to minimize $P_e(W)$ when the input bits are
equiprobable, is defined by 
\begin{equation}\label{eq_Definition of ML decoder} W(y|u) > W(y|u')\quad \forall u' \neq
u \Rightarrow \phi(y) = u.\end{equation}
The ML decoder is not unique, as it does not define how ties are resolved. In the absence of ties,
the ML decoding rule is $\phi(y) = \argmax_u W(y|u)$. We denote by $P_e^{\ML}(W)$ the probability of
error of the ML decoder. 

We now consider two \emph{dependent} binary-input channels, $W_a:\mathcal{U}\to\mathcal{Y}_a$ and
$W_b:\mathcal{U}\to \mathcal{Y}_b$, with joint distribution $W_{a,b}: \mathcal{U}\times\mathcal{U}
\to \mathcal{Y}_a \times \mathcal{Y}_b$. A decoder is a mapping $\phi:\mathcal{Y}_a \times
\mathcal{Y}_b \to \mathcal{U}\times\mathcal{U}$. The joint probability of error of the decoder is,
as above, 
\begin{equation} \begin{split} &P_e(W_{a,b}) \\&= \sum_{u_a,u_b}\sum_{y_a,y_b}
\frac{W_{a,b}(y_a,y_b|u_a,u_b)}{|\mathcal{U}|^2} \Prob{\phi(y_a,y_b) \neq (u_a,u_b)}.
\end{split}\label{eq_joint prob of error} \end{equation}

An optimal decoder for the joint channel considers both outputs together and makes a decision for both inputs jointly, to minimize $P_e(W_{a,b})$. We denote its probability of error by $P_e^{\star}(W_{a,b})$. When the input bits are equiprobable, $P_e^{\star}(W_{a,b}) = P_e^{\ML}(W_{a,b})$. 

 Rather than jointly decoding the input bits based on the joint output, we may opt to decode each marginal channel separately. That is, consider decoders of the form $\phi(y_a,y_b) = (\phi_a(y_a), \phi_b(y_b))$. In words, the decoder of channel $W_a$ bases its decision solely on $y_a$ and completely ignores $y_b$ and vice versa. What are the optimal decoders $\phi_a$ and $\phi_b$? The answer depends on the criterion of optimality. 

Denote by $\mathcal{E}_i$ the error event of channel $W_i$ under some decoder $\phi_i:\mathcal{Y}_i
\to \mathcal{U}$. 
The \emph{Individual Maximum Likelihood }(IML) decoder minimizes each individual marginal channel's
probability of error. That is, we set
$\phi_a$ and $\phi_b$ as ML decoders for their respective marginal channels. 
We denote its joint probability of error by $P_e^{\IML}(W_{a,b})$.
Hence, $P_e^{\IML}(W_{a,b})$ is computed by~\eqref{eq_joint prob of error}, with
$\phi(y_a,y_b) = (\phi_a^{\ML}(y_a), \phi_b^{\ML}(y_b))$, where $\phi_a^{\ML}$  and $\phi_b^{\ML}$ are ML decoders for the
marginal channels
$W_a$ and $W_b$, respectively.

Another criterion is to minimize
$\Prob{\mathcal{E}_a\cup\mathcal{E}_b}$, the probability that at least one of the decoders makes an
error. We call the decoder that minimizes this probability using individual decoders for each
channel the \emph{Individual Minimum Joint Probability of error} (IMJP) decoder. The event
$\mathcal{E}_a \cup \mathcal{E}_b$ is not the same as the error event of the optimal decoder for the
joint channel, even when the individual decoders turn out to be ML decoders. This is because we
decode each input bit separately using only a portion of the joint output. Clearly,
\begin{equation} \label{eq_ordering of error probabilities} P_e^{\star}(W_{a,b}) \leq \min_{\phi_a, \phi_b} \Prob{\mathcal{E}_a \cup \mathcal{E}_b} \leq P_e^{\IML}(W_{a,b}).\end{equation}
We denote 
\[ P_e^{\IMJP}(W_{a,b}) = \min_{\phi_a, \phi_b} \Prob{\mathcal{E}_a \cup \mathcal{E}_b}.\]

The three decoders in~\eqref{eq_ordering of error probabilities} successively use less information for their decisions. The optimal decoder uses both outputs jointly as well as knowledge of the joint probability distribution; the IMJP decoder retains the knowledge of the joint probability distribution, but uses  each output separately; finally, the IML decoder dispenses with the joint probability distribution and operates as if the marginals are independent channels.  

\begin{example}\label{ex_Joint BMS where P(EML1UEML2) is not optimal}
	The conditional distribution $W_{a,b}(y_a, y_b|u_a,u_b)$ of some joint channel is given in \Cref{tab_Example of channel where ML is not the same as minimizing P(E1UE2)}.\footnote{This is not a joint distribution of two synthetic channels that result from polarization. However, the phenomena observed here  hold for joint distributions of two synthetic channels as well, and similar examples may be constructed for the polar case.} The marginals are channels $W_a(y_a|u_a)$ and $W_b(y_b|u_b)$. Three decoders for this channel are shown in \Cref{tab_Example of decoders}. Note that for the IML and IMJP decoders we have $\phi(y_a,y_b) = (\phi_a(y_a),\phi_b(y_b))$.
	
		The  optimal decoder for the joint channel chooses, for each output pair, the input pair with the highest probability. The IML decoder is formed by using an ML decoder for each marginal; the ML decoders of the marginals decide that the input is $0$ when $1$ is received and vice versa.  The IMJP decoder is found by checking all combinations of marginal channel decoders $\phi_a$ and $\phi_b$ and choosing that pair the achieves $\min_{\phi_a, \phi_b}\Prob{\mathcal{E}_a \cup \mathcal{E}_b}$. We then have 
	\begin{align*}
		P_e^{\star}(W_{a,b}) &= 1-(0.44+0.54+0.32+0.62)/4 =  0.52, \\ 
		P_e^{\IML}(W_{a,b})  &= 1-(0.05+0.49+0.01+0.62)/4 = 0.7075, \\
		P_e^{\IMJP}(W_{a,b}) &= 1-(0.22+0.49+0.04+0.62)/4 = 0.6575. 
	\end{align*}
%
As expected,~\eqref{eq_ordering of error probabilities} holds. 	

	We now demonstrate that the probability of error of suboptimal decoders is not ordered by degradation. To this end, we degrade the joint channel in \Cref{tab_Example of channel where ML is not the same as minimizing P(E1UE2)} by merging the output symbols $(0,0), (1,1)$ into a new symbol, $(0',0')$ and $(0,1), (1,0)$ into a new symbol, $(1',1')$. We denote the new joint channel by $W'_{a,b}$ and provide its conditional distribution in \Cref{tab_Example of degraded channel}. For each of the marginals, the ML decoder declares $0$ upon receipt of $0'$, and $1$ otherwise. Hence, for the degraded channel, $P_e^{\IML}(W'_{a,b}) = 1 - (0.92+0.86)/4 = 0.555$, which is \emph{lower} than $P_e^{\IML}(W_{a,b})$. For the degraded channel, the IML decoder is also the optimal decoder. As this is a degraded channel, however, $P_e^{\IML}(W'_{a,b}) = P_e^{\star}(W'_{a,b}) \geq P_e^{\star}(W_{a,b}) = 0.52$.

	\begin{table}[t]
			\centering
				\caption{Conditional distribution $W_{a,b}(y_a,y_b|u_a,u_b)$. In this case, the ML decoders of the marginals do not minimize $\Prob{\mathcal{E}_a\cup\mathcal{E}_b}$.}  \label{tab_Example of channel where ML is not the same as minimizing P(E1UE2)}
	\begin{tabular}{ccccc}
	\toprule \multirow{2}[3]{*}{$(u_a,u_b)$} & \multicolumn{4}{c}{$(y_a,y_b)$}\\ \cmidrule(rl){2-5}
		 & $(0,0)$ & $(0,1)$ & $(1,0)$ & $(1,1)$ \\ \midrule
		$(0,0)$ & $0.30$ & $0.04$ & $0.04$ & $0.62$ \\ 
		$(0,1)$ & $0.44$ & $0.46$ & $0.01$ & $0.09$ \\ 
		$(1,0)$ & $0.22$ & $0.49$ & $0.24$ & $0.05$ \\ 
		$(1,1)$ & $0.05$ & $0.54$ & $0.32$ & $0.09$ \\ \bottomrule
	\end{tabular} 	
	\end{table}		
	\begin{table}[t]
		\centering
			\caption{Various decoders for joint channel $W_{a,b}$ from \Cref{tab_Example of channel where ML is not the same as minimizing P(E1UE2)}. Three decoders are shown: the optimal decoder, the IML decoder, and the IMJP decoder. The leftmost column is the received joint channel output, and the remaining columns depict the decisions of the various decoders.} %
			\label{tab_Example of decoders}
	\begin{tabular}{cccc}
	\toprule \multirow{2}[3]{*}{$(y_a,y_b)$} & \multicolumn{3}{c}{$(\hat{u}_a,\hat{u}_b) = \phi(y_a,y_b)$}\\ \cmidrule(rl){2-4}
		        & optimal & IML     & IMJP  \\ \midrule
		$(0,0)$ & $(0,1)$ & $(1,1)$ & $(1,0)$  \\ 
		$(0,1)$ & $(1,1)$ & $(1,0)$ & $(1,0)$  \\ 
		$(1,0)$ & $(1,1)$ & $(0,1)$ & $(0,0)$  \\ 
		$(1,1)$ & $(0,0)$ & $(0,0)$ & $(0,0)$  \\ \bottomrule
	\end{tabular} 	
	\end{table}	
	\begin{table}[t]
		\centering
			\caption{Channel $W'_{a,b}(y_a,y_b|u_a,u_b)$, degraded from $W_{a,b}$ of \Cref{tab_Example of channel where ML is not the same as minimizing P(E1UE2)}.}  \label{tab_Example of degraded channel}
	\begin{tabular}{ccc}
	\toprule \multirow{2}[3]{*}{$(u_a,u_b)$} & \multicolumn{2}{c}{$(y_a,y_b)$}\\ \cmidrule(rl){2-3}
		        & $(0',0')$ & $(1',1')$ \\ \midrule
		$(0,0)$ & $0.92$    & $0.08$    \\ 
		$(0,1)$ & $0.53$    & $0.47$    \\ 
		$(1,0)$ & $0.27$    & $0.73$    \\ 
		$(1,1)$ & $0.14$    & $0.86$     \\ \bottomrule
	\end{tabular} 	
	\end{table}
\end{example}

\subsection{Polar Coding Setting}\label{subsec_Polar Coding Setting}
Given a joint channel, finding an optimal or IML decoder 
is an easy task. In both cases we use maximum-likelihood decoders; in the first case based on the joint channel, whereas in the second case based on the marginal channels. On the other hand, finding an IMJP decoder requires an exhaustive search, which may be costly. In the polar coding setting, as we now show, the special structure of joint synthetic channels permits finding the IMJP decoder without resorting to a search procedure.

\subsubsection{Joint Distribution of Two Synthetic Channels}

Let $W$ be some BMS channel that undergoes $n$ polarization steps. Let $a$ and $b$ be two indices of synthetic channels, where $b>a$. The synthetic channels are $W_a(y_a|u_a)$ and $W_b(y_b|u_b)$, where $y_a = (y_1^N,u_1^{a-1})$, $y_b = (y_1^N,u_1^{b-1})$, and $N=2^n$. We call them the \emph{a-channel} and the \emph{b-channel}, respectively. Their joint distribution is $W_{a,b}(y_a,y_b|u_a,u_b)$; this is the probability that the output of the a-channel is $y_a$ and the output of the b-channel is $y_b$, given that the inputs to the channels are $u_a$ and $u_b$, respectively. 

With probability $1$, the prefix of $y_b$ is $(y_a,u_a)$. Namely, $y_b$ has the form  
	\[ y_b = ((y_1^N,u_1^{a-1}), u_a, u_{a+1}^{b-1}) \equiv (y_a,u_a,y_{\ba}),\] 
where $y_{\ba}$ denotes the remainder of $y_b$ after removing $y_a$ and $u_a$. 
Thus,  
\begin{equation}\label{eq_Wab and its relationship to Wb} W_{a,b} (y_a,y_b|u_a,u_b) = 2W_b(y_b|u_b)\kindi{y_b = (y_a,u_a,y_{\ba})},\end{equation}
for some arbitrary $y_{\ba}$. The factor $2$ stems from the uniform distribution of $u_a$. With some abuse of notation, we will write
\begin{align*} W_{a,b}(y_a,y_b|u_a,u_b) &= W_{a,b}(y_b|u_a,u_b)\\ &= W_{a,b}(y_a,u_a,y_{\ba}|u_a,u_b). \end{align*}
The rightmost expression makes it clear that the portion of $y_b$ in which the input of the a-channel appears must equal the actual input of the a-channel. 

Observe from~\eqref{eq_Wab and its relationship to Wb}  that we can think of $W_b(y_a,u_a,y_{\ba}|u_b)$ as the joint channel $W_{a,b}$ up to a constant factor. Indeed, we will use $W_b(y_a,u_a,y_{\ba}|u_b)$ to  denote the joint channel where convenient. 

\subsubsection{Decoders for Joint Synthetic Channels}
Which decoders can we consider for joint synthetic channels? The optimal decoder extracts $u_a$ from the output of the b-channel and proceeds to decode $u_b$. This outperforms the SC decoder but is also impractical and does not lend itself to computing the probability that is of interest to us, the probability that \emph{either} of the synthetic channels errs. A natural suggestion is to mimic the SC decoder, i.e., to use an IML decoder. The joint probability of error of this decoder may decrease after stochastic degradation, so we discard this option.  

Consider two decoders $\phi_a$ and $\phi_b$ for channels $W_a$ and $W_b$, respectively. As above,
$\mathcal{E}_i$ is the error event of channel $W_i$ using decoder $\phi_i$, $i=a,b$. We seek a lower bound on $\Prob{\mathcal{E}_a \cup \mathcal{E}_b}$. Therefore, we choose decoders $\phi_a$ and $\phi_b$ that minimize $\Prob{\mathcal{E}_a \cup \mathcal{E}_b}$; this is none other than the IMJP decoder. Its performance lower-bounds that of the IML decoder [see~\eqref{eq_ordering of error probabilities}]. As we shall later see, combined with a suitable degrading channel structure, the probability of error of the IMJP decoder increases after stochastic degradation. Conversely, it decreases under stochastic upgradation; thus, combining the IMJP decoder with a suitable upgrading procedure produces the desired lower bound. 

Multiple decoders may achieve $\min_{\phi_a,\phi_b}\Prob{\mathcal{E}_a \cup \mathcal{E}_b}$. One decoder can be found in a straight-forward manner; we call it \emph{the} IMJP decoder. The following theorem shows how to find it. Its proof is a direct consequence of \Cref{lem_optimal phi2 is ML decoder,lem_optimal phi1 for given phi2} that follow.
\begin{theorem} \label{thm_minimizing phi_a and phi_b for polar channel}
	Let $W_a(y_a|u_a)$ and $W_b(y_b|u_b)$ be two 
	channels with joint distribution $W_{a,b}$ that  
	satisfies~\eqref{eq_Wab and its relationship to Wb}. Then, $\min_{\phi_a,\phi_b}\Prob{\mathcal{E}_a \cup \mathcal{E}_b}$ is achieved by setting $\phi_b$ as an ML decoder for $W_b$ and $\phi_a$ according to \begin{equation} \label{eq_phi1 as argmax} \phi_a(y_a) = \argmax_{u_a} T(y_a|u_a),\end{equation} 
	where
		\begin{equation} \label{eq_def of T(y_a|x_a)} T(y_a|u_a) = \frac{1}{2}\sum_{\substack{u_b,\\y_b}} W_{a,b}(y_a,y_b|u_a,u_b) \Prob{\phi_b(y_b) = u_b}. \end{equation}

\end{theorem}
Note that $T(y_a|u_a)$ is not a conditional distribution; it is non-negative, but its sum over $y_a$ does not necessarily equal $1$. In the right-hand side of~\eqref{eq_def of T(y_a|x_a)}, the dependence on $y_a,u_a$ is via~\eqref{eq_Wab and its relationship to Wb}, as $W_{a,b}(y_a,y_b|u_a,u_b) = 0$ if $y_b \neq (y_a,u_a,y_{\ba})$ for some $y_{\ba}$. 

\begin{corollary}
\Cref{thm_minimizing phi_a and phi_b for polar channel} holds for any two synthetic channels $W_a(y_a|u_a)$ and $W_b(y_b|u_b)$ that result from the same number of polarization steps of a BMS, where index $b$ is greater than $a$.
\end{corollary}
\begin{IEEEproof}
In the polar code case, the joint channel satisfies~\eqref{eq_Wab and its relationship to Wb}, so \Cref{thm_minimizing phi_a and phi_b for polar channel} applies. 	
\end{IEEEproof}	

In what follows, denote \[\varphi_i(y_i,u_i) \triangleq \Prob{\phi_i(y_i) = u_i},\quad  i=a,b.\] 
	
\begin{lemma} \label{lem_optimal phi2 is ML decoder}
	Let $W_a(y_a|u_a)$ and $W_b(y_b|u_b)$ be two dependent binary-input channels with equiprobable inputs and joint distribution $W_{a,b}$ that satisfies~\eqref{eq_Wab and its relationship to Wb}. Let $\phi_a:\mathcal{Y}_a \to \mathcal{U}$ be some decoder for channel $W_a$ with error event $\mathcal{E}_a$. Then, setting $\phi_b$ as an ML decoder for $W_b$ achieves $\min_{\phi_b} \Prob{\mathcal{E}_a \cup \mathcal{E}_b}$.
	\end{lemma}

\begin{IEEEproof}
	Recall that $y_b = (y_a,u_a,y_{\ba})$. 
	Using~\eqref{eq_Wab and its relationship to Wb}, 	
\begin{align*}
& 1-\Prob{\mathcal{E}_a \cup \mathcal{E}_b} \\  & \quad= \frac{1}{4} \sum_{\substack{u_a,\\u_b}} \sum_{\substack{y_a,\\y_b}}  W_{a,b}(y_a,y_b|u_a,u_b) \varphi_a(y_a,u_a)\varphi_b(y_b,u_b)\\
		 & \quad = \frac{1}{2}  \sum_{\substack{u_a,\\y_a,y_b}}\varphi_a(y_a,u_a) \kindi{y_b = (y_a,u_a,y_{\ba})}g(y_b), 
	\end{align*}
	where 
	\[ g(y_b) = \sum_{u_b} \varphi_b(y_b,u_b)W_b(y_b|u_b).\]
	The problem of finding the decoder $\phi_b$ that minimizes $\Prob{\mathcal{E}_a \cup \mathcal{E}_b}$ is separable over $u_a,y_a,y_b$; the terms $\varphi_a(y_a,u_a)$, $\kindi{y_b = (y_a,u_a,y_{\ba})}$ are non-negative and independent of $u_b$.
	Therefore, the optimal decoder $\phi_b$ is given by  
	$ \phi_b(y_b) = \arg \max_{u'_b} W_b(y_b|u'_b).$ 
\end{IEEEproof}
We remark that \Cref{lem_optimal phi2 is ML decoder} holds for \emph{any} a-channel decoder $\phi_a$. Thus, regardless of the selection of $\phi_a$, the optimal decoder for the  b-channel (in the sense of minimizing $\min_{\phi_b} \Prob{\mathcal{E}_a \cup \mathcal{E}_b}$) is an ML decoder. 

\begin{lemma} \label{lem_optimal phi1 for given phi2}
	Let $W_a(y_a|u_a)$ and $W_b(y_b|u_b)$ be two binary-input channels with joint distribution $W_{a,b}(y_a,y_b|u_a,u_b)$ and equiprobable inputs. Let $\phi_b:\mathcal{Y}_b \to \mathcal{U}$ be some decoder for channel $W_b$. Then, the decoder $\phi_a$  
	for channel $W_a$ given by~\eqref{eq_phi1 as argmax}  minimizes $\Prob{\mathcal{E}_a \cup \mathcal{E}_b}$.
\end{lemma}
\begin{IEEEproof}
	Since the input is equiprobable, 
	\begin{align*}
	& 1-\Prob{\mathcal{E}_a \cup \mathcal{E}_b} \\
	&= \frac{1}{4} \sum_{\substack{u_a,\\y_a}} \sum_{\substack{u_b,\\y_b}}  W_{a,b}(y_a,y_b|u_a,u_b) \varphi_a(y_a,u_a)\varphi_b(y_b,u_b)\\
	&= \frac{1}{2} \sum_{\substack{u_a,\\ y_a}} \varphi_a(y_a,u_a) \cdot \frac{1}{2}\sum_{\substack{u_b,\\ y_b}} W_{a,b}(y_a,y_b|u_a,u_b)\varphi_b(y_b,u_b) \\
	&= \frac{1}{2} \sum_{\substack{u_a,\\ y_a}} T(y_a|u_a)\varphi_a(y_a,u_a), 
	\end{align*}
	where the last equality is by~\eqref{eq_def of T(y_a|x_a)}. The problem of finding the decoder $\phi_a$ that minimizes $\Prob{\mathcal{E}_a \cup \mathcal{E}_b}$ is separable over $y_a$; clearly the optimal decoder is the one that sets
	$ \phi_a(y_a) = \arg \max_{u'_a} T(y_a|u'_a). $ 
	\end{IEEEproof}

Using~\eqref{eq_Wab and its relationship to Wb}, if $\phi_b$ is chosen as an ML decoder, as per \Cref{lem_optimal phi2 is ML decoder}, we have the following expression for $T(y_a|u_a)$:
\begin{equation}
	\begin{split}
	T(y_a|u_a) &= \sum_{y_{\ba}}\sum_{u_b} W_b(y_a,u_a,y_{\ba}|u_b)\varphi_b(y_b,u_b) \\ 
			   &= \sum_{y_{\ba}} \max_{u_b} W_b(y_a,u_a,y_{\ba}|u_b). 
	\end{split} \label{eq_expressions for T using yba}
\end{equation}

The IMJP and IML decoders do not coincide in general, although in some cases they may indeed coincide. We demonstrate this in the following example.
\begin{example} Let $W$ be a binary symmetric channel with crossover probability $p$. We perform $n=2$ polarization steps and consider the joint channel $W_{1,4}$, i.e., $W_a = W^{--}$ and $W_b = W^{++}$. When $p=0.4$, we have $ 0.6544 = P_e^{\IMJP}(W_{1,4}) < P_e^{\IML}(W_{1,4}) = 0.6976$. On the other hand, when $p=0.2$, the IMJP and IML decoders coincide, and $P_e^{\IMJP}(W_{1,4}) = P_e^{\IML}(W_{1,4}) = 0.5136$. In either case,~\eqref{eq_ordering of error probabilities} holds. 	
\end{example}

\begin{remark}
In the special case where $W$ is a BEC and $W_a$ and $W_b$ are two of its polar descendants, the IMJP and IML (SC) decoders coincide. This is thanks to a special property of the BEC that erasures for a synthetic channel are determined by the outputs of the $N=2^n$ copies of a BEC, regardless of the inputs of previous synthetic channels. We show this in Appendix~\ref{ap_IMJP for BEC}. 
	\end{remark}

\subsubsection{Proper Degrading Channels}
The IMJP decoder is attractive for joint polar synthetic channels since, by \Cref{thm_minimizing phi_a and phi_b for polar channel}, we can efficiently compute it. This was made possible by the successive form of the joint channel~\eqref{eq_Wab and its relationship to Wb}. Thus, we seek degrading channels that maintain this form. 

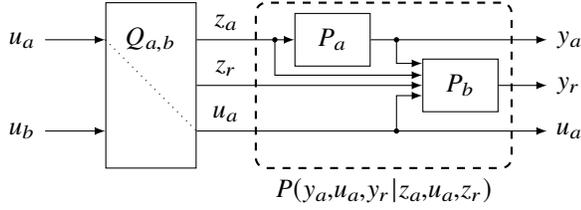
\begin{figure}[t] 
	\centering
	\begin{tikzpicture}[>=latex]	
	\node (Q) [rectangle, draw, minimum height = 2.2 cm, minimum width = 1.2cm] {}; 	
	\node (Pa) [right = 1.3cm of Q.45, rectangle, draw, minimum width = 1cm, minimum height = 0.7cm] {$P_a$};
	\node (Pb) [right = 3 cm of Q, rectangle, draw, minimum width = 1cm, minimum height = 0.7cm] {$P_b$}; 
 
	\node at (Q.center|-Pa) {$Q_{a,b}$};
	
	\draw[<-] (Q.135) -- +(-0.8,0) node[left] {$u_a$}; 
	\draw[<-] (Q.225) -- +(-0.8,0) node[left] {$u_b$}; 
	\draw[dotted] (Q.135) -- (Q.-45); 
	
	\draw[->] (Pa) -- ($(Pa)!0.5!(Pa-|Pb)$) node (PaPb){} -- +(2,0) node(ya){} node [right] {$y_a$};
	\fill (PaPb) circle (1pt); 
	\draw[->] (PaPb.center) |- (Pb.150); 
	\draw[->] (Pb) -- (ya|-Pb) node[right] {$y_r$}; 
	
	\node at ($(Q-|Pa)!0.4!(Pb)$) [rounded corners, dashed, thick, draw, minimum height = 2.2 cm, minimum width = 3.4cm, label=below:$P(y_a{,}u_a{,}y_r|z_a{,}u_a{,}z_r)$] {};
	\draw[->] (Q.45) -- node[pos = 0.3, above] (za) {$z_a$} node[pos=0.8] (QPa) {} (Pa); 
	\draw[->] (Q) -- (Q-|za) node[above] {$z_r$} -- (Pb); 
	\draw[->] (Q.-45) -- (Q.-45-|za) node[above] {$u_a$} -- (ya|-Q.-45) node [right] {$u_a$}; 
	\fill (QPa) circle (1pt); 
	
	\draw[->] (QPa.center) |- (Pb.165); 

	\fill (PaPb |- Q.-45) circle (1pt);
	\draw[->] (PaPb|-Q.-45) |-(Pb.195);
	\end{tikzpicture}
	\caption{The structure of proper degrading channels.}
	\label{fig_joint degrading channel}
\end{figure}

Let $W_{a,b}(y_a,y_b|u_a,u_b)$ be a joint distribution of two synthetic channels and let $Q_{a,b}(z_a,z_b|u_a,u_b) \succcurlyeq W_{a,b}(y_a,y_b|u_a,u_b)$. The marginal channels of $Q_{a,b}$ are $Q_a(z_a|u_a)$ and $Q_b(z_b|u_b)$. The most general degrading channel is of the form
\[ P(y_a,y_b|z_a,z_b) = P_1(y_a|z_a,z_b)\cdot P_2(y_b|z_a,z_b, y_a), \]
where $P_1$ and $P_2$ are probability distributions. 
This form does not preserve the successive structure of joint synthetic channels~\eqref{eq_Wab and its relationship to Wb}. Even if $Q_{a,b}$ satisfies~\eqref{eq_Wab and its relationship to Wb}, the resulting $W_{a,b}$ may not. To this end, we turn to a subset of degrading channels. Recalling that $y_b = (y_a,u_a,y_{\ba})$, we consider degrading channels of the form 
\iftoggle{twocol}{
\begin{equation}
\label{eq_proper form of P} 
\begin{split} &
P(y_a,u_a,y_{\ba}| z_a, u_a,z_{\ba})
\\&\quad = P_a(y_a|z_a) \cdot P_b(y_{\ba} | z_a, u_a, z_{\ba}, y_a).
\end{split}
\end{equation}}
{
\begin{equation}
\label{eq_proper form of P} 
P(y_a,u_a,y_{\ba}| z_a, u_a,z_{\ba})
 = P_a(y_a|z_a) \cdot P_b(y_{\ba} | z_a, u_a, z_{\ba}, y_a).
\end{equation}
}
That is, these degrading channels degrade $z_a$, the output of $Q_a$, to $y_a$, pass $u_a$ unchanged, and degrade $z_{\ba}$, the remainder of $Q_b$'s output, to $y_{\ba}$. For this to be a valid channel, $P_a$ and $P_b$ must be probability distributions. 
This degrading channel structure is illustrated in \Cref{fig_joint degrading channel}. By construction,  degrading channels of the form~\eqref{eq_proper form of P} preserve the form~\eqref{eq_Wab and its relationship to Wb} that is required for efficiently computing the IMJP decoder as in \Cref{thm_minimizing phi_a and phi_b for polar channel}. 
\begin{definition}[Proper degrading channels]
	A degrading channel of the form~\eqref{eq_proper form of P} is called \emph{proper}. We write $Q \properup W$ to denote that channel $Q$ is upgraded from $W$ with a proper degrading channel. We say that an upgrading (degrading) procedure is proper if its degrading channel is proper. 
\end{definition}
	  
By marginalizing the joint channel it is straight-forward to deduce the following for joint synthetic channel distributions.  
\begin{lemma} \label{lem_Qa and Qb are degraded from Wa and Wb}
	If $Q_{a,b}(z_a,u_a,z_{\ba}|u_a,u_b) \properup W_{a,b}(y_a,u_a,y_{\ba}|u_a,u_b)$, 
	then $Q_a(z_a|u_a) \succcurlyeq W_a(y_a|u_a)$ and $Q_b(z_a,u_a,z_{\ba}|u_b) \succcurlyeq W_b(y_a,u_a,y_{\ba}|u_b)$.
\end{lemma}
This lemma is encouraging, but  insufficient for our purposes. It is easy to take degrading channels that are used for degrading a single (not joint) synthetic channel and cast them into a proper degrading channel for joint channels. This, however, is not our goal. Instead, we start with $W_{a,b}$ and seek an \emph{upgraded} $Q_{a,b}$ with smaller output alphabet that can be degraded to $W_{a,b}$ using a proper degrading channel. This is a very different problem than the degrading one, and its solution is not immediately apparent. Plain-vanilla attempts to use upgrading procedures for single channels fail to produce the desired results.  
	Later, we  develop proper upgrading procedures that upgrade one of the marginals without changing the other. 

We now show that the probability of error of the IMJP decoder does not decrease after degradation by proper degrading channels. Intuitively, this is because the decoder for the original channel can simulate the degrading channel. We denote by $\mathcal{E}^W_a$ the error event of channel $W_a$ under some decoder $\phi_a$, and similarly define $\mathcal{E}^Q_a$, $\mathcal{E}^W_b$, and $\mathcal{E}^Q_b$. Further, we denote by $\phi_{i}$ decoders for $W_i$ and by $\psi_i$ decoders for $Q_i$, $i=a,b$.

\begin{lemma}\label{lem_conditions on degrading for conservation of pe}
	Let joint channel $W_{a,b}(y_a,u_a,y_{\ba}|u_a,u_b)$ have marginals $W_a(y_a|u_a)$ and $W_b(y_a,u_a,y_{\ba}|u_b)$. Assume that $Q_{a,b}(z_a,u_a,z_{\ba}|u_a,u_b) \properup W_{a,b}(y_a,u_a,y_{\ba}|u_a,u_b)$, then $\min_{\psi_a,\psi_b} \Prob{\mathcal{E}^Q_a\cup\mathcal{E}^Q_b} \leq  \min_{\phi_a,\phi_b} \Prob{\mathcal{E}^W_a\cup\mathcal{E}^W_b}$.
\end{lemma} 
\begin{IEEEproof}
	The proof follows by noting that for any decoder $\phi_i$, $i=a,b$ we can find a decoder $\psi_i$ with identical performance. First consider the decoder for channel $a$. Denote by $\arg P_a(y_a|z_a)$ the result of drawing $y_a$ with probability $P_a(\cdot|z_a)$. Then, the decoder $\psi_a$ for $Q_a$, defined as $\psi_a(z_a) = \phi_a(\arg P_a(y_a|z_a))$, has performance identical to $\phi_a$ for $W_a$. The decoder $\psi_a$ results from first degrading the a-channel output and only then decoding. Next, consider the decoder for the b-channel. Denote by $\arg P_b(y_{\ba}|z_a, u_a,z_{\ba},y_a)$ the result of drawing $y_{\ba}$ with probability $P_b(\cdot|z_a,u_a,z_{\ba},y_a)$. Then, similar to the a-channel case, the decoder $\psi_b$ for $Q_b$, defined as $\psi_b(z_a,u_a,z_{\ba}) = \phi_b(\arg P_a(y_a|z_a), u_a,\arg P_b(y_{\ba}|z_a, u_a,z_{\ba},y_a))$, has performance identical to $\phi_b$ for $W_b$. Hence, the best decoder pair $\psi_a,\psi_b$ cannot do worse than the best decoder pair $\phi_a,\phi_b$. 
\end{IEEEproof}

Let $W$ be a BMS channel that undergoes $n$ polarization steps. The probability of error of a polar code with non-frozen set $\mathcal{A}$ under SC decoding is given by $P_e^{\SC}(W) = \Prob{\bigcup_{a \in \mathcal{A}} \mathcal{E}_a^{\ML}},$ where $\mathcal{E}_a^{\ML}$ is the error probability of synthetic channel $W_a$ under ML decoding. Obviously, for any $\mathcal{A}' \subseteq \mathcal{A}$, 
\begin{equation} P_e^{\SC}(W) \geq \Prob{\bigcup_{a \in \mathcal{A}'} \mathcal{E}_a^{\ML}}.\label{eq_lower bound on a union using a subset}\end{equation}  
We have already mentioned the simplest such lower bound, $P_e^{\SC}(W) \geq \max_{a \in \mathcal{A}} \Prob{\mathcal{E}_a^{\ML}}$. We now show that the IMJP decoder provides a tighter lower bound.   
To this end, recall that  
$ P_e^{\IMJP}(W_{a,b}) = \min_{\phi_a, \phi_b} \Prob{\mathcal{E}_a \cup \mathcal{E}_b},$ where
$\mathcal{E}_i$ is the probability of error of channel $i$ under decoder $\phi_i$, $i=a,b$.
\begin{lemma} \label{lem_IMJP provides a tighter lower bound than max Pe}
	Let $W$ be a BMS channel that undergoes $n$ polarization steps, and let $\mathcal{A}$ be the
    non-frozen set. Then, 
\begin{equation} P_e^{\SC}(W) \geq \max_{a,b \in \mathcal{A}} P_e^{\IMJP}(W_{a,b}) \geq \max_{a \in \mathcal{A}}\Prob{\mathcal{E}_a^{\ML}}. \label{eq_string of Pe inequalities}\end{equation} 
\end{lemma}
\begin{IEEEproof}
	Using~\eqref{eq_lower bound on a union using a subset}, $P_e^{\SC}(W) \geq \max_{a,b\in\mathcal{A}} \Prob{\mathcal{E}_a^{\ML} \cup \mathcal{E}_b^{\ML}}$. By definition, the IMJP decoder seeks decoders $\phi_a$ and $\phi_b$ that minimize the joint probability of error of synthetic channels with indices $a$ and $b$. Therefore, for any two indices $a$ and $b$ we have 
	$\Prob{\mathcal{E}_a^{\ML} \cup \mathcal{E}_b^{\ML}} \geq P_e^{\IMJP}(W_{a,b}).$ In particular, this holds for the indices $a,b$ that maximize the right-hand side.  This establishes the leftmost inequality of~\eqref{eq_string of Pe inequalities}. 
 
To establish the rightmost inequality of~\eqref{eq_string of Pe inequalities}, we first show that for any $a,b$,
\begin{equation} P_e^{\IMJP}(W_{a,b}) \geq \max\{\Prob{\mathcal{E}_{a}^{\ML}},\Prob{\mathcal{E}_{b}^{\ML}}\}.\label{eq_IMJP better than marginals} \end{equation} 
To see this, first recall that the IMJP decoder performs ML decoding on the b-channel, yielding $ P_e^{\IMJP}(W_{a,b}) \geq \Prob{\mathcal{E}_{b}^{\ML}}$. Next, we construct $W'_{a,b} \properup W_{a,b}$ in which the b-channel is noiseless, by augmenting the $y_{\ba}$ portion of the output of $W_{a,b}$ with $u_b$, i.e., 
\iftoggle{twocol}{
\begin{multline*}
	 W_{a,b}'(y_a,u_a, (y_{\ba}, v_b)|u_{a},u_b) \\=  W_{a,b}(y_a,u_a,y_{\ba}| u_{a},u_b) \kindi{v_b = u_b}.\end{multline*}}
{\[	 W_{a,b}'(y_a,u_a, (y_{\ba}, v_b)|u_{a},u_b) =  W_{a,b}(y_a,u_a,y_{\ba}| u_{a},u_b) \kindi{v_b = u_b}.\]}
Channel $W'_{a,b}$ can be degraded to $W_{a,b}$ using a proper degrading channel by omitting $v_b$ from the $y_{\ba}$ portion of the output and leaving $y_a$ unchanged.
 Thus, 
$P_e^{\IMJP}(W_{a,b}) \geq P_e^{\IMJP}(W'_{a,b}) = \Prob{\mathcal{E}_{a}^{\ML}}$. 
	
	Finally, denote $a_0  = \argmax_{a \in \mathcal{A}} \Prob{\mathcal{E}_a^{\ML}}$. By~\eqref{eq_IMJP better than marginals}, for any $c>a_0>d$ we have $P_e^{\IMJP}(W_{a_0,c}) \geq \Prob{\mathcal{E}_{a_0}^{\ML}}$ and $P_e^{\IMJP}(W_{d,a_0}) \geq \Prob{\mathcal{E}_{a_0}^{\ML}}$. Since $\max_{a,b \in \mathcal{A}} P_e^{\IMJP}(W_{a,b}) \geq \max_{c,d}\{P_e^{\IMJP}(W_{a_0,c}),P_e^{\IMJP}(W_{d,a_0})\}$ we obtain the proof.  
\end{IEEEproof}

\Cref{lem_conditions on degrading for conservation of pe,lem_IMJP provides a tighter lower bound than max Pe} are instrumental for our lower bound, which combines upgrading operations and the IMJP decoder.  

\section{Properties of Joint Synthetic Channels}	\label{sec_properties of joint bit channels}
In this section, we study the properties of joint synthetic channels. We begin by bringing the joint synthetic channel into an equivalent form where the b-channel's ML decision is immediately apparent. We then explain how to jointly polarize synthetic channels. Finally, we describe some consequences of symmetry on joint channels and on the IMJP decoder. 

\subsection{Representation of Joint Synthetic Channel Distribution using $D$-values} \label{subsec_d value representation}
Two channels $W$ and $W'$ with the same input alphabet but possibly different output alphabets  are called \emph{equivalent} if $W\succcurlyeq W'$ and $W'\succcurlyeq W$. We denote this by $W \equiv W'$. Channel equivalence can cast a channel in a more convenient form. For example, if $W$ is a BMS, one can transform it to an equivalent channel whose output is a sufficient statistic, such as a $D$-value (see Appendix~\ref{ap_Definition of D values}), in which case the ML decoder's decision is immediately apparent. 

Let $W_{a,b}(y_a,u_a,y_{\ba}|u_a,u_b)$ be a joint synthetic channel. Since the joint distribution is
determined by the distribution of $W_b$,  we can transform $W_{a,b}$ to an equivalent channel in
which the b-channel $D$-value\footnote{By ``b-channel $D$-value'' we mean the $D$-value computed for
channel $W_b$. Instead of $D$-values, other sufficient statistics of the b-channel could have been
used. In fact, for practical implementation (see \Cref{sec_implementation}), we recommend to use likelihood ratios, which offer
a superior dynamic range. Our use of $D$-values in the exposition was prompted by their bounded
range: $[-1,1]$. This simplifies many of the expressions that follow.} of
symbol $(y_a,u_a,y_{\ba})$ is immediately apparent. 

\begin{definition}[$D$-value representation] Joint channel $W_{a,b}(y_a,u_a,d_b|u_a,u_b)$ is in $D$-value representation if the marginal $W_b$ satisfies
\[ d_b = \frac{W_{b}(y_a,u_a,d_b|0) - W_{b}(y_a,u_a,d_b|1)}{W_{b}(y_a,u_a,d_b|0)+W_{b}(y_a,u_a,d_b|1)}.\] 
\end{definition}

We use the same notation $W_{a,b}$ for both the regular and the $D$-value representations of the joint channel due to their equivalence. The discussion of the various representations of joint channels in \Cref{subsec_Polar Coding Setting} applies here as well. In particular, we will frequently use $W_b(y_a,u_a,d_b|u_b)$ to denote the joint synthetic channel distribution. 

The following lemma affords a more convenient description of the joint channel, in which, in line with the IMJP decoder, the b-channel's ML decision is immediately apparent. Moreover, this description greatly simplifies the expressions that follow. 
\begin{lemma} Channels $W_{a,b}(y_a,u_a,y_{\ba}|u_a,u_b)$ and $W_{a,b}(y_a,u_a,d_b|u_a,u_b)$ are equivalent and the degrading channels from one to the other are proper. \label{lem_Representation of Joint Bit-Channel Distribution using $D$-values}
\end{lemma}
\begin{IEEEproof}
	To establish equivalence we show that each channel is degraded from the other using proper degrading channels. The only portion of interest in~\eqref{eq_proper form of P} is $P_b$, as in either direction $y_a$ and $u_a$ are unchanged by the degrading channel. Denote by $\Dyauadb$ the set of all symbols $y_{\ba}$ such that the b-channel $D$-value of $(y_a,u_a,y_{\ba})$ is $d_b$, for fixed $y_a,u_a$. 
	Then, 
\iftoggle{twocol}{
		\begin{align*}
		&W_{a,b}(y_a,u_a,d_b|u_a,u_b)\\ &\quad = \sum_{\mathclap{\Dyauadb}} W_{a,b}(y_a,u_a,y_{\ba}|u_a,u_b) \\ &\quad = \sum_{y_{\ba}} W_{a,b}(y_a,u_a,y_{\ba}|u_a,u_b)\cdot P_b(d_b|y_{\ba},y_a,u_a),
		\end{align*}}{
		\begin{align*}
		W_{a,b}(y_a,u_a,d_b|u_a,u_b) & = \sum_{\mathclap{\Dyauadb}} W_{a,b}(y_a,u_a,y_{\ba}|u_a,u_b) \\ &= \sum_{y_{\ba}} W_{a,b}(y_a,u_a,y_{\ba}|u_a,u_b)\cdot P_b(d_b|y_{\ba},y_a,u_a),
		\end{align*}}
	where
\[	P_b(d_b|y_{\ba},y_a,u_a) = \kindi{y_{\ba} \in \Dyauadb}. \]
	Clearly, the b-channel $D$-value of $(y_a,u_a,d_b)$ is $d_b$. 
	
	On the other hand, by~\eqref{eq_Wab and its relationship to Wb} and since all symbols in $\Dyauadb$ share the same b-channel $D$-value,
\iftoggle{twocol}{
	\begin{multline*}
		W_{a,b}(y_a,u_a,y_{\ba}|u_a,u_b)\\ = \sum_{d_b} W_{a,b}(y_a,u_a,d_b|u_a,u_b)\cdot P_b'(y_{\ba}|d_b,y_a,u_a),
		\end{multline*}}{
\[	W_{a,b}(y_a,u_a,y_{\ba}|u_a,u_b) = \sum_{d_b} W_{a,b}(y_a,u_a,d_b|u_a,u_b)\cdot P_b'(y_{\ba}|d_b,y_a,u_a),\]}
	where
	\[ P_b'(y_{\ba}|d_b,y_a,u_a) =  \frac{W_b(y_a,u_a,y_{\ba})}{\displaystyle \sum_{\mathclap{\Dyauadb}}W_b(y_a,u_a,y_{\ba})} \kindi{y_{\ba} \in \Dyauadb},\]
	and $W_b(y_a,u_a,y_{\ba}) = \frac{1}{2}\sum_{u_b} W_b(y_a,u_a,y_{\ba}|u_b)$.
\end{IEEEproof}

\begin{remark}
In  \Cref{sec_polarization for joint bit channels} we will show how to jointly polarize a joint channel $W_{a,b}$. Even if $W_{a,b}$ is given in $D$-value representation, the jointly polarized version is not. However, this lemma enables us to convert the jointly polarized distribution to $D$-value representation.
This is possible because \Cref{lem_Representation of Joint Bit-Channel Distribution using $D$-values} holds for any representation of $W_{a,b}(y_a,u_a,y_{\ba}|u_a,u_b)$ in which $u_a, y_a$ are the input and output, respectively, of the a-channel, $u_b$ is the input of the b-channel, and $(y_a,u_a,y_{\ba})$ is the output of the b-channel. In particular, $y_{\ba}$ need not consist of inputs to channels $W_{a+1}, \ldots, W_{b-1}$.  
\end{remark} 

\begin{remark}
	At this point the reader may wonder why we have stopped here and not converted the a-channel output to its $D$-value. The reason is that this constitutes a degrading operation, which is the opposite of what we need. Two a-channel symbols with the same a-channel $D$-value may have very different meanings for the IMJP decoder. Thus, we cannot combine them to a single symbol without incurring loss.  
\end{remark}

When the joint channel is in $D$-value representation, proper degrading channels admit the form
\begin{equation}\label{eq_proper form of degrading channels, d version}
P(y_a,u_a,d_b|z_a,u_a,z_b) = P_a(y_a|z_a) P_b(d_b|z_a,y_a,u_a,z_b).
\end{equation}
It is obvious that all properties obtained from degrading channels of the form~\eqref{eq_proper form of P} are retained for degrading channels of the form~\eqref{eq_proper form of degrading channels, d version}. 
By \Cref{lem_Representation of Joint Bit-Channel Distribution using $D$-values}, we may assume that the degraded channel is also in $D$-value representation.

\subsection{Polarization for Joint Synthetic Channels}\label{sec_polarization for joint bit channels} 
Let $W_{a,b}(y_a,u_a,d_b|u_a,u_b)$ be some joint synthetic channel distribution in $D$-value representation. Recall that $a$ and $b$ are indices of synthetic channels. For $\alpha ,\beta \in\{-,+\}$, we denote by $a^{\alpha}$ and $b^{\beta}$ the indices of the synthetic channels that result from polar transforms of $W_a$ and $W_b$ according to $\alpha$ and $\beta$. That is, 
\[ a^{\alpha} = \begin{cases} 2a-1, & \alpha = - \\ 2a, & \alpha = + \end{cases}\] and a similar relationship holds for $b^{\beta}$. 
The resulting joint channel is, thus,  $W_{a^{\alpha},b^{\beta}}$.


 Even though $W_{a,b}$ is in $D$-value representation, after a polarization transform this is no longer the case.  Of course, one can always bring the polarized joint channel to an equivalent $D$-value representation as in \Cref{lem_Representation of Joint Bit-Channel Distribution using $D$-values}.
 
\begin{figure}[t]
\centering
		\begin{tikzpicture}[>=latex,
			channel/.style = {rectangle, minimum height = 1.5cm, minimum width = 2cm, draw},
			sumnode/.style = {circle, inner sep = 0pt, minimum size = 5mm, draw},
			fulldot/.style = {circle, fill, inner sep = 0pt, minimum size = 3pt}]
		\node[channel] (W1) {$W_{a,b}$}; 
		\node[channel, below = 0.6cm of W1] (W2) {$W_{a,b}$};

		\coordinate (W1top) at ($(W1.west)!0.5!(W1.north west)$); 
		\coordinate (W1bot) at ($(W1.west)!0.5!(W1.south west)$);
		\coordinate (W2top) at ($(W2.west)!0.5!(W2.north west)$); 
		\coordinate (W2bot) at ($(W2.west)!0.5!(W2.south west)$);
	
		\node[sumnode, left = 0.7 cm of W1top] (A1) {$+$};
		\node[sumnode, left = 1.3 cm of W1bot] (A2) {$+$};

		\coordinate (topdot) at (A1|-W2top);
		\coordinate (botdot) at (A2|-W2bot);

		\coordinate[left = of A2] (uat);
		\coordinate (A2r) at ($(uat)!0.7!(A2.west)$) ;
		\coordinate (A2l) at ($(uat)!0.3!(A2.west)$) ;		
		
		\draw[-{Triangle[open]}, thick] (topdot) -- (A1); 
		\draw[->] (botdot) -- (A2);
		
		\draw[white, line width = 2pt] ($(A2.east)+(0.1,0)$) -- (W1bot); 
		\draw[-{Triangle[open]}, thick] (A1) -- (W1top); 
		\draw[->] (A2) -- (W1bot);
 
		\draw[-{Triangle[open]}, thick] (uat|-A1)node[left]{$u_a$} -- (A1); 
		\draw[->] (uat|-W2bot) node[left]{$\nu_b$}  -- (W2bot); 
		
		\draw[white, line width = 2pt] ($(A2r|-topdot)+(0.2,0)$) -- ($(topdot)+(-0.2,0)$); 
		\draw[-{Triangle[open]}, thick] (uat) node[left]{$\nu_a$} -- (A2l) -- (A2r|-topdot) -- (W2top); 	
		\draw[white, line width = 2pt] (A2l|-W2top) -- (A2r); 
		\draw[->] (uat|-W2top) node[left]{$u_b$} -- (A2l|-W2top) -- (A2r) -- (A2); 

		\coordinate (W2r2) at ($(W2.south east)!0.14!(W2.north east)$);
		\coordinate (W2r4) at ($(W2.south east)!0.38!(W2.north east)$);
		\coordinate (W2r6) at ($(W2.south east)!0.62!(W2.north east)$);
		\coordinate (W2r8) at ($(W2.south east)!0.86!(W2.north east)$);
		\coordinate[right = 1.3cm of W2r2] (dout);
		
		\coordinate (W1r2) at ($(W1.south east)!0.2!(W1.north east)$);
		\coordinate (W1r6) at ($(W1.south east)!0.5!(W1.north east)$);
		\coordinate (W1r8) at ($(W1.south east)!0.8!(W1.north east)$);

		\path[name path = l1] (W1r2) -- ++(0.3,0) -- ++(-75:2.5cm);
		\path[name path = l2] (dout|-W2r4) -- ++(-1,0);
		\path[name intersections={of = l1 and l2}]; 
		\coordinate (a) at (intersection-1); 
		\draw[->] (W1r2) -- ++(0.3,0) -- (a) -- (dout|-W2r4) node[right]{$d_b$}; 

		\path[name path = l1] (W1r6) -- ++(0.4,0) -- ++(-75:2.5cm);
		\path[name path = l2] (dout|-W2r8) -- ++(-1,0);
		\path[name intersections={of = l1 and l2}]; 
		\coordinate (b) at (intersection-1); 
		\draw[->] (W1r6) -- ++(0.4,0) -- (b) -- (dout|-W2r8) node[right]{$u_a\oplus \nu_a$}; 
		
		\draw[white, line width = 2pt] ($(W2r8)+(0.3,0)$) -- (W1r2-|b); 
		\draw[-{Triangle[open]}, thick] (W2r8) -- ++(0.3,0) -- (W1r2-|b) -- (dout|-W1r2) node[right]{$\eta_a$};
		
		\draw[-{Triangle[open]}, thick] (W1r8) -- (dout|-W1r8) node[right]{$y_a$}; 

		\draw[->] (W2r2) -- (dout) node[right]{$\delta_b$};
		\draw[white, line width = 2pt] ($(W2r6)+(0.3,0)$) -- (dout|-W2r6); 
		\draw[->] (W2r6) -- (dout|-W2r6) node[right]{$\nu_a$};
		
		\node[fulldot] at (topdot) {};
		\node[fulldot] at (botdot) {};

		\end{tikzpicture}
		\caption{Two independent copies of joint channel $W_{a,b}$ combined using a $(u \oplus v, v)$ construction. 
		The a-channel input and output for each copy are denoted using thicker arrows with hollow tips.} 
		\label{fig_joint channel polarization_db version}
\end{figure}
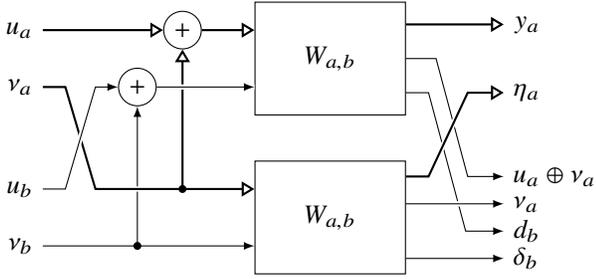

 The polar construction is shown in \Cref{fig_joint channel polarization_db version}. Here, two independent copies of the joint channel $W_{a,b}$ (in $D$-value representation) are combined. The inputs and outputs of the a-channel of each copy are denoted explicitly using thicker arrows with hollow tips (\tikz{\draw[-{Triangle[open]}, thick] (0,0) -- (0.5,0)}). For example, for the bottom copy of $W_{a,b}$, the a-input is $\nu_a$ and the a-output is $\eta_a$, whereas the b-input is $(\nu_b)$ and the b-output is $(\eta_a,\nu_a, \delta_b)$. 

The input $u_{a^{\alpha}}$ and output $y_{a^{\alpha}}$ of $W_{a^{\alpha}}$ are given by 
\begin{align*}
	u_{a^{\alpha}} &= \begin{cases}
		u_a, & \alpha = - \\ \nu_a, & \alpha = +,
	\end{cases} \\ 	
	y_{a^{\alpha}} &= \begin{cases}
		(y_a,\eta_a), & \alpha = - \\ (y_a,\eta_a,u_a), & \alpha = +.
	\end{cases}
\end{align*}
The input $u_{b^{\beta}}$ and output $y_{b^{\beta}}$ of $W_{b^{\beta}}$ are given by 
\begin{align*}
	u_{b^{\beta}} &= \begin{cases}
		u_b, & \beta = - \\ \nu_b, & \beta = +,
	\end{cases}\\y_{b^{\beta}} &= \begin{cases}
						(y_a, \eta_a, u_a, \nu_a, d_b,\delta_b), & \beta = - \\
					    (y_a, \eta_a, u_a, \nu_a, u_b,d_b,\delta_b), & \beta = +.
	\end{cases}
\end{align*}
Note that $y_{a^{\alpha}}$ and $u_{a^{\alpha}}$ are contained in $y_{b^{\beta}}$. That is, 
$ y_{b^{\beta}} = (y_{a^{\alpha}}, u_{a^{\alpha}}, y_r)$, where
\[ y_r = \begin{cases} \nu_a, d_b, \delta_b,        & \alpha = -, \beta = - \\
 					   \nu_a, u_b,   d_b, \delta_b, & \alpha = -, \beta = +\\
 					   d_b, \delta_b,               & \alpha = +, \beta = -\\
 					   u_b, d_b, \delta_b,        & \alpha = +, \beta = +.	
 \end{cases}\]
Thus, the joint output of both channels is $y_{b^{\beta}}$.

The distribution of the jointly polarized channel is given by 
\iftoggle{twocol}{
\begin{equation}
\begin{split}
		&W_{a^{\alpha},b^{\beta}}(y_{a^{\alpha}},y_{b^{\beta}}|u_{a^{\alpha}},u_{b^{\beta}})\\ 
		&\quad= 2 W_{b^{\beta}}(y_{b^{\beta}}|u_{b^{\beta}}) \kindi{y_{b^{\beta}} = (y_{a^{\alpha}},u_{a^{\alpha}},y_r)}\\
		& \quad= \sum_{B_{\beta}} \Big( W_b(y_a,u_a\oplus \nu_a, d_b|u_b \oplus \nu_b) W_{b}(\eta_a,\nu_a, \delta_b|\nu_b)\Big),
\end{split}\label{eq_polarizing Wab, general db version}
\end{equation}}{
\begin{equation}
\begin{split}
		W_{a^{\alpha},b^{\beta}}(y_{a^{\alpha}},y_{b^{\beta}}|u_{a^{\alpha}},u_{b^{\beta}}) 
		&= 2 W_{b^{\beta}}(y_{b^{\beta}}|u_{b^{\beta}}) \kindi{y_{b^{\beta}} = (y_{a^{\alpha}},u_{a^{\alpha}},y_r)} \\
		&= \sum_{B_{\beta}} \Big( W_b(y_a,u_a\oplus \nu_a, d_b|u_b \oplus \nu_b) W_{b}(\eta_a,\nu_a, \delta_b|\nu_b)\Big),
\end{split}\label{eq_polarizing Wab, general db version}
\end{equation}}
where
\[\sum_{B_{\beta}} \equiv \begin{dcases}  \sum_{\nu_b}, & \beta = - \\ \text{No sum}, & \beta = +.  	\end{dcases}\] 

We have shown how to generate $W_{a^{\alpha},b^{\beta}}$ from $W_{a,b}$. Another case of interest is generating $W_{a^-,a^+}$ from $W_a$. Denote the output of $W_{a^-}$ by $y_{a^-}$. The output of $W_{a^+}$ is $(y_{a^-},u_a)$. From~\eqref{eq_Wab and its relationship to Wb}, we need only compute $W_{a^+}$ to find $W_{a^-,a^+}$. This is accomplished by~\eqref{eq_plus transform}.  

If two channels are ordered by degradation, so are their polar transforms~\cite[Lemma 4.7]{korada}. That is, if $Q \succcurlyeq W$ then $Q^- \succcurlyeq W^-$ and $Q^+ \succcurlyeq W^+$. This is readily extended to joint channels. To this end, for BMS channel $W$ we denote the joint channel formed by its `$-$'- and `$+$'-transforms by $W_{-,+}$.  
\begin{lemma}\label{lem_degradation is preserved for joint distributions-+}
	Let BMS channel $Q \succcurlyeq W$. Then $Q_{-,+} \properup W_{-,+}$. 
\end{lemma}
\begin{IEEEproof}
	Using~\eqref{eq_definition of degradation} and the definition of $W_{-,+}$ we have 
\iftoggle{twocol}{
	\begin{align*}
		& W_{-,+}((y_1,y_2),u_1|u_1,u_2) \\
        &\quad = 2W_{+}((y_1,y_2),u_1|u_2) \\ 
									     &\quad =W(y_1|u_1\oplus u_2)W(y_2|u_2)\\
									     &\quad= \sum_{z_1,z_2} Q(z_1|u_1\oplus u_2)P(y_1|z_1) Q(z_2|u_2)P(y_2|z_2)\\
									     &\quad= \sum_{z_1,z_2} Q_{-,+}((z_1,z_2),u_1|u_1,u_2)P_a(y_1,y_2|z_1,z_2),
	\end{align*}}{
	\begin{align*}
		 W_{-,+}((y_1,y_2),u_1|u_1,u_2) 
									     & = \frac{1}{2}W(y_1|u_1\oplus u_2)W(y_2|u_2)\\
									     &= \sum_{z_1,z_2} \frac{1}{2}Q(z_1|u_1\oplus u_2)P(y_1|z_1) Q(z_2|u_2)P(y_2|z_2)\\
									     &= \sum_{z_1,z_2} Q_{-,+}((z_1,z_2),u_1|u_1,u_2)P_a(y_1,y_2|z_1,z_2),
	\end{align*}}
where $P_a(y_1,y_2|z_1,z_2) = P(y_1|z_1)P(y_2|z_2)$ is a proper degrading channel. 
\end{IEEEproof}
\begin{lemma} \label{lem_degradation is preserved for joint distributions}
	If $Q_{a,b}(z_a,z_b|u_a,u_b) \properup W_{a,b}(y_a,y_b|u_a,u_b)$, then, for $\alpha,\beta \in \{-,+\}$, $ Q_{a^{\alpha},b^{\beta}} \properup W_{a^{\alpha},b^{\beta}}$. 
\end{lemma}

\begin{IEEEproof}
The proof follows similar lines to the proof of \Cref{lem_degradation is preserved for joint distributions-+}. Expand $W_{a^{\alpha},b^{\beta}}$ using~\eqref{eq_polarizing Wab, general db version} and expand again using the definition of joint degradation with a proper degrading channel. Using the one-to-one mappings between the outputs of the polarized channels and the inputs and outputs of non-polarized channels, the desired results are obtained. The details are mostly technical, and are omitted.  
\end{IEEEproof}

The operational meaning of \Cref{lem_degradation is preserved for joint distributions} is that to compute an upgraded approximation of $W_{a^{\alpha},b^{\beta}}$ we may start with $Q_{a,b}$, an upgraded approximation of $W_{a,b}$, and polarize it. The result $Q_{a^{\alpha},b^{\beta}}$ is an upgraded approximation of $W_{a^{\alpha},b^{\beta}}$. This enables us to iteratively compute upgraded approximations of joint synthetic channels. Whenever the joint synthetic channel exceeds an allotted size, we upgrade it to a joint channel with a smaller alphabet size and continue from there. We make sure to use proper upgrading procedures; this preserves the special structure of the joint channel and enables us to compute a lower bound on the probability of error. In \Cref{sec_upgrading procedures for joint bit channels} we  derive such upgrading procedures. 

Since a sequence of polarization and proper upgrading steps is equivalent to proper upgrading of the overall polarized joint channel, using \Cref{lem_conditions on degrading for conservation of pe,lem_IMJP provides a tighter lower bound than max Pe} we  obtain that the IMJP decoding error of a joint channel that has undergone multiple polarization and proper upgrading steps lower-bounds the SC decoding error of the joint channel that has undergone only the same polarization steps (without upgrading steps).

\subsection{Double Symmetry for Joint Channels}
A binary input channel $W(y|u)$ is called \emph{symmetric} if for every output $y$ there exists a conjugate output $\bar{y}$ such that $W(y|0) = W(\bar{y}|1)$. We now extend this to joint synthetic channels. 

\begin{definition}[Double symmetry] \label{def_double symmetry db version}
	Joint channel $W_{b}(y_a,u_a,d_b|u_b)$ exhibits double symmetry if for every $y_a$, $d_b$ there exist $\aconj{y}_a$, $\bconj{y}_a$,  $\abconj{y}_a$ such that 
\begin{equation}
		\begin{split}
			W_b(y_a,u_a,d_b|u_b) &= W_b(\aconj{y}_a,\bar{u}_a,d_b|u_b)\\
			&= W_b(\bconj{y}_a,u_a,-d_b|\bar{u}_b) \\&= W_b(\abconj{y}_a,\bar{u}_a,-d_b|\bar{u}_b).
		\end{split}  
	\label{eq_definition of double symmetry for joint channels (Wb db version)}
\end{equation}
\end{definition}
We call $\aconj{(\cdot)}$ the a-conjugate; $\bconj{(\cdot)}$ the b-conjugate; and $\abconj{(\cdot)}$ the ab-conjugate. We can also cast this definition using the regular (non-$D$-value) representation of joint channels in a straight-forward manner, which we omit here. 

\begin{example}\label{ex_conjugates for W-+}
	Let $W$ be a BMS channel and denote by $W_{-,+}$ the joint channel formed by its `$-$'- and `$+$'-transforms. What are the a-, b-, and ab-conjugates of the a-channel output $y_a$? Recall that the output of the a-channel $W^-$ consists of the outputs of two copies of $W$. Denote $y_a = (y_1,y_2)$, where $y_1$ and $y_2$ are two possible outputs of $W$ with conjugates $\bar{y}_1, \bar{y}_2$, respectively. We then have
	\begin{align*} W_{-,+}(y_a,u_a|u_a,u_b) &= 2W^+(y_a,u_a|u_b) \\ &= W(y_1|u_a\oplus u_b) W(y_2|u_b). \end{align*} 
	By symmetry of $W$ we obtain $\aconj{y}_a = (\bar{y}_1,y_2)$, $\bconj{y}_a = (\bar{y}_1,\bar{y}_2)$, and $\abconj{y}_a = (y_1,\bar{y}_2)$. Indeed, 
		\begin{align*}
			W^+(y_a,u_a|u_b) & = W^+(\aconj{y}_a,\bar{u}_a|u_b) \\
							 & = W^+(\bconj{y}_a,u_a|\bar{u}_b) \\
							 & = W^+(\abconj{y}_a,\bar{u}_a|\bar{u}_b).
	\end{align*}
 	We leave it to the reader to show that~\eqref{eq_definition of double symmetry for joint channels (Wb db version)} holds for the $D$-value representation of the joint channel. 
\end{example}

Pairs of polar synthetic channels exhibit double symmetry. One can see this directly from symmetry properties of polar synthetic channels, see~\cite[Proposition 13]{Arikan_2009}. Alternatively, one can use induction to show directly that the polar construction preserves double symmetry; we omit the details.  
This implies the following Proposition. 
\begin{proposition}\label{prop_double symmetry for Wab}
	Let $W_{a,b}$ be the joint distribution of two synthetic channels $W_a$ and $W_b$ that result from $n$ polarization steps of BMS channel $W$. Then, $W_{a,b}$ exhibits double symmetry. 
\end{proposition}

The following is a direct consequence of double symmetry. 
\begin{lemma} \label{lem_double symmetry and d values}
	Let $W_{a,b}(y_a,u_a,d_b|u_a,u_b)$ be a joint channel in $D$-value representation that exhibits double symmetry. Then 
	\begin{enumerate}
		\item For the b-channel, $(y_a,u_a,d_b)$ and $(\aconj{y}_a,\bar{u}_a,d_b)$ have the same b-channel $D$-value $d_b$.
		\item For the a-channel, $y_a$ and $\bconj{y}_a$ have the same a-channel $D$-value $d_a$, and $\aconj{y}_a$ and $\abconj{y}_a$ have the same a-channel $D$-value $-d_a$. 
	\end{enumerate}
\end{lemma}
\begin{IEEEproof}
	The first item is obvious from~\eqref{eq_definition of double symmetry for joint channels (Wb db version)}. 		For the second item, note that   \begin{align*}
 	W_a(y_a|u_a) &= \sum_{d_b}\sum_{u_b}W_b(y_a,u_a,d_b|u_b) \\ 
 				 &\overset{\mathclap{(a)}}{=} \sum_{d_b}\sum_{u_b}W_b(\bconj{y}_a,u_a,-d_b|\bar{u}_b) \\
 				 &= \sum_{-d_b}\sum_{\bar{u}_b}W_b(\bconj{y}_a,u_a,-d_b|\bar{u}_b)\\
 				 &= W_a(\bconj{y}_a|u_a),
 \end{align*}
where $(a)$ is by~\eqref{eq_definition of double symmetry for joint channels (Wb db version)}.
In the same manner,  $\aconj{y}_a$ and $\abconj{y}_a$ have the same a-channel $D$-value, $-d_a$.
\end{IEEEproof}
\Cref{lem_double symmetry and d values} implies that an SC decoder does not distinguish between $y_a$ and $\bconj{y}_a$ when making its decision for the a-channel. We now show that a similar conclusion holds for the IMJP decoder. 

\begin{lemma}\label{lem_symmetry of T}
	Let $y_a$ be some output of $W_a$. Then
	\[
		T(y_a|u_a) = T(\bconj{y}_a|u_a) = T(\aconj{y}_a|\bar{u}_a) = T(\abconj{y}_a|\bar{u}_a). 
	\]
\end{lemma}
\begin{IEEEproof}
	\Cref{thm_minimizing phi_a and phi_b for polar channel} holds for joint channels given in $D$-value representation, $W_{a,b}(y_a,u_a,d_b|u_a,u_b)$. This is easily seen by following the proof with minor changes. Under the $D$-value representation,~\eqref{eq_expressions for T using yba} becomes
\begin{equation}
\begin{split} T(y_a|u_a) &= \frac{1}{2} \sum_{d_b} \max_{u_b} W_{a,b}(y_a,u_a,d_b|u_a,u_b)\\
					      &= \sum_{d_b}\max_{u_b} W_b(y_a,u_a,d_b|u_b).	
\end{split}\label{eq_expression for T using db}
\end{equation}
	 The remainder of the proof hinges on double symmetry and follows along similar lines to the proof of \Cref{lem_double symmetry and d values}, with $W_a$ replaced with $T$ and accordingly the sum over $u_b$ replaced with a maximum operation over $u_b$. 
\end{IEEEproof}
\Cref{lem_symmetry of T} implies that the IMJP decoder does not distinguish between $y_a$ and $\bconj{y}_a$.
\begin{corollary}\label{cor_IMJP decoder makes the same decision for ya and bconj ya}
	Let $\phi_a$ be the IMJP decoder for the a-channel. Then 
	$ \phi_a(y_a) = \phi_a(\bconj{y}_a) = 1-\phi_a(\aconj{y}_a) = 1-\phi_a(\abconj{y}_a).$
\end{corollary}

\section{Symmetrized Joint Synthetic Channels}\label{sec_Symmetrized Joint Bit-Channels}
In this section we introduce the symmetrizing transform. The resultant channel is \emph{degraded} from the original joint channel yet has the same probability of error. Its main merit is to decouple the a-channel from the b-channel. This simpler structure is the key to upgrading the a-channel, as we shall see in \Cref{sec_upgrading procedures for joint bit channels}.

\subsection{Symmetrized Joint Channel}
The SC decoder observes marginal distributions and makes a decision based on the $D$-value of each synthetic channel's output. In particular, by \Cref{lem_double symmetry and d values}, the SC decoder makes the same decision for the a-channel whether its output was $y_a$ or $\bconj{y}_a$ and the b-channel decision is based on $d_b$ without regard to $y_a$. By \Cref{cor_IMJP decoder makes the same decision for ya and bconj ya}, the IMJP decoder acts similarly. That is, the IMJP decoder makes the same decision for the a-channel whether its output is $y_a$ or $\bconj{y}_a$, and the decision for the b-channel is based solely on $d_b$. 

We conclude that if the a-channel were told only whether its output was one of $\{y_a,\bconj{y}_a\}$, it would make the same decision had it been told its output was, say, $y_a$. This is true for either the SC or IMJP decoder. Consequently, either decoder's probability of error is unaffected by obscuring the a-channel output in this manner. 

This leads us to define a \emph{symmetrized} version of the joint synthetic channel distribution, $\symm{W}_{a,b}$, as follows. Let\footnote{The order of elements in $\symm{y}_a$ and $\bar{\symm{y}}_a$ does not matter. That is, $\{y_a,\bconj{y}_a\}$ is a \emph{set} containing both $y_a$ and $\bconj{y}_a$.}
\begin{align*} 
	\symm{y}_a &\triangleq \{y_a,\bconj{y}_a\}, \\
	\bar{\symm{y}}_a &\triangleq \{\aconj{y}_a,\abconj{y}_a\}
\end{align*}
and define 
\begin{equation}
\begin{split}
	\symm{W}_{a,b}(\symm{y}_a,u_a,d_b|u_a,u_b) ={}& W_{a,b}(y_a,u_a,d_b|u_a,u_b)\\
										&+ W_{a,b}(\bconj{y}_a,u_a,d_b|u_a,u_b), \\
	\symm{W}_{a,b}(\bar{\symm{y}}_a,u_a,d_b|u_a,u_b) ={}& W_{a,b}(\aconj{y}_a,u_a,d_b|u_a,u_b) \\
											  &+ W_{a,b}(\abconj{y}_a,u_a,d_b|u_a,u_b).
\end{split} \label{eq_symmetrized channel definition}	
\end{equation}

\begin{lemma} \label{lem_symmetrized and non symmetrized channels have the same pe}
	Let $W_{a,b}$ be a joint synthetic channel distribution, and let $\symm{W}_{a,b}$ be its symmetrized version. Then, the probability of error under SC (IMJP) decoding of either channel is identical.
\end{lemma}
\begin{IEEEproof}
	 By \Cref{lem_double symmetry and d values} for the SC decoder or \Cref{cor_IMJP decoder makes the same decision for ya and bconj ya} for the IMJP decoder, if the decoder for the symmetrized channel makes an error for some symbol $\symm{y}_a$ then the decoder for the non-symmetrized channel makes an error for both $y_a$ and $\bconj{y}_a$, and vice-versa. Therefore, denoting by $\mathcal{E}$ the error indicator of the decoder,
	\begin{align*}
		P_e(\symm{W}_{a,b}) &= \frac{1}{4} \smashoperator[l]{\sum_{u_a,u_b \vphantom{\symm{y}_a}}}\smashoperator[r]{\sum_{\symm{y}_a,d_b}} \symm{W}_{a,b}(\symm{y}_a,u_a,d_b|u_a,u_b) \mathcal{E} \\ 
			&\stackrel{\mathclap{(a)}}{=} \frac{1}{4} \smashoperator[l]{\sum_{u_a,u_b\vphantom{d_b}}} \smashoperator[r]{\sum_{y_a,d_b}} W_{a,b}(y_a,u_a,d_b|u_a,u_b) \mathcal{E}\\
			&= P_e(W_{a,b}),
	\end{align*}	
	where $(a)$ is by~\eqref{eq_symmetrized channel definition}.
\end{IEEEproof}

The marginal synthetic channels $\symm{W}_a$ and $\symm{W}_b$ are given by 
\begin{align*}
		\symm{W}_a(\symm{y}_a|u_a) &= \sum_{u_b,d_b}\symm{W}_{a,b}(\symm{y}_a,u_a,d_b|u_a,u_b),\\
		\symm{W}_b(\symm{y}_a,u_a,d_b|u_b) &= \frac{1}{2}\symm{W}_{a,b}(\symm{y}_a,u_a,d_b|u_a,u_b).
\end{align*}
Note that by double symmetry
\begin{equation}
\begin{split}
	\symm{W}_a(\symm{y}_a|u_a) &= \symm{W}_a(\bar{\symm{y}}_a|\bar{u}_a), \\	
	\symm{W}_b(\symm{y}_a,u_a,d_b|u_b) &= \symm{W}_b(\bar{\symm{y}}_a,\bar{u}_a,d_b|u_b) \\
	&= \symm{W}_b(\symm{y}_a,u_a,-d_b|\bar{u}_b) \\
	&= \symm{W}_b(\bar{\symm{y}}_a,\bar{u}_a,-d_b|\bar{u}_b).
\end{split}	\label{eq_symmetrized channel properties}
\end{equation}
\begin{definition}[Symmetrized distribution]
	A joint channel whose marginals satisfy~\eqref{eq_symmetrized channel properties} is called \emph{symmetrized}.
\end{definition} 
The name `symmetrized' stems from comparison of~\eqref{eq_symmetrized channel properties} and~\eqref{eq_definition of double symmetry for joint channels (Wb db version)}. We note that \Cref{thm_minimizing phi_a and phi_b for polar channel} holds for $\symm{W}_{a,b}$.

A symmetrized joint channel remains symmetrized upon polarization. That is, if $\symm{W}_{a,b}$ is a symmetrized joint channel and $\symm{W}_{a^{\alpha},b^{\beta}}$, $\alpha,\beta \in \{-,+\}$ is the result of jointly polarizing it (without applying a further symmetrization operation), then the marginals $\symm{W}_{a^{\alpha}}$ and $\symm{W}_{b^{\beta}}$ satisfy~\eqref{eq_symmetrized channel properties}. This is easily seen from~\eqref{eq_polarizing Wab, general db version} and~\eqref{eq_symmetrized channel properties}. 

Clearly, $\symm{W}_{a,b}$ is \emph{degraded} with respect to $W_{a,b}$, exactly the opposite of our main thrust.  Nevertheless, as established in \Cref{lem_symmetrized and non symmetrized channels have the same pe}, both channels have the same probability of error under SC (IMJP) decoding. Moreover, if we upgrade the symmetrized version of the channel, its probability of error under IMJP decoding lower-bounds the probability of error of the non-symmetrized channel under either SC or IMJP decoding. 

What is not immediately obvious, however, is what happens after polarization. That is, if we take a joint channel, symmetrize it, and then polarize it, how does its probability of error compare to the original joint channel that has just undergone polarization? Furthermore, what happens if the symmetrized version undergoes an upgrading transform?  

In the following proposition, we provide an answer. To this end, a \emph{joint polarization step} is a pair $(\alpha,\beta)\in \{-,+\}^2$ that denotes which transforms the a-channel and b-channel undergo. For example, the result of joint polarization step $(-,+)$ on joint channel $W_{a,b}$ is the joint channel $W_{a^-,b^+}$. A sequence $\mathsf{t}$ of such pairs is called a sequence of joint polarization steps. The joint polarization steps are applied in succession: the result of joint polarization of $W_{a,b}$ according to the sequence $\mathsf{t}=\{(\alpha_1,\beta_1),(\alpha_2,\beta_2),(\alpha_3,\beta_3),\ldots, (\alpha_k,\beta_k)\}$ is the same as the result of joint polarization of $W_{a^{\alpha_1},b^{\beta_1}}$ according to the sequence $\mathsf{t}'=\{(\alpha_2,\beta_2),(\alpha_3,\beta_3),\ldots,(\alpha_k,\beta_k)\}$. 

\begin{proposition}\label{prop_Symmetrizing the joint distribution yields a lower bound}
	Let $W_{a,b}$ be a joint distribution of two synthetic channels and let $W_{a,b}^{\mathsf{t}}$ denote this joint distribution after a sequence $\mathsf{t}$ of joint polarization steps. Then $P_e^{\IMJP}(W_{a,b}^{\mathsf{t}}) \geq P_e^{\IMJP}(\symm{Q}_{a,b}^{\mathsf{t}})$, where $\symm{Q}_{a,b}^{\mathsf{t}}$ is the distribution of $\symm{W}_{a,b}$ after the same sequence of polarization steps and any number of proper upgrading transforms along the way.   
\end{proposition}
\begin{IEEEproof}
Let $W_{a,b}$ be a joint channel with symmetrized version $\symm{W}_{a,b}$. For $\alpha, \beta \in \{-,+\}$, denote by $W_{a^{\alpha},b^{\beta}}$ and $\symm{W}_{a^{\alpha},b^{\beta}}$ the polarized versions of $W_{a,b}$ and $\symm{W}_{a,b}$, respectively. For the $b^{\beta}$-channel, the decoder makes the same decision for either $W_{a^{\alpha},b^{\beta}}$ or $\symm{W}_{a^{\alpha},b^{\beta}}$. This is because the decision is based on the b-channel $D$-value, which is unaffected by symmetrization [see~\eqref{eq_symmetrized channel definition}]. 

Next, for the $a^{\alpha}$ channel, using on~\eqref{eq_polarizing Wab, general db version} a derivation similar to the proof of \Cref{lem_symmetry of T}, $T(y_{a^{\alpha}}|u_{a^{\alpha}}) = T(y'_{a^{\alpha}}|u_{a^{\alpha}})$, where $y'_{a^{\alpha}}$ is any combination of an element of $\symm{y}_a$ and an element of $\symm{\eta}_a$. That is, $y'_{a^{\alpha}}$ is any one of $\{y_a,\eta_a\}$, $\{\bconj{y}_a,\eta_a\}$, $\{y_a,\bconj{\eta}_a\}$, and $\{\bconj{y}_a,\bconj{\eta}_a\}$. 
Thus, the IMJP decoder makes the same decision for the $a^{\alpha}$-channel for either $W_{a^{\alpha},b^{\beta}}$ or $\symm{W}_{a^{\alpha},b^{\beta}}$.

We compare the channels obtained by the following two procedures. 
\begin{itemize}
	\item \emph{Procedure 1:} Joint channel $W_{a,b}$ goes through sequence $\mathsf{t}$ of polarization steps. 
	\item \emph{Procedure 2:} Joint channel $W_{a,b}$ is symmetrized to form $\symm{W}_{a,b}$. It goes through sequence $\mathsf{t}$ of polarization steps (without any further symmetrization operations). 
\end{itemize}
We iteratively apply the above reasoning and conclude in a similar manner to \Cref{lem_symmetrized and non symmetrized channels have the same pe} that both channels have the same performance under IMJP decoding.  
Next, we modify Procedure 2. 
\begin{itemize}
	\item \emph{Procedure 2a:} Joint channel $W_{a,b}$ is symmetrized to form $\symm{W}_{a,b}$. It goes through sequence $\mathsf{t}$ of polarization steps (without any further symmetrization operations), but at some point mid-sequence, it undergoes a proper upgrading procedure.
\end{itemize}
Since polarizing and proper upgrading is equivalent to proper upgrading and polarizing (see \Cref{lem_degradation is preserved for joint distributions}) we can assume that the upgrading happens after the entire sequence of polarization steps. Thus, under IMJP decoding, the probability of error of the channel that results from Procedure 2a lower-bounds the probability of error of the channels resulting from Procedures 1 and 2. Similarly, multiple upgrading transforms can also be thought of as occurring after all polarization steps. 
\end{IEEEproof} 

\begin{corollary}\label{cor_procedure leads to lower bound}
	Let $W$ be a BMS channel that undergoes $n$ polarization steps. Let $W_{a,b}$ be the joint channel of two of its polar descendants such that $a,b \in \mathcal{A}$, and let $\symm{Q}_{a,b} \properup \symm{W}_{a,b}$. Then $P_e^{\SC}(W) \geq P_e^{\IMJP}(\symm{Q}_{a,b})$.
\end{corollary}
\begin{IEEEproof}
	A direct consequence of \Cref{lem_IMJP provides a tighter lower bound than max Pe,lem_conditions on degrading for conservation of pe} combined with \Cref{prop_Symmetrizing the joint distribution yields a lower bound}.
	\end{IEEEproof}
We emphasize that, by \Cref{prop_Symmetrizing the joint distribution yields a lower bound}, it does not matter how we arrive at $\symm{Q}_{a,b}$. So long as $\symm{Q}_{a,b} \properup \symm{W}_{a,b}$ and $a,b \in \mathcal{A}$, we can use $\symm{Q}_{a,b}$ to obtain a lower bound on $P_e^{\SC}(W)$. A practical way to obtain $\symm{Q}_{a,b}$ is via multiple proper upgrading operations that we perform after joint polarization operations. This is the route we take in \Cref{sec_lower bound procedures}. 

Due to \Cref{prop_Symmetrizing the joint distribution yields a lower bound}, we henceforth assume that joint channel $W_{a,b}$ is symmetrized, and no longer distinguish symmetrized channels or symbols by the $(\symm{\cdot})$ symbol. Replacing the joint channel with its symmetrized version need only be performed once, at the first instance the two channels go through different polarization transforms. 

\emph{Implementation:} Since symmetrization is performed only once, and since this invariably happens when 
converting a channel $W$ to $W_{-,+}$, we find the a-, b-, and ab-conjugates using the results of \Cref{ex_conjugates for W-+}. We then form the symmetrized channel using~\eqref{eq_symmetrized channel definition}. Note that it is sufficient to find just the b-conjugates and use the first equation of~\eqref{eq_symmetrized channel definition}. 

\subsection{Decomposition of Symmetrized Joint Channels}
Let the joint channel be $W_b(y_a,u_a,d_b|u_b)$, which, as mentioned above, we assume to be symmetrized. We have %
\begin{equation} 
\begin{split} 
	W_b(y_a,u_a,d_b|u_b)  &= \Prob{y_a,u_a|u_b}\Prob{d_b|u_b,y_a,u_a} \\
							&= \Prob{u_a}\Prob{y_a|u_a,u_b}\Prob{d_b|u_b,y_a,u_a}\vphantom{\frac{1}{2}} \\
							&= \frac{1}{2}W_1(y_a|u_a,u_b)\cdot W_2(d_b|u_b;y_a,u_a),
\end{split}
\label{eq_decomposition of Wb, db version} \end{equation} 
in which we used the independence and uniformity of the input bits $u_a$ and $u_b$. The distribution $W_1$ is given by 
$ W_1(y_a|u_a,u_b) = 2\sum_{d_b} W_b(y_a,u_a,d_b|u_b).$ Whenever $W_1(y_a|u_a,u_b)$ is nonzero,
distribution $W_2(d_b|u_b;y_a,u_a)$ is obtained by dividing $W_b(y_a,u_a,d_b|u_b)$ by
$W_1(y_a|u_a,u_b)/2$. Our notation $W_2(d_b|u_b;y_a,u_a)$ (with a semicolon, as opposed to
$W_2(d_b|y_a,u_a,d_b)$) reminds us that for fixed $y_a,u_a$, channel $W_2$ is a binary-input channel with input $u_b$ and output $d_b$.  If $W_1(y_{a0}|u_{a0},u_b) = 0$ for some $y_{a0}, u_{a0}$, we define $W_2(d_b|u_b;y_{a0},u_{a0})$ to be some arbitrary BMS channel, to ensure it is always a valid channel. 

Since the joint channel is symmetrized, by~\eqref{eq_symmetrized channel properties} we have $W_1(y_a|u_a,u_b) = W_1(y_a|u_a,\bar{u}_b)$. Hence, for any $u_b$,
\begin{equation}
W_a(y_a|u_a) = \sum_{u'_b} W_1(y_a|u_a,u'_b)\Prob{u'_b} = W_1(y_a|u_a,u_b). \label{eq_decoupling}
\end{equation}
That is, a consequence of symmetrization is that given $u_a$, output $y_a$ becomes \emph{independent} of $u_b$. This is not true in the general case where the joint channel is not symmetrized. 

The decomposition of~\eqref{eq_decomposition of Wb, db version} essentially decouples the symmetrized joint channel to a product of two distributions.
\begin{lemma} \label{lem_decomposition of symmetrized distribution}
Let $W_b(y_a,u_a,d_b|u_b)$ be a symmetrized joint channel. It admits the decomposition
\begin{equation}
W_b(y_a,u_a,d_b|u_b) = \frac{1}{2}W_a(y_a|u_a) W_2(d_b|u_b;y_a,u_a).\label{eq_decomposition of W_b to W1W2}	
\end{equation}	
For any $y_a,u_a$, channel $W_2$ is a BMS channel with input $u_b$ and output $d_b$, i.e., 
\begin{equation} \label{eq_W2 is a BMS} W_2(d_b|u_b;y_a,u_a) = W_2(-d_b|\bar{u}_b;y_a,u_a).\end{equation} 
Moreover, $W_2$ satisfies
\begin{equation}\label{eq_symmetry for W2}
	W_2(d_b|u_b;y_a,u_a) = W_2(d_b|u_b;\bar{y}_a,\bar{u}_a). 
\end{equation}
\end{lemma}
\begin{IEEEproof}
	Using~\eqref{eq_decoupling} in~\eqref{eq_decomposition of Wb, db version} yields~\eqref{eq_decomposition of W_b to W1W2}. The remainder of this lemma is readily obtained by using~\eqref{eq_symmetrized channel properties} in~\eqref{eq_decomposition of W_b to W1W2}.
\end{IEEEproof}

\begin{definition}[Decoupling decomposition] \label{def_decoupling decomposition}
	A decomposition of the form~\eqref{eq_decomposition of W_b to W1W2} for a symmetrized joint channel is called a \emph{decoupling decomposition}. 
	Channel $W_a$ is obtained by marginalization, i.e., 
    \begin{align*}
        W_a(y_a|u_a) &= \smashoperator{\sum_{u_b,d_b}} W_b(y_a,u_a,d_b|u_b) \\
                 &=  2\sum_{d_b}W_b(y_a,u_a,d_b|u_b), 
\end{align*}
    where the latter equality, which is due to symmetry, holds for any $u_b$. Then, we 
    compute channel $W_2(d_b|u_b;y_a,u_a)$ using~\eqref{eq_decomposition
    of W_b to W1W2}. The special case where $W_a(y_a|u_a)=0$ requires special attention. Such a case
    invariably happens for perfect symbols --- that is, symbols for which $W_a(y_a|u_a)>0$ but
    $W_a(y_a|\bar{u}_a) = 0$ for some $u_a \in \{0,1\}$. Specifically,
    we ensure that $W_2$ is a well-defined BMS channel even in this case, so we set it to an
    arbitrary BSC. Thus,
    \begin{equation} \label{eq_formula for W2} 
        W_2(d_b|u_b; y_a, u_a) = \begin{cases}
            \dfrac{W_b(y_a,u_a,d_b|u_b)}{W_a(y_a|u_a)/2}, & W_a(y_a|u_a) >0 \\ 
        \text{An arbitrary BSC}, & \text{otherwise.} \end{cases} 
    \end{equation}
    When setting to an arbitrary BSC, we make sure not to add new b-channel $D$-values. One possible
    choice is to set to a BSC whose output has the highest b-channel $D$-value. 
\end{definition}

We use decoupling decompositions of symmetrized joint channels in the sequel. We shall see in
\Cref{subsec_Upgrading $W_a$} that $W_2$ plays a central role in the a-channel upgrading procedure. 
    
%
We conclude this section with an example that compares a joint channel and its symmetrized
version. In particular, we demonstrate the decoupling decomposition for the symmetrized joint
channel.
\begin{example}
	Let $W$ be a BSC with crossover probability $0.2$ and consider $W_{-,+}$, the joint synthetic channel of the `$-$'- and `$+$'-transforms of $W$. 
In $D$-value representation, the a-channel has four possible outputs $y_a \in \{00,01,10,11\}$ and there are three values of $d_b$: $d_b \in \{-\frac{15}{17},0,\frac{15}{17}\}$	. 
\Cref{tab_W-+ ua0} contains the probability table of this joint synthetic channel for $u_a = 0$ and varying $y_a,u_b,d_b$.
When $y_a = 00$ and $u_a = 0$, the b-channel input $u_b$ is more likely to be $1$ than $0$. 
Similarly, when $y_a = 11$ and $u_a = 0$, the b-channel input $u_b$ is more likely to be $0$ than $1$. Thus, the channel in \Cref{tab_W-+ ua0} does not satisfy~\eqref{eq_decomposition of W_b to W1W2}. 

After symmetrization, the a-channel output is either $\symm{0} = \{00,11\}$ or $\symm{1} = \{01,10\}$. 
The probability table for the symmetrized channel with $u_a = 0$ is shown in \Cref{tab_symm W-+ ua0}. 
Here, when $u_a = 0$ and $\symm{0}$ is received at the a-channel, $u_b = 0$ or $1$ are equally likely. Indeed, $W^{-}$ is a BSC with crossover probability $2p(1-p) = 0.32$, and the channel in \Cref{tab_symm W-+ ua0} satisfies~\eqref{eq_decomposition of W_b to W1W2}.

	\begin{table}[t]
			\centering
				\caption{Probability table of joint synthetic channel $W_{-,+}$ derived from a BSC with crossover probability $0.2$. Only the case where $u_a = 0$ is shown.}  \label{tab_W-+ ua0}
	\begin{tabular}{c@{\hspace{4mm}}cccc@{\hspace{4mm}}ccc}
        \textcolor{gray}{($u_a = 0$)} & \multicolumn{3}{c}{$u_b=0$} && \multicolumn{3}{c}{$u_b=1$}\\
	\toprule \multirow{2}[3]{*}{$y_a$} & \multicolumn{3}{c}{$d_b$} && \multicolumn{3}{c}{$d_b$}\\ \cmidrule(r){2-4} \cmidrule(r){6-8}
		     & $-\frac{15}{17}$ & $0$ & $\frac{15}{17}$ && $-\frac{15}{17}$ & $0$ & $\frac{15}{17}$ \\ \midrule
		$00$ & $0.02$ & $0$    & $0$    && $0.32$ & $0$    &  $0$   \\ 
		$01$ & $0$    & $0.08$ & $0$    && $0$    & $0.08$ &  $0$   \\ 
		$10$ & $0$    & $0.08$ & $0$    && $0$    & $0.08$ &  $0$  \\ 
		$11$ & $0$    & $0$    & $0.32$ && $0$    & $0$    &  $0.02$ \\ \bottomrule
	\end{tabular} 	
	\end{table}	
	
	\begin{table}[t]
			\centering
				\caption{Probability table of the symmetrized version of the channel from \Cref{tab_W-+ ua0}. Only the case where $u_a = 0$ is shown.}  \label{tab_symm W-+ ua0}
	\begin{tabular}{c@{\hspace{4mm}}cccc@{\hspace{4mm}}ccc}
	\textcolor{gray}{($u_a = 0$)}& \multicolumn{3}{c}{$u_b=0$} && \multicolumn{3}{c}{$u_b=1$}\\
	\toprule \multirow{2}[3]{*}{$y_a$} & \multicolumn{3}{c}{$d_b$} && \multicolumn{3}{c}{$d_b$}\\ \cmidrule(r){2-4} \cmidrule(r){6-8}
		     & $-\frac{15}{17}$ & $0$ & $\frac{15}{17}$ && $-\frac{15}{17}$ & $0$ & $\frac{15}{17}$ \\ \midrule
		$\symm{0}$ & $0.02$ & $0$     & $0.32$  && $0.32$ & $0$    & $0.02$   \\ 
		$\symm{1}$ & $0$    & $0.16$  & $0$     && $0$    & $0.16$ & $0$   \\  \bottomrule
	\end{tabular} 	
	\end{table}	

\end{example}

\section{Upgrading Procedures for Joint Synthetic Channels}\label{sec_upgrading procedures for joint bit channels}
In this section, we introduce proper upgrading procedures for joint synthetic channels. The overall goal is to reduce the alphabet size of the joint channel. The upgrading procedures we develop enable us to reduce the alphabet size of each of the marginals without changing the distribution of the other; there is a different procedure for each marginal. As an intermediate step, we further couple the marginals by increasing the alphabet size of one of them. 

The joint channel $W_{a,b}$ is assumed to be symmetrized and in $D$-value representation. The upgrading procedures will maintain this. As discussed in \Cref{sec_Symmetrized Joint Bit-Channels}, we do not distinguish symmetrized channels with any special symbol. The upgrading procedure of \Cref{subsec_Upgrading $W_a$} hinges on symmetrization. The upgrading procedure of \Cref{subsec_upgrading $W_b$} does not require symmetrization and holds for non-symmetrized channels without change. However, we shall see that symmetrization simplifies the resulting expressions.

\subsection{Upgrading Channel $W_a$}\label{subsec_Upgrading $W_a$} 
We now introduce a theorem that enables us to deduce an upgrading procedure that upgrades $W_a$ and reduces its output alphabet size. Let symmetrized joint channel $W_b(y_a,u_a,d_b|u_b)$ admit decoupling decomposition~\eqref{eq_decomposition of W_b to W1W2}. Let $Q_b(z_a,u_a,z_b|u_b)$ be another symmetrized joint channel, where $z_b$ represents the $D$-value of the b-channel output. It also admits a decoupling decomposition,  
\begin{equation} Q_b(z_a,u_a,z_b|u_b) = \frac{1}{2}Q_a(z_a|u_a)Q_2(z_b|u_b;z_a,u_a).\label{eq_decomposition of Q_b to Q1Q2}\end{equation} 

\begin{theorem}\label{thm_upgrading Wa}
	Let $W_b$ and $Q_b$ be symmetrized joint channels with decoupling decompositions~\eqref{eq_decomposition of W_b to W1W2} and~\eqref{eq_decomposition of Q_b to Q1Q2}, respectively. Then, $Q_b \properup W_b$ if  
	\begin{enumerate}
		\item  $Q_a(z_a|u_a) \succcurlyeq W_a(y_a|u_a)$ with degrading channel $P_a(y_a|z_a)$. 
		\item  $Q_2(z_b|u_b;z_a,u_a) \succcurlyeq W_2(d_b|u_b;y_a,u_a)$  for all $u_a,y_a,z_a$ such that $P_a(y_a|z_a) > 0$.  
	\end{enumerate}
\end{theorem}
Before going into the proof, some comments are in order. 
First, we do not claim that any $Q_b$ that is upgraded from $W_b$ must satisfy this theorem. 
Second, the meaning of the second item is that, for fixed $z_a,u_a$, BMS channel $Q_2(z_b|u_b;z_a,u_a)$ with binary input $u_b$ is upgraded from a set of BMS channels $\{W_2(d_b|u_b;y_a,u_a)\}_{y_a}$ with the same binary input. 	

\begin{IEEEproof}
Using decoupling decompositions~\eqref{eq_decomposition of W_b to W1W2} and~\eqref{eq_decomposition of Q_b to Q1Q2} and the structure of a proper degrading channel~\eqref{eq_proper form of degrading channels, d version},  $Q_b \properup W_b$ 
if and only if there exist $P_a'$ and $P_b'$ such that 
\iftoggle{twocol}{
\begin{multline}
\sum_{z_a} Q_a(z_a|u_a) P_a'(y_a|z_a) V(d_b|z_a,y_a,u_a,u_b) \\= W_a(y_a|u_a) W_2(d_b|u_b;y_a,u_a), \label{eq_degrading Q1Q2 to W1W2}
\end{multline}}{
\begin{equation}
\sum_{z_a} Q_a(z_a|u_a) P_a'(y_a|z_a) V(d_b|z_a,y_a,u_a,u_b) = W_a(y_a|u_a) W_2(d_b|u_b;y_a,u_a), \label{eq_degrading Q1Q2 to W1W2}
\end{equation}}
where 
\iftoggle{twocol}{
\begin{multline}
	 V(d_b|z_a,y_a,u_a,u_b)\\ = \sum_{z_b}Q_2(z_b|u_b;z_a,u_a) P_b'(d_b|y_a,z_a,u_a,z_b). \label{eq_degrading Q2 to V}
	\end{multline}}{ 
\begin{equation}
	 V(d_b|z_a,y_a,u_a,u_b) = \sum_{z_b}Q_2(z_b|u_b;z_a,u_a) P_b'(d_b|y_a,z_a,u_a,z_b). \label{eq_degrading Q2 to V}
	\end{equation}}
We now find $P_a'$ and $P_b'$ from the conditions of the theorem. 

The first condition of the theorem implies that there exists a channel $P_a(y_a|z_a)$ such that
\begin{equation}
    \sum_{z_a} Q_a(z_a|u_a)P_a(y_a|z_a) = W_a(y_a|u_a). \label{eq_relationship between Qa and Wa} 
\end{equation} 
The second condition of the theorem implies that for each $y_a,u_a,z_a$ there exists a channel $P_b(d_b|y_a,z_a,u_a,z_b)$ such that
\iftoggle{twocol}{
\begin{multline}  \sum_{z_b}Q_2(z_b|u_b;z_a,u_a)P_b(d_b|y_a,z_a,u_a,z_b) \\= W_2(d_b|u_b;y_a,u_a)\cdot \kindi{\Prob{y_a|z_a}>0}.\label{eq_Q2 is upgraded from many W2 channels}\end{multline}}{
\begin{equation}  \sum_{z_b}Q_2(z_b|u_b;z_a,u_a)P_b(d_b|y_a,z_a,u_a,z_b) = W_2(d_b|u_b;y_a,u_a)\cdot \kindi{\Prob{y_a|z_a}>0}.\label{eq_Q2 is upgraded from many W2 channels}\end{equation}}
We set 
 \begin{align*} P_a'(y_a|z_a) 			  &= P_a(y_a|z_a), \\
    P_b'(d_b|y_a,z_a,u_a,z_b) &= P_b(d_b|y_a,z_a,u_a,z_b).\end{align*} 
Using~\eqref{eq_Q2 is upgraded from many W2 channels} in~\eqref{eq_degrading Q2 to V}, we have 
 \[ V(d_b|z_a,y_a,u_a,u_b) = W_2(d_b|u_b;y_a,u_a) \cdot \kindi{P_a(y_a|z_a)>0 }.\] 
It is easily verified that~\eqref{eq_degrading Q1Q2 to W1W2} is satisfied by $P_a'=P_a$ and this $V$, completing the proof. 
\end{IEEEproof}

\begin{remark} \label{rem_W2 doesn't matter} Recall from~\eqref{eq_formula for W2} that when $W_a(y_a|u_a) = 0$, we set $W_2$ to
    an arbitrary BSC. At this point, the reader may wonder what effect --- if any --- does this have
    on the resulting joint channel. We now show that there is no effect. To see this, observe
    from~\eqref{eq_relationship between Qa and Wa} that if $W_a(y_a|u_a) = 0$ and $P_a(y_a|z_a)>0$,
    then necessarily $Q_a(z_a|u_a) = 0$. Hence, by~\eqref{eq_decomposition of Q_b to Q1Q2},
    $Q_b(z_a,u_a, z_b|u_b) = 0$. This latter equality is the same regardless of how we had set $W_2(d_b|u_b;
    y_a,u_a)$.  
\end{remark}

How might one use \Cref{thm_upgrading Wa} to upgrade the a-channel? A naive way would be to first upgrade the marginal $W_a$ to $Q_a$ using some known method (e.g., the methods of~\cite{Tal_2013}, see Appendix~\ref{ap_BMS channel upgrades}). This yields degrading channel $P_a$ by which one can find channel $Q_2$ that satisfies~\eqref{eq_Q2 is upgraded from many W2 channels}. With $Q_a$ and $Q_2$ at hand, one forms the product~\eqref{eq_decomposition of Q_b to Q1Q2} to obtain $Q_b$. If the reader were to attempt to do this, she would find out that it often changes the b-channel. Moreover, this change may be radical: the resulting b-channel may be so upgraded to become almost noiseless, which boils down to an uninteresting bound, the trivial lower bound~\eqref{eq_trivial lower bound}. 
It \emph{is} possible to upgrade the a-channel without changing the b-channel; this requires an additional transform  we now introduce. 

The \emph{upgrade-couple} transform enables upgrading the a-channel without changing the b-channel. The idea is to split each a-channel symbol to several classes, according to the possible b-channel outputs. Symbols within a class have the same $W_2$ channel, so that confining upgrade-merges to operate within a class inherently satisfies the second condition of \Cref{thm_upgrading Wa}. Thus, we circumvent changes to the b-channel.  This results in only a modest increase to the number of output symbols of the overall joint channel.

Let channel $W_b$ have $2B$ possible $D$-values, $\pm d_{b1}, \pm d_{b2},\ldots, \pm d_{bB}$. We assume that erasure symbols are duplicated,\footnote{That is, there is a ``positive'' and a ``negative'' erasure, see~\cite[Lemma 4]{Tal_2013}.} and 
	$0 \leq d_{b1} \leq d_{b2} \leq \cdots \leq d_{bB} \leq 1$. For each a-channel symbol $y_a$ we
    define $B^2$ upgrade-couple symbols  $y_a^{i,j}$, $i,j \in\{1,2,\ldots B\}$. The new symbols
    \emph{couple} the outputs of the a- and b-channels (whence the name of the upgrade-couple transform). Namely, if the a-channel output is $y_a^{i,j}$ and $u_a = 0$, the b-channel output can only be $\pm d_{bi}$; if the a-channel output is $y_a^{i,j}$ and $u_a = 1$, the b-channel output can only be $\pm d_{bj}$. 

The upgrade-couple channel $\upgradec{W}_b(y_a^{i,j},u_a,d_b|u_b)$ is defined by
\begin{equation}	\upgradec{W}_b(y_a^{i,j},u_a,d_b|u_b) \triangleq W_b(y_a,u_a,d_b|u_b)\cdot
S_{i,j}(y_a,u_a,d_b), \label{eq_upgrade split definition}
 \end{equation}
where 
\begin{align*} S_{i,j}(y_a,u_a,d_b) 
      &= \begin{dcases}
    \smashoperator[r]{\sum_{u_b}} W_2(d_{bj}|u_b;y_a, 1), & \begin{aligned} u_a&=0,
    \\[-0.1cm]  d_b&=\pm d_{bi}\end{aligned} \\ 
    \smashoperator[r]{\sum_{u_b}} W_2(d_{bi}|u_b;y_a, 0),  &  \begin{aligned}
        u_a&=1,
    \\[-0.1cm]  d_b&=\pm d_{bj}\end{aligned}\\
    0,
    & \text{otherwise,}
\end{dcases}
 \end{align*} and  $W_2(d_b|u_b;y_a,u_a)$ is derived from the decoupling decomposition of $W_b$,
 see~\eqref{eq_formula for W2}. 

As intuition for the factor $S_{i,j}(y_a,u_a,d_b)$, observe that it ensures that
$\upgradec{W}_b(y_a^{i,j},u_a=0,d_b|u_b)=0$ for $d_b \not\in \pm d_{bi}$ and that 
$\upgradec{W}_b(y_a^{i,j},u_a=1,d_b|u_b)=0$ for $d_b \not\in \pm d_{bj}$. Crucially, 
it \emph{does not} upgrade the marginal channels (see \Cref{cor_Wbstar and Wbhatstar are the
    same,cor_Wastar and Wahatstar are the same}). In particular, as shown in \Cref{lem_properties of
upgrade-couple}, the factor $S_{i,j}(y_a,u_a,d_b)$ ensures that symbols $y_a$ of channel $W_a$ and
$y_a^{i,j}$ of channel $\upgradec{W}_a$ share the same a-channel $D$-value. 
\begin{remark}
    For the original joint channel there may be a-channel symbols $y_a$ for which $W_a(y_a|0) > 0$ but $W_a(y_a|1) = 0$.
    For the upgrade-couple channel $\upgradec{W}_b$, the symbol $y_a^{i,j}$ determines the possible values for the
    b-channel output when $u_a = 0$ or when $u_a = 1$. The symbol $y_a$ never appears with
    positive probability if $u_a = 1$, yet, because it may appear with positive 
    probability if $u_a = 0$, we still need to map it to some $y_a^{i,j}$. The
    upgrade-couple transform is well defined even in this case, thanks to our definition of $W_2$,
    see~\eqref{eq_formula for W2}.  
    In particular, if $y_a$ never occurs with positive probability with $u_a = 1$, say, then
    $y_a^{i,j}$ for the upgrade-couple channel also never occurs with positive probability with $u_a
    =1$ (see~\Cref{lem_properties of upgrade-couple}, item 2). 
\end{remark}


A parameter that is related to $S_{i,j}$ and will be useful in the sequel is 
\begin{equation} \label{eq_definition of alphaij}
    \alpha_{i,j}(y_a) = S_{i,j}(y_a,u_a=0,d_{bi})\cdot S_{i,j}(y_a,u_a=1,d_{bj}).
\end{equation}
For every $y_a,u_a,d_b$, there must exist some $i,j$ such that $S_{i,j}(y_a,u_a,d_b) > 0$. The
following lemma makes this clear. 
\begin{lemma}
    For any $y_a,u_a,d_b$ we have 
\begin{align}
    \sum_{i,j} S_{i,j}(y_a,u_a,d_b) &= 1, \label{eq_sum Sij=1}\\ 
    \sum_{i,j} \alpha_{i,j}(y_a) &= 1.   \label{eq_sum alphaij=1}
\end{align}
\end{lemma}
\begin{IEEEproof}
    Without loss of generality, we shall show~\eqref{eq_sum Sij=1} for $u_a = 0$ and $d_b = + d_{b1}$.  
    Observe that $S_{i,j}(y_a,0,d_{b1}) = 0$ for all $i>1$. Thus, $\sum_{i,j}S_{i,j}(y_a,0,d_{b1})
    = \sum_j S_{1,j}(y_a,0,d_{b1})$. Next, by~\eqref{eq_W2 is a BMS},
    \begin{align*}
        \sum_{j} S_{1,j}(y_a,0,d_{b1}) &= \sum_j \sum_{u_b} W_2(+d_{bj}|u_b; y_a,1) \\ 
                                       &= \sum_{d_b} W_2(d_b|0;y_a,1) \\
                                       &=1, 
    \end{align*}
    where the latter equality is because $W_2$ is a valid BMS channel. 

    To see~\eqref{eq_sum alphaij=1}, observe that
    \begin{equation*} \alpha_{i,j}(y_a) = \left( \sum_{u_b} W_2(d_{bj}|u_b;y_a,1) \right)\cdot \left( \sum_{u_b'}
    W_2(d_{bi}|u_b';y_a,0) \right). \end{equation*}
    Summing over $i,j$ and using~\eqref{eq_W2 is a BMS} yields the result.
\end{IEEEproof}

As we now show, since $W_b$ is symmetrized, so is $\upgradec{W}_b$. 
\begin{lemma}\label{lem_upgrade couple channel is symmetrized}
	Let $W_b(y_a,u_a,d_b|u_b)$ be a symmetrized joint channel. Then, $\upgradec{W}_b(y_a^{i,j},u_a,d_b|u_b)$, defined as in~\eqref{eq_upgrade split definition}, is also symmetrized. 
\end{lemma}
\begin{IEEEproof}
	To establish the lemma, we need to show that~\eqref{eq_symmetrized channel properties} holds for
    the upgrade-couple channel. For the a-channel $W_a$, let symbols $y_a, \bar{y}_a$ be conjugates,
    i.e., $W_a(y_a|u_a) = W_a(\bar{y}_a|\bar{u}_a)$. Channel $W_b$ is symmetrized, so, 
    by~\eqref{eq_symmetry for W2},
    $S_{i,j}(y_a,u_a,d_b)= S_{j,i}(\bar{y}_a,\bar{u}_a,d_b)$. Furthermore, by definition, $S_{i,j}(y_a,u_a,d_b) = S_{i,j}(y_a,u_a,-d_b)$. Thus, 
	\begin{align*}
		\upgradec{W}_b(y_a^{	i,j},u_a,d_b|u_b) &= \upgradec{W}_b(\bar{y}_a^{j,i},\bar{u}_a,d_b|u_b) \\
											  &= \upgradec{W}_b(y_a^{i,j},u_a, -d_b|\bar{u}_b) \\ 
											  &= \upgradec{W}_b(\bar{y}_a^{j,i},\bar{u}_a -d_b|\bar{u}_b).
	\end{align*}
	Next, recall that $\upgradec{W}_a(y_a^{i,j}|u_a) = \sum_{d_b,u_b} \upgradec{W}_b(y_a^{	i,j},u_a,d_b|u_b)$, so that $ \upgradec{W}_a(y_a^{i,j}|u_a) = \upgradec{W}_a(\bar{y}_a^{j,i}|\bar{u}_a)$. Thus,~\eqref{eq_symmetrized channel properties} holds as required.  
\end{IEEEproof}
In the proof of \Cref{lem_upgrade couple channel is symmetrized} we have seen that the conjugate symbol of $y_a^{i,j}$ is $\bar{y}_a^{j,i}$ (with the order of $i$ and $j$ flipped). We summarize this in the following corollary. 
\begin{corollary}\label{cor_conjugate symbols of upgrade-couple channel}
	If  $W_a(\bar{y}_a|\bar{u}_a) = W_a(y_a|u_a)$ then $\upgradec{W}_a(\bar{y}_a^{j,i}|\bar{u}_a) = \upgradec{W}_a(y_a^{i,j}|u_a)$.
\end{corollary}

Since $\upgradec{W}_b$ is symmetrized, it admits decoupling decomposition
\begin{equation} \upgradec{W}_b(y_a^{i,j},u_a,d_b|u_b) = \frac{1}{2}\upgradec{W}_a(y_a^{i,j}|u_a)\upgradec{W}_2(d_b|u_b;y_a^{i,j},u_a).\label{eq_decoupling decomposition of check Wb}	
\end{equation}
Denote by $\text{BSC}(p)$ a binary symmetric channel with crossover probability $p$. In \Cref{lem_properties of upgrade-couple} we derive $\upgradec{W}_a$ [see~\eqref{eq_definition of Wahat}] and establish that for every $y_a$, 
\begin{equation} \upgradec{W}_2(d_b|u_b;y_a^{i,j},u_a) = \begin{dcases} \text{BSC}\left(\frac{1-d_{bi}}{2}\right), & u_a = 0 \\[0.2cm]
														 \text{BSC}\left(\frac{1-d_{bj}}{2}\right), & u_a = 1.
										   \end{dcases}\label{eq_W2 for decoupling decomposition}\end{equation}
That is, when $u_a=0$ we have $\upgradec{W}_2(\pm d_{bi}|u_b;y_a^{i,j},u_a) = (1\pm
(-1)^{u_b}d_{bi})/2$, when $u_a = 1$ we have $\upgradec{W}_2(\pm d_{bj}|u_b;y_a^{i,j},u_a) = (1\pm
(-1)^{u_b}d_{bj})/2$, and $\upgradec{W}_2(d_b|u_b;y_a^{i,j},u_a)$ is zero for any other $d_b$. We
emphasize that we define $\upgradec{W}_2(d_b|u_b;y_a^{i,j},u_a)$ using~\eqref{eq_W2 for decoupling decomposition} even if $\upgradec{W}_a(y_a^{i,j}|u_a) = 0$. 										   
 
 \begin{lemma}\label{lem_properties of upgrade-couple}
 	Let $W_b(y_a,u_a,d_b|u_b)$ be a symmetrized joint channel and let $\upgradec{W}_b(y_a^{i,j},u_a,d_b|u_b)$ be defined as in~\eqref{eq_upgrade split definition}, with decoupling decomposition~\eqref{eq_decoupling decomposition of check Wb}.   
 	Then 
 	\begin{enumerate}
 		\item Joint channel $\upgradec{W}_b$ is upgraded from joint channel $W_b$ with a proper degrading channel that deterministically maps $y_a^{i,j}$ to $y_a$.
        \item We have 
            \begin{equation} \upgradec{W}_a(y_a^{i,j}|u_a) = W_a(y_a|u_a) \cdot
            \alpha_{i,j}(y_a).\label{eq_definition of Wahat} \end{equation} 
 		Moreover, symbols $y_a$ of channel $W_a$ and $y_a^{i,j}$ of channel $\upgradec{W}_a$ have the same a-channel $D$-value for every $i,j$ such that $\upgradec{W}_b(y_a^{i,j},u_a,d_b|u_b)>0$.  
 		\item For every $y_a$, BMS channel $\upgradec{W}_2(d_b|u_b;y_a^{i,j},u_a)$ with input $u_b$ and output $d_b$ is $\text{BSC}((1-d_{bi})/2)$ if $u_a = 0$ and $\text{BSC}((1-d_{bj})/2)$ if $u_a = 1$. 
 	\end{enumerate}
 \end{lemma}
\begin{IEEEproof}
	For the first item, we sum~\eqref{eq_upgrade split definition} over $i,j$ and obtain,
    using~\eqref{eq_sum Sij=1}, 
	\[ W_b(y_a,u_a,d_b|u_b) = \sum_{i,j} \upgradec{W}_b(y_a^{i,j},u_a,d_b|u_b).\]
	That is, joint channel $\upgradec{W}_b$ is upgraded from $W_b$ with degrading channel $P_a$ that deterministically maps $y_a^{i,j}$ to $y_a$. This is a proper degrading channel. 
	
    For the second item, we marginalize $\upgradec{W}_b$ over $d_b$ and $u_b$.
    Using~\eqref{eq_decomposition of W_b to W1W2} in the right-hand-side of~\eqref{eq_upgrade split
    definition}, we obtain~\eqref{eq_definition of Wahat}, where $\alpha_{i,j}(y_a)$ is given
    in~\eqref{eq_definition of alphaij}.  
    Whenever $\upgradec{W}_b(y_a^{i,j},u_a,d_b|u_b)>0$, we have, by~\eqref{eq_upgrade split
    definition},  $\alpha_{i,j}(y_a)>0$. Thus,
\[ \frac{\upgradec{W}_a(y_a^{i,j}|0) - \upgradec{W}_a(y_a^{i,j}|1)}{\upgradec{W}_a(y_a^{i,j}|0) + \upgradec{W}_a(y_a^{i,j}|1)} = \frac{W_a(y_a|0) - W_a(y_a|1)}{W_a(y_a|0) + W_a(y_a|1)},\] implying that $y_a$ and $y_a^{i,j}$ have the same a-channel $D$-value for their respective channels. 

For the final item, if $\upgradec{W}_a(y_a^{i,j}|u_a) = 0$, we are free to set $\upgradec{W}_2(d_b|u_b;y_a^{i,j},u_a)$ as we please, so we set it as per the item. Otherwise, there are only two values of $d_b$ for which $S_{i,j}(y_a,u_a,d_b)$ is nonzero. Hence, $\upgradec{W}_b$ can output only two b-channel $D$-values for fixed $y_a^{i,j}$ and $u_a$. 
	Thus, $\upgradec{W}_2$ is a BMS channel with only two possible outputs, or, in other words, a BSC. A BSC that outputs $D$-values $\pm d$, $0 \leq d\leq 1$, has crossover probability $(1-d)/2$. This establishes the item. 
\end{IEEEproof} 
 
\begin{definition}[Canonical channel]
	The canonical channel $W^*(d|u)$ of channel $W(y|u)$ has a single entry for each $D$-value. That is, denoting by $\Dd$ the set of symbols $y$ whose $D$-value is $d$, we have
	$W^*(d|u) = \sum_{\Dd} W(y|u).$ It can be shown that a channel is equivalent to its canonical form, i.e., each form can be degraded from the other. 
\end{definition}

\begin{corollary}\label{cor_Wbstar and Wbhatstar are the same}
	The canonical b-channels of $\upgradec{W}_b(y_a^{i,j},u_a,d_b|u_b)$ and $W_b(y_a,u_a,d_b|u_b)$ coincide. 
\end{corollary} 
\begin{IEEEproof}
This is a direct consequence of the first item of \Cref{lem_properties of upgrade-couple}: 
  \begin{equation*}
   \begin{IEEEeqnarraybox}{rCl}
	\upgradec{W}_b^*(d_b|u_b) &=& \sum_{y_a,u_a} \sum_{i,j} \upgradec{W}_b(y_a^{i,j},u_a,d_b|u_b)\\ 
			&=& \sum_{y_a,u_a} 	W_b(y_a,u_a,d_b|u_b) \\
			&=& W_b^*(d_b|u_b).
  \end{IEEEeqnarraybox}
  \IEEEQEDhereeqn
  \end{equation*}
\end{IEEEproof}

 \begin{corollary} \label{cor_Wastar and Wahatstar are the same}
	The canonical a-channels of $\upgradec{W}_b(y_a^{i,j},u_a,d_b|u_b)$ and $W_b(y_a,u_a,d_b|u_b)$ coincide. 
\end{corollary} 
\begin{IEEEproof}
    This follows from  the second item of \Cref{lem_properties of upgrade-couple}, \eqref{eq_sum
    alphaij=1}, and~\eqref{eq_definition of Wahat}.
\end{IEEEproof}

\begin{definition}[Class]
	The \emph{class} $C_{i,j}$ is the set of symbols $y_a^{i,j}$ with fixed $i,j$. 
\end{definition}
There are $B^2$ classes. The size of each class is the number of symbols $y_a$. By~\eqref{eq_W2 for decoupling decomposition}, 
	$\upgradec{W}_2(d_b|u_b;y_a^{i,j},u_a)$ is the \emph{same} BSC for all symbols of class $C_{i,j}$ and fixed $u_a$. Thus, the second item of \Cref{thm_upgrading Wa} becomes trivial and is immediately satisfied if we use an upgrading procedure that upgrade-merges several symbols of the same class $C_{i,j}$. 

To determine which upgrading procedures may be used, we turn to the degrading channel. So long as
the degrading channel does not mix a symbol and its conjugate, the upgrading procedure can be
confined to a single class. This is because conjugate symbols belong to different classes, as
established in \Cref{cor_conjugate symbols of upgrade-couple channel}. Thus, of the upgrading
procedures of~\cite{Tal_2013} (see Appendix~\ref{ap_BMS channel upgrades}) we can use either
upgrade-merge-3 without restriction or upgrade-merge-2 provided that the two symbols to be merged
have the same a-channel $D$-value. 

\begin{theorem}\label{thm_upgrade split}
	Let $W_b(y_a,u_a,d_b|u_b)$ be some joint channel with marginals $W_a(y_a|u_a), W_b^*(d_b|u_b)$ and upgrade-couple counterpart  $\upgradec{W}_b(y_a^{i,j},u_a,d_b|u_b)$.  Let $Q_a(z_a|u_a) \succcurlyeq W_a(y_a|u_a)$ obtained by an upgrade-merge-3 procedure. Then there exists joint channel $\upgradec{Q}_b(z_a^{i,j},u_a,d_b|u_b) \properup \upgradec{W}_b(y_a^{i,j},u_a,d_b|u_b)$ with canonical marginals $\upgradec{Q}_a^*(z_a|u_a), \upgradec{Q}_b^*(d_b|u_b)$ such that $\upgradec{Q}_a^* = Q_a^*$ and $\upgradec{Q}_b^* = W_b^*$. 
\end{theorem}
\begin{IEEEproof}
	The idea is to confine the upgrading procedures to work within a class, utilizing \Cref{thm_upgrading Wa} over each class separately. 
	
	Assume that the upgrading procedure from $W_a$ to $Q_a$ replaces symbols $y_{a1},y_{a2},y_{a3}$ with symbols $z_{a1},z_{a3}$. 
	We obtain $\upgradec{Q}_b$ by using \Cref{thm_upgrading Wa} for each class $C_{i,j}$ of
    $\upgradec{W}_b$ separately. The a-channel upgrade procedure for class $C_{i,j}$ is
    upgrade-merge-3 from $\upgradec{W}_a$ to $\upgradec{Q}_a$ that replaces symbols
    $y_{a1}^{i,j},y_{a2}^{i,j},y_{a3}^{i,j}$ with symbols $z_{a1}^{i,j},z_{a3}^{i,j}$. As the
    upgrade is confined to symbols of the same class, the channel $\upgradec{W}_2$ --- given
    by~\eqref{eq_W2 for decoupling decomposition} --- is the same regardless of $y_a$, as
    established in \Cref{lem_properties of upgrade-couple}, item 3. Hence, the second item of
    \Cref{thm_upgrading Wa} is automatically satisfied within a class $C_{i,j}$, with 
	\begin{equation} \upgradec{Q}_2(d_b|u_b;z_a^{i,j},u_a) = \upgradec{W}_2(d_b|u_b;y_a^{i,j},u_a)\label{eq_within a class Q2 is the same}\end{equation} for all $y_a,z_a$. Channel $\upgradec{Q}_b$ is then obtained by the product of $\upgradec{Q}_a$ and $\upgradec{Q}_2$ as per~\eqref{eq_decoupling decomposition of check Wb}:
	\begin{equation} \upgradec{Q}_b(z_a^{i,j},u_a,d_b|u_b) = \frac{1}{2} \upgradec{Q}_a(z_a^{i,j}|u_a)\upgradec{Q}_2(d_b|u_b;z_a^{i,j},u_a).\label{eq_decoupling decomposition of check Qb} \end{equation}
	
	By properties of upgrade-merge-3 (see~\eqref{eq_upgrade_merge_3} in Appendix~\ref{ap_upgrade merge 3}) we have
	$ \sum_{z_a} \upgradec{Q}_a (z_a^{i,j}|u_a) = \sum_{y_a} \upgradec{W}_a(y_a^{i,j}|u_a).$ 
	Therefore, 
	\begin{align*}
		\upgradec{Q}_b^*(d_b|u_b) &= \sum_{i,j,u_a}\sum_{z_a}\upgradec{Q}_b(z_a^{i,j},u_a,d_b|u_b) \\ 
							 &\overset{\mathclap{(a)}}{=} \sum_{i,j,u_a}\sum_{z_a}\frac{1}{2} \upgradec{Q}_2(d_b|u_b;z_a^{i,j},u_a) \upgradec{Q}_a(z_a^{i,j}|u_a) \\
							 &\overset{\mathclap{(b)}}{=} \sum_{i,j,u_a}\frac{1}{2} \upgradec{W}_2(d_b|u_b;y_{a1}^{i,j},u_a) \sum_{z_a}\upgradec{Q}_a(z_a^{i,j}|u_a) \\
							 &= \sum_{i,j,u_a}\frac{1}{2} \upgradec{W}_2(d_b|u_b;y_{a1}^{i,j},u_a) \sum_{y_a}\upgradec{W}_a(y_a^{i,j}|u_a) \\
							 &\overset{\mathclap{(c)}}{=} \sum_{i,j,u_a}\sum_{y_a}\frac{1}{2} \upgradec{W}_2(d_b|u_b;y_{a}^{i,j},u_a) \upgradec{W}_a(y_a^{i,j}|u_a) \\
							 &\overset{\mathclap{(d)}}{=} W_b^*(d_b|u_b), 
	\end{align*}
	where in $(a)$ we used the decoupling decomposition~\eqref{eq_decoupling decomposition of check Qb}; $(b)$ and $(c)$ are by \Cref{lem_properties of upgrade-couple}, item 3 and by~\eqref{eq_within a class Q2 is the same}; finally, $(d)$ is due to \Cref{cor_Wbstar and Wbhatstar are the same}. 

To see that the canonical a-channel marginals coincide, note that by \Cref{lem_properties of upgrade-couple}, item 2, for any fixed $z_a$, the symbols $\{z_a^{i,j}\}_{i,j}$ all have the same a-channel $D$-value. Let $d_a$ be some a-channel $D$-value, and let $D_{d_a}$ be the set of a-channel outputs $z_a$ whose a-channel $D$-value is $d_a$. Then, 
\begin{align*}
	\upgradec{Q}_a^*(d_a|u_a) &= \sum_{z_a \in \Dda} \sum_{i,j} \sum_{d_b,u_b}\upgradec{Q}_b(z_a^{i,j},u_a,d_b|u_b) \\
							  &= \sum_{z_a \in \Dda} \sum_{i,j}\upgradec{Q}_a(z_a^{i,j}|u_a) \\
							  &\overset{\mathclap{(a)}}{=} \sum_{z_a \in \Dda} Q_a(z_a|u_a)\\
							  &= Q_a^*(d_a|u_a),
\end{align*}
where $(a)$ is a direct consequence of the expressions for upgrade-merge-3 and our construction of upgrading each class separately. 
\end{IEEEproof}

To use \Cref{thm_upgrade split}, one begins with a design parameter $A$	that controls the output alphabet size. Working one class at a time, one then applies upgrade operations in succession to reduce the class size to $2A$. The resulting channel, therefore, will have $2AB^2$ symbols overall. The canonical a-channel marginal that results from this operation will have at most $2A$ symbols. 

\begin{remark}
The upgrade-merge-3 procedure replaces three conjugate symbol pairs with two conjugate symbol pairs. Recall from \Cref{cor_conjugate symbols of upgrade-couple channel} that after the upgrade-couple transform, conjugate symbols belong to different classes. In particular, if $y_a$ and $\bar{y}_a$ are a conjugate pair of the a-channel before the upgrade-couple transform, then  $y_a^{i,j} \in C_{i,j}$ and $\bar{y}_a^{j,i} \in C_{j,i}$ are a conjugate pair of the a-channel after the upgrade-couple transform. Therefore, when one uses \Cref{thm_upgrade split} 
	to replace the symbols \[\left\{y_{a1}^{i,j},y_{a2}^{i,j},y_{a3}^{i,j}\right\} \to \left\{z_{a1}^{i,j},z_{a3}^{i,j}\right\},\] one must also replace their conjugates\[\left\{\bar{y}_{a1}^{j,i},\bar{y}_{a2}^{j,i},\bar{y}_{a3}^{j,i}\right\} \to \left\{\bar{z}_{a1}^{j,i},\bar{z}_{a3}^{j,i}\right\}.\] We still always operate within a class as nowhere do we mix symbols from different classes.
	Alternatively, one may upgrade only classes $C_{i,j}$ with $i \geq j$ and then use channel symmetry to obtain the upgraded forms of classes $C_{j,i}$. 
\end{remark}

There is one case where it is possible to use upgrade-merge-2, as stated in the following corollary. 
\begin{corollary}\label{cor_upgrade split}
	\Cref{thm_upgrade split} also holds  if the a-channel upgrade procedure is upgrade-merge-2 applied to two symbols of the same a-channel $D$-value.
\end{corollary}
\begin{IEEEproof}
	While in general the upgrade-merge-2 procedure mixes a symbol and its conjugate, when the two symbols to be merged have the same a-channel $D$-value this is no longer the case (see Appendix~\ref{ap_upgrade merge 2}), and we can follow along the lines of the proof of \Cref{thm_upgrade split}. We omit the details. 
\end{IEEEproof}

The reason that~\cite{Tal_2013} introduced both the upgrade-merge-2 and upgrade-merge-3 procedures despite the superiority of the latter stems from numerical issues. To implement upgrade-merge-3 we must divide by the difference of the extremal $D$-values to be merged. If these are very close this can lead to numerical errors. Upgrade-merge-2 is not susceptible to such errors. On the other hand, upgrade-merge-2  cannot be used in the manner stated above; it requires us to mix symbols from two classes $C_{i,j}$ and $C_{j,i}$ that may have wildly different $\upgradec{Q}_2$ channels. Thus, this  will undesirably upgrade the b-channel. 

In practice, however, we may be confronted with a triplet of symbols with very close, but not identical, a-channel $D$-values. To avoid numerical issues, we utilize a fourth nearby symbol. Say that our triplet\footnote{To simplify notation, we omit the dependence on the class; it is clear that we do this for each class separately.} is $y_{a1},y_{a2},y_{a3}$ with a-channel $D$-values $d_{a1}\leq d_{a2}<d_{a3}$ such that $d_{a3}-d_{a1} < \epsilon$, for some ``closeness'' threshold $\epsilon$. Let $y_{a4}$ have a-channel $D$-value $d_{a4}$ such that $d_{a4}-d_{a1} > \epsilon$. Then, we apply upgrade-merge-3 twice: first for $y_{a1},y_{a2},y_{a4}$ obtaining $z_{a1}, z_{a4}$ with a-channel $D$-values $d_{a1},d_{a4}$ and then for $z_{a1},y_{a3},z_{a4}$, ending up with $z'_{a1},z'_{a4}$ with a-channel $D$-values $d_{a1},d_{a4}$. In this example we have chosen a fourth symbol with a greater a-channel $D$-value than $d_{a4}$, but we could have similarly chosen a fourth symbol with a smaller a-channel $D$-value than $d_{a1}$ instead.

\subsection{Upgrading Channel $W_b$} \label{subsec_upgrading $W_b$}

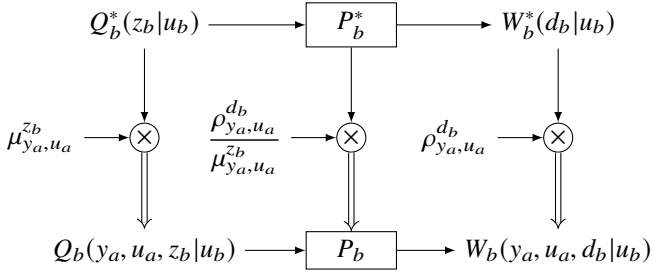
\begin{figure}[t] 
	\centering
	\begin{tikzpicture}[>=latex]
	
	\node (Qb*) at (-2,3) {$Q_b^*(z_b|u_b)$};
	\node (Wb*) at ( 3.5,3) {$W_b^*(d_b|u_b)$};
	\node (Qb) at (-2,0) {$Q_b(y_a,u_a,z_b|u_b)$};
	\node (Wb) at ( 3.5,0) {$W_b(y_a,u_a,d_b|u_b)$};
	
	\node (rect1) at ($(Qb*)!0.5!(Wb*)$) [rectangle, draw, inner sep = 3pt, minimum width = 1.2cm] {$P_b^*$};
	\node (rect2) at ( $(Qb)!0.5!(Wb)$ ) [rectangle, draw, inner sep = 3pt, minimum width = 1.2cm] {$P_b$};		
	\node[draw,circle, inner sep = 1pt] (A1) at ($(Qb*)!0.5!(Qb)$) {$\times$};
	\node[draw,circle, inner sep = 1pt] (A2) at ($(Wb*)!0.5!(Wb)$) {$\times$};
	\node[draw,circle, inner sep = 1pt] (A3) at ($(rect1)!0.5!(rect2)$) {$\times$};
	
	\draw[<-] (A1) -- ++(-0.8,0) node[left] {$\muyauazb$};
	\draw[<-] (A2) -- ++(-0.8,0) node[left] {$\rhoyauadb$}; 
	\draw[<-] (A3) -- ++(-0.8,0) node[left] {$\displaystyle \frac{\rhoyauadb}{\muyauazb}$};			
	\draw[->] (Qb*) -- (rect1); \draw[->] (rect1) -- (Wb*);
	\draw[->] (Qb)  -- (rect2); \draw[->] (rect2) -- (Wb); 
	
	\draw[->] (Qb*) -- (A1); \draw[double equal sign distance, -implies] (A1) -- (Qb); 
	\draw[->] (Wb*) -- (A2); \draw[double equal sign distance, -implies] (A2) -- (Wb); 
	\draw[->] (rect1) -- (A3); \draw[double equal sign distance, -implies] (A3) -- (rect2); 	
	\end{tikzpicture}
	\caption{Illustration of how to transform an upgrading procedure from $W_b^*$ to $Q_b^*$ to an upgrading procedure from $W_b$ to $Q_b$. The double arrows represent splitting to multiple outputs.}
	\label{fig_upgragding channel b}
\end{figure}

We now show how to upgrade $W_{a,b}(y_a,u_a,d_b|u_a,u_b)$ to channel $Q_{a,b}(y_a,u_a,z_b|u_a,u_b)$ such that  $Q_b \succcurlyeq W_b$ and  $Q_a = W_a$. 
The idea is to begin with $W_b^*$, a channel equivalent to $W_b$ in which $y_a$ and $u_a$ are not explicit in the output. The channel $W_b^*$ is given by $W_b^*(d_b|u_b) = \sum_{y_a,u_a} W_b(y_a,u_a,d_b|u_b)$.  
We upgrade $W_b^*$ to $Q_b^*$ using some known method, such that channel $P_b^*$ degrades $Q_b^*$ to $W_b^*$. To form upgraded channel $Q_b$, we ``split'' the outputs of $Q_b^*$ to include $y_a$ and $u_a$ and find a degrading channel that degrades $Q_b$ to $W_b$. We shall see that the upgraded channel $Q_b$ is given by
\[ Q_b(y_a,u_a,z_b|u_b) = Q_b^*(z_b|u_b) \sum_{d_b} \frac{P_b^*(d_b|z_b)W_b(y_a,u_a,d_b)}{W_b^*(d_b)},\]
where $W_b(y_a,u_a,d_b)$ and $W_b^*(d_b)$ are defined in~\eqref{eq_definition of Wbstar and and Wb}, below.
 Finally, we form the joint channel $Q_{a,b}$ using~\eqref{eq_Wab and its relationship to Wb}. We illustrate this in \Cref{fig_upgragding channel b}. 

\begin{theorem} \label{thm_Wb can be upgraded to Qb with ya and va intact}
	Let $W_b(y_a,u_a,d_b|u_b)$ be a joint channel where $d_b$ is the $D$-value of the b-channel's output. Let $W_b^*(d_b|u_b)$ be a channel equivalent to $W_b$, and let $Q_b^*(z_b|u_b) \succcurlyeq W_b^*(d_b|u_b)$ with degrading channel $P_b^*(d_b|z_b)$. Then there exists joint channel $Q_b(y_a,u_a,z_b|u_b)$ such that $Q_b(y_a,u_a,z_b|u_b) \properup W_b(y_a,u_a,d_b|u_b)$ and $\sum_{\omegaa} Q_b(y_a,u_a,z_b|u_b) = Q_b^*(z_b|u_b)$. 
\end{theorem}
\begin{IEEEproof}
	We shall explicitly find $Q_b$ and an appropriate degrading channel. The degrading channel will be of the form $P_b(d_b|y_a,u_a,z_b)$, i.e., $y_a$ and $u_a$ pass through the degrading channel unchanged. Such degrading channels are proper. Since $Q_b^* \succcurlyeq W_b^*$ we have, for any $d_b$ and $u_b$,
	\begin{equation} W_b^*(d_b|u_b) = \sum_{z_b} P_b^*(d_b|z_b) Q_b^*(z_b|u_b).\label{eq_Qb* is upgraded from Wb* using Pb*}\end{equation}
	
	Denote 
	\begin{equation}
		\begin{split} 
			W_b(y_a,u_a,d_b) &= \frac{1}{2}\sum_{u_b} W_b(y_a,u_a,d_b|u_b), \\
			W_b^*(d_b) &= \frac{1}{2}\sum_{u_b} W_b^*(d_b|u_b).
		\end{split}\label{eq_definition of Wbstar and and Wb}
	\end{equation}
	We assume that $W_b^*(d_b) >0$, for otherwise output $d_b$ never appears with positive probability and may be ignored, and define 
	\begin{align} \rhoyauadb &\triangleq  \frac{W_b(y_a,u_a,d_b)}{W_b^*(d_b)}. \nonumber \\
	\intertext{We have $\rhoyauadb \geq 0$, $\sum_{\omegaa}\rhoyauadb = 1$ for any $d_b$, and, for any $u_b$,}   
	 W_b(y_a,u_a,d_b|u_b) &= \rhoyauadb W_b^*(d_b|u_b).\label{eq_connection between Wb and Wbstar} 
	 \end{align}
	
	For each $z_b$, we will shortly define constants $\muyauazb$ such that $\muyauazb \geq 0 $ and $\sum_{\omegaa}\muyauazb = 1$. Similar to~\eqref{eq_connection between Wb and Wbstar}, we use these constants to define channel $Q_b$ by
	\begin{equation} Q_b(y_a,u_a,z_b|u_b) = \muyauazb Q_b^*(z_b|u_b).\label{eq_connection between Qb and Qbstar} \end{equation}
	Indeed,  $\sum_{\omegaa} Q_b(y_a,u_a,z_b|u_b) = Q_b^*(z_b|u_b)$. We now find the constants $\muyauazb$ and an appropriate degrading channel $P_b(d_b|y_a,u_a,z_b)$ such that 
	\begin{equation} W_b(y_a,u_a,d_b|u_b) = \sum_{z_b} P_b(d_b|y_a,u_a,z_b) Q_b(y_a,u_a,z_b|u_b),\label{eq_Qb is upgraded from Wb using Pb}\end{equation} which will establish our goal. 
	
	Let $\omegaa$, and $d_b$ be such that the left-hand side of~\eqref{eq_Qb is upgraded from Wb using Pb} is positive\footnote{Since $W_b^*(d_b)>0$, there will always be at least one selection of $\omegaa$ for which the left-hand side of~\eqref{eq_Qb is upgraded from Wb using Pb} is positive.}, so that $\rhoyauadb >0$. We shall see that the resulting expressions hold for the zero case as well.  Using~\eqref{eq_connection between Wb and Wbstar} and~\eqref{eq_connection between Qb and Qbstar}, we can rewrite~\eqref{eq_Qb is upgraded from Wb using Pb} as 
	\[ W_b^*(d_b|u_b) = \sum_{z_b} \left(\frac{P_b(d_b|y_a,u_a,z_b)\muyauazb}{\rhoyauadb} \right) Q_b^*(z_b|u_b).\]
	Comparing this with~\eqref{eq_Qb* is upgraded from Wb* using Pb*}, we set 
	\begin{align} P_b^*(d_b|z_b) &= \frac{P_b(d_b|y_a,u_a,z_b)\muyauazb}{\rhoyauadb}.\label{eq_Pb* and its relation to Pb mu and rho}\\
	\intertext{Since $P_b$ is a probability distribution, by rearranging and summing over $d_b$ we obtain}
	 \muyauazb &= \sum_{d_b} P_b^*(d_b|z_b) \rhoyauadb.\label{eq_formula for mu}
	 \end{align} 
	It is easily verified that $\muyauazb \geq 0$ and $\sum_{\omegaa}\muyauazb = 1$.  
	Using the expression for $\muyauazb$ in~\eqref{eq_Pb* and its relation to Pb mu and rho} yields
	\begin{equation} P_b(d_b|y_a,u_a,z_b) = \frac{P_b^*(d_b|z_b)\rhoyauadb}{\sum_{d_b'}
    P_b^*(d_b'|z_b) \rho_{\omegaa}^{d_b'}}.\label{eq_formula for Pb in bchannel upgrade}
\end{equation} This is a valid probability distribution. We remark that~\eqref{eq_Qb is upgraded
from Wb using Pb} is satisfied by~\eqref{eq_formula for mu} and~\eqref{eq_formula for Pb in bchannel upgrade} even when $\rhoyauadb = 0$. 
	We have found $Q_b$ and a proper degrading channel $P_b$ as required. 
\end{IEEEproof}

\begin{corollary}
	In \Cref{thm_Wb can be upgraded to Qb with ya and va intact}, the marginal a-channels of $Q_b$ and $W_b$ coincide.
\end{corollary}
\begin{IEEEproof}
	By construction, the degrading channel from $Q_b$ to $W_b$ does not change the a-channel output, implying that the a-channel marginal remains the same. 
\end{IEEEproof}

To use \Cref{thm_Wb can be upgraded to Qb with ya and va intact}, one begins with design parameter
$B$ that controls the output alphabet size. The channel $Q_b^*$, with output alphabet of size $2B$,
is obtained from $W_b^*$ using a sequence of upgrade operations. To obtain upgraded joint channel
$Q_b$, one uses the Theorem to turn them into a sequence of upgrade operations to be performed on
channel $W_b$. If one uses the techniques of~\cite{Tal_2013}, the upgrade operations will consist of
upgrade-merge-2 and upgrade-merge-3 operations (see Appendix~\ref{ap_BMS channel upgrades}). In the following examples we apply \Cref{thm_Wb can be upgraded to Qb with ya and va intact} specifically to these upgrades.

For brevity, we will use the following notation:  
\begin{equation}
	\begin{split}
		\bchanpi{\omegaa}{d_b} &\triangleq \sum_{u_b} W_b(y_a,u_a,d_b|u_b), \\
			   \bchanpi{}{d_b} &\triangleq \sum_{u_b} W_b^*(d_{b}|u_b).
	\end{split}	\label{eq_definition of pidyaua}
\end{equation}

\begin{example}[Upgrading $W_b$ Based on Upgrade-Merge-2]\label{ex_Upgrading $W_b$ Based on Upgrade-Merge-2}
The upgrade-merge-2 procedure of~\cite{Tal_2013} selects two conjugate symbols pairs and replaces them with a single conjugate symbol pair. The details of the transformation, in our notation, appear in Appendix~\ref{ap_upgrade merge 2}. 

Let joint channel $W_b(y_a,u_a,d_b|u_b)$ have b-channel marginal $W_b^*(d_b|u_b)$, in which all symbols with the same $D$-value are combined to a single symbol. We select symbols $d_{bj}, d_{bk}$ and their respective conjugates $\bar{d}_{bj} = -d_{bj}, \bar{d}_{bk}=-d_{bk}$, such that $d_{bk} \geq d_{bj} > 0$ and upgrade $W_b^*(d_b|u_b)$ to $Q_b^*(z_b|u_b)$ given by~\eqref{eq_Formula for simple upgrade merge 2}  (Appendix~\ref{ap_upgrade merge 2}). We denote by $\mathcal{D}_b$ the output alphabet of $W_b^*$  and by $\mathcal{D}_{z_{bk}}$ the set 
\[ \mathcal{D}_{z_{bk}} \triangleq \{d_{bk},d_{bj},\bar{d}_{bj},\bar{d}_{bk}\}.\]
The output alphabet of $Q_b^*$ is $\mathcal{Z} = (\mathcal{D}_b \setminus \mathcal{D}_{z_{bk}} )\cup (z_{bk},\bar{z}_{bk})$; outputs of $Q_b^*$ represent $D$-values. In particular, the $D$-values of $z_{bk}$ and $\bar{z}_{bk}$ are $d_{bk}$ and $-d_{bk}$, respectively.

Using \Cref{thm_Wb can be upgraded to Qb with ya and va intact}, we form channel $Q_b(y_a,u_a,z_b|u_b)$ by
\[ Q_b(y_a,u_a,z_b|u_b) = \begin{cases} \muyauazbk Q_b^*(z_{bk}|u_b), & z_b = z_{bk} \\[0.1cm]
										\muyauazbkbar Q_b^*(\bar{z}_{bk}|u_b), & z_b = \bar{z}_{bk}\\[0.1cm]
										W_b(y_a,u_a,z_b|u_b), & \text{otherwise,} 
										\end{cases}\] 
where by~\eqref{eq_formula for mu},
\begin{align*}
\muyauazbk &= \frac{\sum_{d \in \mathcal{D}_{z_{bk}}}\left(\bchanpi{\omegaa}{d}\cdot (d_{bk} + d)\right)}{2(\bchanpi{}{d_{bj}} + \bchanpi{}{d_{bk}})d_{bk}}, \\[0.1cm]
\muyauazbkbar &= \frac{\sum_{d \in \mathcal{D}_{z_{bk}}}\left(\bchanpi{\omegaa}{d}\cdot( d_{bk} - d)\right)}{2(\bchanpi{}{d_{bj}} + \bchanpi{}{d_{bk}})d_{bk}}. 
\end{align*} 

We can simplify this when $W_b$ is a symmetrized channel. In this case, $\bchanpi{\omegaa}{d_{b}} = \bchanpi{\omegaa}{\bar{d}_{b}}$, yielding 
	\[ \muyauazbk = \muyauazbkbar = \frac{\bchanpi{\omegaa}{d_{bj}} + \bchanpi{\omegaa}{d_{bk}}}{\bchanpi{}{d_{bj}} + \bchanpi{}{d_{bk}}}.\]
Therefore, the upgraded joint channel becomes
\[ Q_b(y_a,u_a,z_b|u_b) = \begin{dcases} \bchanPi{y_a,u_a}{z_{bk}}\left(\frac{1+(-1)^{u_b}d_{bk}}{2} \right),	& z_b = z_{bk} \\[0.1cm]
\bchanPi{y_a,u_a}{z_{bk}} \left(\frac{1-(-1)^{u_b}d_{bk}}{2} \right),	& z_b = \bar{z}_{bk} \\[0.1cm] 
W_b(y_a,u_a,z_b|u_b), & \text{otherwise,} \end{dcases}\]
where
\[ \bchanPi{y_a,u_a}{z_{bk}} = (\bchanpi{\omegaa}{d_{bj}} + \bchanpi{\omegaa}{d_{bk}}).\] 
\end{example}

\begin{example}[Upgrading $W_b$ Based on Upgrade-Merge-3]\label{ex_Upgrading $W_b$ Based on Upgrade-Merge-3}
The upgrade-merge-3 procedure replaces three conjugate symbols pairs with two conjugate symbol pairs. The details of the transformation, in our notation, appear in Appendix~\ref{ap_upgrade merge 3}. 

As above, let joint channel $W_b(y_a,u_a,d_b|u_b)$ have b-channel marginal $W_b^*(d_b|u_b)$. For the upgrade procedure we select symbols $d_{bi}, d_{bj},d_{bk}$ and their respective conjugates, such that $0 \leq d_{bi} < d_{bj} \leq d_{bk}$.\footnote{We could have also selected them such that $0 \leq d_{bi} \leq d_{bj} < d_{bk}$. At least one of the inequalities $d_{bi}\leq d_{bj}$ or $d_{bj}\leq d_{bk}$ must be strict.} We upgrade $W_b^*(d_b|u_b)$ to $Q_b^*(z_b|u_b)$ given by~\eqref{eq_formula for upgrade merge 3}  (Appendix~\ref{ap_upgrade merge 3}). We denote by $\mathcal{D}_b$ the output alphabet of $W_b^*$  and by $\mathcal{D}_{z_{bk},z_{bi}}$ the set 
\[ \mathcal{D}_{z_{bk},z_{bi}} \triangleq \{d_{bk},d_{bj},d_{bi},\bar{d}_{bi},\bar{d}_{bj},\bar{d}_{bk}\}.\]
The output alphabet of $Q_b^*$ is $\mathcal{Z} = (\mathcal{D}_b \setminus \mathcal{D}_{z_{bk},z_{bi}} )\cup (z_{bk},z_{bi},\bar{z}_{bi},\bar{z}_{bk})$; outputs of $Q_b^*$ represent $D$-values. In particular, the $D$-values of $z_{bk}$ and $z_{bi}$ are $d_{bk}$ and $d_{bi}$, respectively.

Assuming that $W_b$ is symmetrized, we form channel $Q_b(y_a,u_a,z_b|u_b)$ using \Cref{thm_Wb can be upgraded to Qb with ya and va intact} as
\[ Q_b(y_a,u_a,z_b|u_b) = \begin{cases} \muyauazbk Q_b^*(z_{bk}|u_b), & z_b = z_{bk} \\[0.1cm]
										\muyauazbi Q_b^*(z_{bi}|u_b), & z_b = z_{bi} \\[0.1cm]
										\muyauazbibar Q_b^*(\bar{z}_{bi}|u_b), & z_b = \bar{z}_{bi} \\[0.1cm]
										\muyauazbkbar Q_b^*(\bar{z}_{bk}|u_b), & z_b = \bar{z}_{bk}\\[0.1cm]
										W_b(y_a,u_a,z_b|u_b), & \text{otherwise,} 
										\end{cases}\] 
where by~\eqref{eq_formula for mu},
\begin{align*}
\muyauazbk &= \frac{\bchanpi{y_a,u_a}{d_{bk}} + \left(\frac{d_{bj}-d_{bi}}{d_{bk}-d_{bi}} \right)\bchanpi{y_a,u_a}{d_{bj}}}{\bchanpi{}{d_{bk}} + \left(\frac{d_{bj}-d_{bi}}{d_{bk}-d_{bi}} \right)\bchanpi{}{d_{bj}}}, \\[0.2cm]
\muyauazbi &= \frac{\bchanpi{y_a,u_a}{d_{bi}} + \left(\frac{d_{bk}-d_{bj}}{d_{bk}-d_{bi}} \right)\bchanpi{y_a,u_a}{d_{bj}}}{\bchanpi{}{d_{bi}} + \left(\frac{d_{bk}-d_{bj}}{d_{bk}-d_{bi}} \right)\bchanpi{}{d_{bj}}}, 
\end{align*} 
and $\muyauazbkbar = \muyauazbk$, $\muyauazbibar = \muyauazbi$. The latter two equalities are due to our assumption that $W_b$ is symmetrized. 

Denoting 
\begin{align*}
	\bchanPi{y_a,u_a}{z_{bk}} &= 	\bchanpi{y_a,u_a}{d_{bk}} + \left(\frac{d_{bj}-d_{bi}}{d_{bk}-d_{bi}} \right)\bchanpi{y_a,u_a}{d_{bj}}\\
	\bchanPi{y_a,u_a}{z_{bi}} &= 	\bchanpi{y_a,u_a}{d_{bi}} + \left(\frac{d_{bk}-d_{bj}}{d_{bk}-d_{bi}} \right)\bchanpi{y_a,u_a}{d_{bj}}\\
						 &= 	\bchanpi{y_a,u_a}{d_{bi}} + \left(1-\frac{d_{bj}-d_{bi}}{d_{bk}-d_{bi}} \right)\bchanpi{y_a,u_a}{d_{bj}},
\end{align*}
 the upgraded joint channel is given by 
\[ Q_b(y_a,u_a,z_b|u_b) = \begin{dcases} \bchanPi{y_a,u_a}{z_{bk}}\left(\frac{1+(-1)^{u_b}d_{bk}}{2} \right),	& z_b = z_{bk} \\[0.1cm] 
 \bchanPi{y_a,u_a}{z_{bi}}\left(\frac{1+(-1)^{u_b}d_{bi}}{2} \right)	,& z_b = z_{bi} \\[0.1cm]
 \bchanPi{y_a,u_a}{z_{bi}}\left(\frac{1-(-1)^{u_b}d_{bi}}{2} \right)	,& z_b = \bar{z}_{bi} \\[0.1cm]
 \bchanPi{y_a,u_a}{z_{bk}}\left(\frac{1-(-1)^{u_b}d_{bk}}{2} \right)	,& z_b = \bar{z}_{bk} \\[0.1cm]
W_b(y_a,u_a,z_b|u_b), & \text{otherwise.} \end{dcases}\]
\end{example}
   
\begin{remark}
	We observe from these examples an interesting parallel between the a-channel and b-channel upgrading procedures. In the former case, we confine upgrade operations to a single class, in which the b-channel $D$-values are fixed. In light of the above examples, the latter case may be viewed as confining upgrade procedures to ``classes'' in which $y_a$ and $u_a$ are fixed. 
\end{remark}   

\section{Lower Bound Procedure} \label{sec_lower bound procedures}
The previous sections have introduced several ingredients for building an overall procedure for obtaining a lower bound on the probability of error of polar codes under SC decoding. We now combine these ingredients and present the overall procedure. First, we lower-bound the probability of error of two synthetic channels. Then, we show how to use lower bounds on channel pairs to obtain better lower bounds on the union of many error events.

\subsection{Lower Bound on the Joint Probability of Error of Two Synthetic Channels}

We now present an upgrading procedure for $W_{a,b}$ that results in channel $Q_{a,b}$ with a smaller alphabet size. The procedure leverages the recursive nature of polar codes. 

The input to our procedure is BMS channel $W$, the number of polarization steps $n$, the indices $a$
and $b$ of the a-channel and b-channel, respectively, and parameters $A$ and $B$ that control the
output alphabet sizes of the a- and b-channels, respectively. The binary expansions of $a-1$ and
$b-1$ are $\bv{a} = \langle \alpha_1,\alpha_2,\ldots,\alpha_m\rangle$ and $\bv{b} = \langle
\beta_1,\beta_2,\ldots,\beta_m\rangle$, respectively. These expansions specify the order of polarization transforms to be performed, where $0$ implies a `$-$'-transform and $1$ implies a `$+$'-transform. 

The algorithm consists of a sequence of polarization and upgrading steps. After each polarization step, we bring the channel to $D$-value representation, as described in \Cref{subsec_d value representation}. A side effect of polarization is increase in alphabet size. The upgrading steps prevents the alphabet size of the channels from growing beyond a predetermined size. After the final upgrading step we obtain joint channel $Q_{a,b}$, which is properly upgraded from $W_{a,b}$. We compute $P_e^{\IMJP}(Q_{a,b})$, which serves as a lower bound to $P_e^{\IML}(W_{a,b})$. We recall that $P_e^{\IML}(W_{a,b})$ is the probability of error under SC decoding of the joint synthetic channel $W_{a,b}$. This, in turn, lower-bounds $P_e^{\SC}(W)$ (see \Cref{cor_procedure leads to lower bound}).  

\Cref{alg_lowerbound} provides a high-level description of the procedure. We begin by determining
the first index $m$ for which $\alpha_m$ and $\beta_m$ differ (i.e. $\alpha_{\ell} = \beta_{\ell}$ for $\ell<m$ and
$\alpha_m\neq \beta_m$). The first $m-1$ polarization steps are of a single channel, as the a-channel and
b-channel indices are the same. Since these are single channels, we utilize the upgrading procedures
of~\cite{Tal_2013} to reduce the output alphabet size. At the $m$th polarization step, the a- and
b-channels differ. We perform joint polarization described in \Cref{sec_polarization for joint bit
channels} and symmetrize the joint channel using~\eqref{eq_symmetrized channel definition}. This
symmetrization need only be performed once as subsequent polarizations maintain symmetrization
(\Cref{prop_Symmetrizing the joint distribution yields a lower bound}). We then perform the
b-channel upgrading procedure (\Cref{subsec_upgrading $W_b$}), which reduces the b-channel alphabet
size to $2B$. Following that, we upgrade the a-channel. As discussed in \Cref{subsec_Upgrading
$W_a$}, this consists of two steps. First, we upgrade-couple the channel, to generate $B^2$ classes.
Second, for each class separately, we use the a-channel upgrade procedure until each class has at
most $2A$ elements (see \Cref{thm_upgrade split} and \Cref{cor_upgrade split}). We confine the
a-channel upgrade procedure to the class by utilizing only upgrade-merge-3 operations. 
We continue to polarize and upgrade the joint channel in this manner, until $\ell = n$. After the
final polarization and upgrading operation, we compute the probability of error of the IMJP decoder
for the resulting channel. 

\begin{algorithm}
\SetKwFunction{algfirstdiff}{first\_difference}
\SetKwFunction{algsinglepolarize}{single\_polarize}
\SetKwFunction{algsingleupgrade}{single\_upgrade}
\SetKwFunction{algjointpolarize}{jointly\_polarize}
\SetKwFunction{algDValue}{D-Value\_representation}
\SetKwFunction{algsymmetrize}{symmetrize}
\SetKwFunction{algaupgrade}{a-channel\_upgrade}
\SetKwFunction{algbupgrade}{b-channel\_upgrade}
\SetKwFunction{algupgradecouple}{upgrade\_couple}
\SetKwData{algclass}{class}
\DontPrintSemicolon

\caption{A lower bound on the probability of error under SC decoding of a joint synthetic channel}
\label{alg_lowerbound}
\KwIn{BMS channel $W$, number of polarization steps $n$, channel indices $a$,$b$, and alphabet-size
control parameters $A$, $B$. The binary representations of $a-1$ and $b-1$ are $\bv{a} = \langle
\alpha_1,\alpha_2,\ldots,\alpha_n\rangle$ and $\bv{b} = \langle \beta_1,\beta_2,\ldots,\beta_n\rangle$, respectively.}
\KwOut{A lower bound on the probability of error $W_{a,b}$.}  
$ m \leftarrow \algfirstdiff(\bv{a},\bv{b})$\;
$ Q \leftarrow \algsingleupgrade(W,\max\{A,B\})$\;
\For{$\ell=1,2,\ldots,n$}
{
	\eIf{$\ell<m$}
	{
		$Q \leftarrow \algsinglepolarize(Q,\alpha_{\ell})$\;
		$Q \leftarrow \algDValue(Q)$\; 
		$Q \leftarrow \algsingleupgrade(Q,\max\{A,B\})$\;
	}
	{
		$Q \leftarrow \algjointpolarize(Q,\alpha_{\ell},\beta_{\ell})$\;
		$Q \leftarrow \algDValue(Q)$\; 
		\If{$\ell=m$}
		{	
			$Q \leftarrow \algsymmetrize(Q)$\; 
		}
		\tcp*[r]{b-channel upgrade:}
		$Q \leftarrow \algbupgrade(Q,B)$\;
		\tcp*[r]{a-channel upgrade:}
		$Q \leftarrow \algupgradecouple(Q)$\; 
		\ForEach{\algclass $\in Q$}
		{
			$Q \leftarrow \algaupgrade(Q,A,\algclass)$\;
			\tcc{Confine to class by using only upgrade-merge-3.}
		}
	}
}
	\Return{$P_e^{\IMJP}(Q)$}

\end{algorithm}

The lower bound of this procedure compares favorably with the trivial lower bound, $\max\{\Prob{\mathcal{E}_a}, \Prob{\mathcal{E}_b}\}$. This is because our upgrading procedure only ever changes one marginal, keeping the other intact. Since it leverages upgrading transforms that can be used on single channels, the marginal channels obtained are the same as would be obtained on single channels using the same upgrading steps. Thus, by \Cref{lem_IMJP provides a tighter lower bound than max Pe} this lower bound is at least as good as $\max\{\Prob{\mathcal{E}_a}, \Prob{\mathcal{E}_b}\}$.

\begin{remark}
	When the BMS $W$ is a BEC, we can recover the bounds of~\cite{Mori_Tanaka_2009} and \cite{Parizi_2013}  using our upgrading procedure. Only a-channel upgrades are required, as the b-channel, in $D$-value representation, remains a BEC. For each a-channel symbol, the channel $W_2$ in \eqref{eq_decomposition of Wb, db version} is either a perfect channel or a pure-noise channel (see \Cref{lem_for BEC erasures are based on channel outputs only} in Appendix~\ref{ap_IMJP for BEC}). Thus, the upgrade-couple procedure splits the a-channel symbols to those that see a perfect channel regardless of $u_a$  and those that see a pure-noise channel regardless of $u_a$. Merging a-channel symbols of the same class is equivalent to merging a-channel symbols for which $\upgradec{W}_2$ is the same type of channel. We thus merge a-channel symbols of the same a-channel $D$-value that ``see'' the same type of b-channel. This corresponds to keeping track of the correlation between erasure events of the two channels. 
\end{remark}

\begin{remark}
	An initial step of \Cref{alg_lowerbound} is to upgrade the channel $W$, even before any polarization operations. This step enables us to apply our algorithm on continuous-output channels, see~\cite[Section VI]{Tal_2013}.
\end{remark}

\subsection{Lower Bound for More than Two Synthetic channels}\label{subsec_lower bound for more than two channels}
Recall that the probability of error of polar codes under SC decoding may be expressed as
$\Prob{\bigcup_{a\in \mathcal{A}} \mathcal{E}_a}$. In the previous section, we developed a lower
bound on $\Prob{\mathcal{E}_a \cup \mathcal{E}_b}$, $a<b$, which lower bounds $\Prob{\bigcup_{a\in
\mathcal{A}} \mathcal{E}_a}$. This lower bound may be strengthened by considering several pairs of synthetic channels and using~\eqref{eq_inclusion exclusion lower bound}. We now show how this can be done.

\begin{lemma}\label{lem_lower bound on union using unions of two events}
The probability of error of a union of $M$ events, $\cup_{a=1}^M\mathcal{E}_a$ is lower bounded by
\[ \Prob{\bigcup_{a=1}^{M} \mathcal{E}_a} \geq \sum_{a<b}\Prob{\mathcal{E}_a\cup\mathcal{E}_b} - (M-2)\sum_a \Prob{\mathcal{E}_a}.\]
\end{lemma}
\begin{IEEEproof}
The proof hinges on using the identity $\Prob{\mathcal{E}_a\cap\mathcal{E}_b} =\Prob{\mathcal{E}_a}+\Prob{\mathcal{E}_b}- \Prob{\mathcal{E}_a\cup\mathcal{E}_b}$ in~\eqref{eq_inclusion exclusion lower bound}. Note that any set of $M$ numbers $\{p_1,p_2,\ldots,p_M\}$ satisfies
\begin{align*} 2M\sum_a p_a &= \sum_{a,b} (p_a + p_b)\\
								        &= \sum_{a<b}(p_a+p_b) + \sum_{a=b}(p_a+p_b) + \sum_{a>b}(p_a+p_b)\\
									    &= 2\sum_{a<b}(p_a+p_b) + 2\sum_a p_a, 	
\end{align*}
so that 
\[ \sum_{a<b} (p_a + p_b) = (M-1)\sum_a p_a.\] 
Therefore,  
\begin{align*}
\sum_{a<b}\Prob{\mathcal{E}_a\cap\mathcal{E}_b} &= \sum_{a<b}\left( \Prob{\mathcal{E}_a}+\Prob{\mathcal{E}_b}- \Prob{\mathcal{E}_a\cup\mathcal{E}_b}  \right) \\
&= (M-1)\sum_a \Prob{\mathcal{E}_a} - \sum_{a<b} \Prob{\mathcal{E}_a\cup\mathcal{E}_b}. 
\end{align*}
Using this in~\eqref{eq_inclusion exclusion lower bound} yields the desired bound. 
\end{IEEEproof}

In practice, we combine the lower bound of \Cref{lem_lower bound on union using unions of two events} with~\eqref{eq_lower bound on a union using a subset}. That is, we compute lower bounds on $\Prob{\mathcal{E}_a \cup \mathcal{E}_b}$ for all pairs of channels in some subset $\mathcal{A}'$ of the non-frozen set, and use \Cref{lem_lower bound on union using unions of two events} over this subset.  

Such bounds are highly dependent on the selection of the subset $\mathcal{A}'$. One possible
strategy is as follows. Let $\mathcal{B}$ be the set of $k$ worst synthetic channels in the
non-frozen set for some $k$. For each channel pair in $\mathcal{B}$, compute a lower bound on the
joint probability of error using \Cref{alg_lowerbound}. Then, form all possible subsets of
$\mathcal{B}$ (there are $2^k$ such subsets) and use \Cref{lem_lower bound on union using unions of
two events} for each subset. Choose the subset with the highest upper bound as $\mathcal{A}'$. The
reason for going over all possible subsets is that  bounds based on the inclusion-exclusion
principle are not guaranteed to be higher than the highest pairwise probability,
see~\cite{Schwager_1984}.

\section{Implementation} \label{sec_implementation}
Our implementation of \Cref{alg_lowerbound}, in C++, is available for download at  \cite{STCode}.
In this section we provide some details on the implementation.

A naive implementation of \Cref{alg_lowerbound} is to perform all steps successively at each iteration. 
That is, first jointly polarize the joint channel, then bring the channel to $D$-value
representation, followed by the b-channel upgrade procedure and the upgrade-couple procedure, and
finally perform the a-channel upgrade procedure. One quickly finds out, however, a limitation
posed by this approach: the memory required to store the outcomes of these stages becomes
prohibitively large when the alphabet-size control parameters $A$ and $B$ grow. 

Observe, however, that the total required memory at the end of each iteration of
\Cref{alg_lowerbound} is actually quite small. We need only store the values of
$\upgradec{W}_a(y_a^{i,j}|0)$ for each value of $y_a,i,j$ (a total of $2A\cdot B^2$ combinations),
a mapping between $y_a$ and its conjugate $\bar{y}_a$, and a list of size $B$ that stores the
possible b-channel $D$-values. Then, we can
compute $\upgradec{W}_b(y_a^{i,j},u_a,d_b|u_b)$ using~\eqref{eq_decoupling decomposition of check Wb},~\eqref{eq_W2
for decoupling decomposition}, and \Cref{cor_conjugate symbols of upgrade-couple channel}. 
Thus, our data structure for an upgrade-coupled joint channel utilizes a three-dimensional matrix
 of size $(2A)\times B\times B$ to store $\upgradec{W}_a(y_a^{i,j}|0)$ (specifically, we use the cube
 data structure provided by~\cite{armadillo}). As for the mapping between $y_a$ and its conjugate,
 if $\upgradec{W}_a(y_a^{i,j}|0)$ is stored in element
 \texttt{(y,i,j)} of the matrix, and \texttt{y} is even, then $\upgradec{W}_a(\bar{y}_a^{i,j}|0)$ is
 stored in element \texttt{(y+1,i,j)}. 
We store the absolute values of the b-channel
$D$-values in a vector of length $B$. 

The second key observation is that each upgrading procedure only ever changes one marginal. That is,
the a-channel upgrading procedure leaves the marginal b-channel unchanged, and the b-channel
upgrading procedure does not affect the marginal a-channel. Thus, since our upgrading procedure
leverage upgrading procedures for single channels, we can pre-compute the upgraded marginal
channels. In essence, given a target upgraded marginal
channel --- computed beforehand using the techniques of \cite{Tal_2013} --- our upgrading procedures ``split'' the
probability of a output symbol among two absorbing symbols. The ``splitting'' factors are functions of the
$D$-values of the three symbols (see appendix~\ref{ap_BMS channel upgrades}). Indeed, we compute
beforehand the polarized and upgraded marginal channels.

The joint polarization step maps each pair of symbols, $y_{a_1}^{i_1,j_1}$ and $y_{a_2}^{i_2,j_2}$ to up
to four polarized counterparts (see \Cref{sec_polarization for joint bit channels}). Knowing
beforehand what the upgraded marginal channels should be, we can directly split each polarized symbol
into the relevant absorbing symbols. We incorporate the upgrade-couple operation into this by
utilizing the factor $\alpha_{i,j}$ from~\eqref{eq_definition of alphaij}. 

Thus, in our implementation, rather than performing each step of an iteration in its entirety,
we perform all steps in one fell swoop. This sidesteps the memory-intensive step of computing the
upgrade-coupled jointly polarized channel.  The interested reader is urged to look at our source code
for further details.

\begin{remark}
    The description here was given in terms of $D$-values, in line with the exposition in this paper. However, for numerical purposes we
    recommend --- and use --- likelihood ratios in practical implementation. Likelihood ratios have a greater dynamic
    range than that of $D$-values, and therefore offer better numerical precision.\footnote{As an
        example, two very different likelihood ratios: $\lambda_1 = 10^{20}$ and $\lambda_2
    = 10^{30}$, cannot be differentiated in double precision upon conversion to
    $D$-values.}
    There is a one-to-one correspondence between $D$-values and likelihood ratios (see
    appendix~\ref{ap_Definition of D values}), and all $D$-value based formulas are easily
    translated to their likelihood ratio counterparts. 
\end{remark}

\section{Numerical Results} \label{sec_numerical results}
\begin{figure}[t]
\centering

\definecolor{mycolor1}{rgb}{1.00000,0.00000,1.00000}%
\begin{tikzpicture}

\begin{axis}[%
width=7cm,
height=7cm,
at={(1.011111in,0.641667in)},
scale only axis,
grid = major,
xmin=0.08,
xmax=0.18,
ymode=log,
ymin=1e-07,
ymax=1,
yminorticks=true,
xlabel={Crossover probability},
ylabel={Probability of error},
legend style={at={(0.02,0.99)},anchor=north west,legend cell align=left,align=left,fill=white, draw=none}
]

\addplot [color=cyan,solid,mark=x,mark options={solid}]
	table[row sep=crcr]{%
0.08	8.82250494144101e-07\\
0.09	5.04844602362815e-06\\
0.1	2.44972367069049e-05\\
0.11	0.000107496599218954\\
0.12	0.000449200020787486\\
0.13	0.00184562015588825\\
0.14	0.00749906609747055\\
0.15	0.0293520647326518\\
0.16	0.105616253923844\\
0.17	0.335763903395806\\
0.18	0.926414891054271\\
};
\addlegendentry{\small Upper Bound};

\addplot [color=blue,solid,mark=square,mark options={solid}]
  table[row sep=crcr]{%
0.08	2.639787e-07\\
0.09	1.477638e-06\\
0.1	6.751064e-06\\
0.11	2.610095e-05\\
0.12	8.792738e-05\\
0.13	0.0002628447\\
0.14	0.0007092876\\
0.15	0.001747507\\
0.16	0.003974171\\
0.17	0.009449577\\
0.18	0.02214191\\
};\addlegendentry{\small Trivial Lower Bound};

\addplot [color=red,solid,mark=o,mark options={solid}]
  table[row sep=crcr]{%
0.08   8.703783218479992e-07 \\      
0.09   4.892176750349999e-06 \\
0.10   2.306111679089992e-05\\ 
0.11    9.517960251400020e-05\\
0.12    3.576489400799994e-04\\
0.13    0.001232854885330 \\ 
0.14    0.004042925069200\\ 
0.15    0.012600137223100\\ 
0.16    0.033952105686000\\
0.17    0.086413934220000 \\ 
0.18    0.181440178360001 \\
};

\addlegendentry{\small New Lower Bound};

%


\end{axis}
\end{tikzpicture}
\caption{Bounds on the probability of error of a rate $0.25$ polar code of length $2^{11} = 2048$
designed for a BSC with crossover probability $0.18$. The code was used over BSCs with a range of
crossover probabilities. The upper bound is based on~\cite{Tal_2013}. The trivial lower bound is
a lower bound on $\max_{a \in \mathcal{A}} P_e^{\ML}(W_a)$. The new lower bound was computed using the techniques of this paper.}
\label{fig_bounds2}
\end{figure}
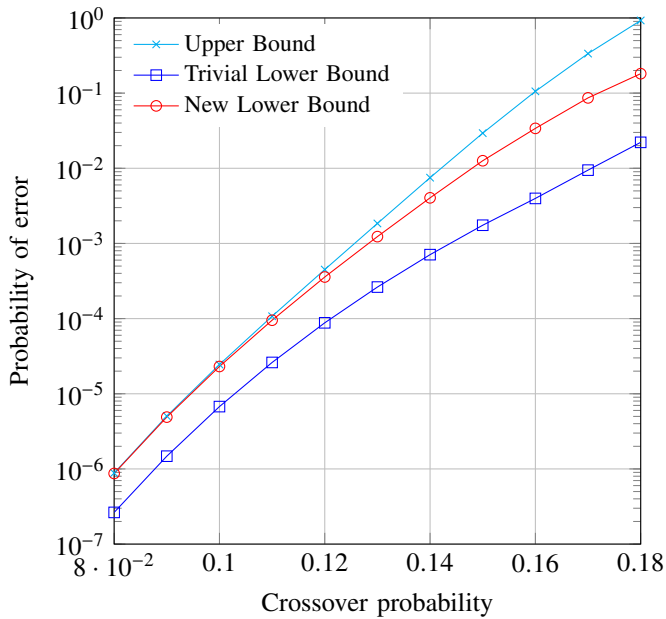

\Cref{fig_bounds,fig_bounds2} present numerical results of our bound for two cases. In both cases,
we designed a polar code for a specific BSC, and then assessed its performance when used over
different BSCs. Specifically:
\begin{itemize}
    \item \emph{\Cref{fig_bounds}:} A code of length $N=2^{10}=1024$, rate $R=0.1$, designed for a BSC with
        crossover probability $0.2$. 
    \item \emph{\Cref{fig_bounds2}:} A code of length $N=2^{11} = 2048$, rate $R=0.25$, designed for a BSC
        with crossover probability $0.18$. 
\end{itemize}
The codes were designed using the techniques of~\cite{Tal_2013} with $128$ quantization levels. The
non-frozen set $\mathcal{A}$ consisted of the $\lfloor NR\rfloor$ channels with smallest probability
of error. This non-frozen set was fixed. 

For each code, we plot three bounds on the probability of error, when used over specific BSCs: an
upper bound on the probability of error, the trivial lower bound on the probability of error, and
the new lower bound on the probability of error presented in this paper. 

For the upper bound, we computed an upper bound on $\sum_{a\in \mathcal{A}} P_e^{\ML}(W_a)$, and for
the trivial lower bound we computed a lower bound on $\max_{a\in\mathcal{A}}P_e^{\ML}(W_a)$; upper
and lower bounds on the probability of error of single channels (i.e., on $P_e^{\ML}(W_a)$) were
obtained using the techniques of~\cite{Tal_2013}.
The new lower bound is based on the IMJP decoder, as described in this paper. We computed the IMJP
decoding error, with $2A=2B=32$ for all possible pairs of the $20$ worst channels in the non-frozen
set.\footnote{Note that there is a different set of $20$ worst channels for each crossover
    probability. For each crossover probability, we selected the $20$ channels in the (fixed)
    non-frozen set with the highest upper bound on decoding error when used over a BSC with \emph{that} 
crossover probability.} We
    then used \Cref{lem_lower bound on union using unions of two events}, computed for the subset of
these $20$ channels that yielded the highest bound; this provides a significantly improved bound over the
bound given by the worst-performing pair. 
The computation utilized \cite{parallel} for
parallel computation of the IMJP decoding error over different channel pairs. 
 
As one may observe, our bounds improve upon the previously known lower bound~\eqref{eq_trivial lower
bound}. In fact, they are quite close to the upper bound on the probability of error. This provides
strong numerical evidence that error events of channel pairs dominate the error probability of polar
codes under SC decoding. 

\section{Discussion and Outlook} \label{sec_discussion}
This research was inspired by~\cite{Parizi_2013}, which showed that --- for the BEC --- the union bound on the probability of error of polar codes under SC decoding is asymptotically tight. The techniques of~\cite{Parizi_2013} hinged on the property that a polarized BEC is itself a BEC. Or, put another way, that the family of binary erasure channels is closed under the polar transform. This property enabled the authors to directly track the joint probability of erasure during the polarization process and bound its rate of decay. Unfortunately, this property is not shared by other channel families. 

Design of polar codes for channel coding is based on selecting a set of indices to be frozen. One
design rule is to select the worst-performing indices as the frozen set. For example, for a code of
length $N$ and rate $R$, choose the $N(1-R)$ indices with the highest probability of error (such
channels can be identified using the techniques of~\cite{Tal_2013}). This design rule optimizes the
union bound on the probability of error of polar codes,~\eqref{eq_union bound}. As Parizi and
Telatar have shown in~\cite{Parizi_2013}, for the BEC such a design rule is essentially optimal. It
is an open question whether a similar claim can be made for other BMS channel families. 

 As our numerical results show, below a certain
crossover probability the upper bound and our lower bound all but coincide, with a significant gap to the
trivial lower bound. 
Thus, we conjecture that the ratio between the union bound and the actual probability of
error approaches $1$ asymptotically for \emph{any} BMS channel. This will imply the essential optimality of the 
the union bound as a design rule. 
 Moreover, we believe that the tools developed in this research are key to proving this
conjecture.

One possible approach is to track analytically the evolution of joint error probabilities during the
polarization process. The symmetrization transformation and the resultant decoupling decomposition
bring joint channels to a form more amenable to analysis. One may look at, for example, the
Bhattacharyya parameter of the channel $W_2$ from~\eqref{eq_decomposition of W_b to W1W2}, when
$u_a,y_a$ are fixed,
\[Z_{b|y_a,u_a} = \sum_{d_b}\sqrt{W_2(d_b|0;y_a,u_a) W_2(d_b|1;y_a,u_a)}.\]     
This quantity, together with the Bhattacharyya parameters of the a-channel, may be used to bound
$\Prob{\mathcal{E}_a \cap \mathcal{E}_b}$. Tracking the evolution of these parameters --- or bounds
on them --- may enable the study of the decay of $\Prob{\mathcal{E}_a \cap \mathcal{E}_b}$ (if indeed there is such decay). In fact, it can be shown that applying the above suggestion to the BEC coincides with the approach of~\cite{Parizi_2013}.   

Interestingly, our bounds are tight despite the various manipulations they perform on the joint channel. The
joint channels that result from our procedure are very different from the actual joint channel, yet
have no effect on the marginal distributions. This curious outcome merits further research on the
upgrade-couple transform and its effect on the joint channel. 
%

%
%

There are several additional avenues of further research. These include:
\begin{itemize}
\item Our results apply only to BMS channels. It would be interesting to extend them to richer settings, such as channels with non-binary input, or non-symmetric channels.
\item This research has concentrated on SC decoding. Can it be expanded/applied to other decoding methods for polar codes (e.g., successive cancellation list (SCL) decoding~\cite{Tal_2015})? A logical first step in analyzing SCL decoding is to look at pairs of error events, as done here. 
\end{itemize}

\section*{Acknowledgment}
The assistance of Ina Talmon is gratefully acknowledged. 
 
\appendices

\section{The IMJP decoder for a BEC}\label{ap_IMJP for BEC}
In the special case where $W$ is a BEC and $W_a$ and $W_b$ are two of its polar descendants, we have the following.  
\begin{proposition}	
 \label{thm_for BEC P(E1UE2) is identical to P(EML1UEML2)}
Let $W_a(y_a|u_a)$ and $W_b(y_a,u_a,y_{\ba}|u_b)$ be two polar descendants of a BEC in the same tier. Then, the IMJP and the IML (SC) decoders coincide.
\end{proposition} 
To prove this, we first show that for the BEC erasures are determined by the received channel symbols, $y_1^{2^n}$, and not previous bit decisions. 
This implies that for fixed $y_a$, regardless of $y_{\ba}$ and in particular $u_a$, either channel $W_b$ always experiences an erasure, or always experiences a non-erasure. If $W_b$ experiences an erasure, it doesn't matter what $\phi_a$ decides in terms of the IMJP decoder -- it may as well use an ML decoder; if $W_b$ does not experience an erasure, then the best bet of $W_a$ is to use an ML decoder. This suggests that the IML and IMJP decoders coincide. 

\begin{lemma}\label{lem_for BEC erasures are based on channel outputs only}
	Let $W_a(y_1^{2^n},u_1^{a-1}|u_a)$ be a polar descendant of a BEC, $W$. Then, there exists a set $E_n$, dependent only on $a$, such that $W_a$ has an erasure if and only if $y_1^{2^n} \in E_n$. 
\end{lemma}

\begin{IEEEproof}Here, $y_1^{2^n}$ are the received channel symbols, and $u_1^{a-1}$ the previous
    bit decisions that are part of $W_a$'s output. Let $\langle \alpha_1,\alpha_2,\ldots, \alpha_n
    \rangle$ be the binary expansion of $a-1$, with $\alpha_1$ the MSB. Recall that channel $W_a$ is
    the result of $n$ polarization steps determined by $\alpha_1,\alpha_2,\ldots, \alpha_n$, where
    $\alpha_j = 0$ is a `$-$'-transform and $\alpha_j = 1$ is a `$+$'-transform. 

Consider first the case where $n=1$, i.e., $a-1 = \alpha_1$. If $\alpha_1 = 0$ then $W_a = W^-$ has
an erasure if and only if at least one of $y_1,y_2$ is an erasure, i.e., if and only if $y_1^2 \in
E_1$, $E_1 =  \{y_1^2| y_1 = e \text{ or } y_2 = e\}$. If $\alpha_1 = 1$ then $W_a = W^+$ has an erasure if and only if both $y_1$ and $y_2$ are erasures, i.e., if and only if $y_1^2 \in E_1$, $E_1 = \{y_1^2 |y_1 = e \text{ and } y_2 = e\}$. Therefore, the claim is true for $n=1$.

We proceed by induction. Let the claim be true for $n-1$: for $a'-1=\langle
\alpha_1,\alpha_2,\ldots, \alpha_{n-1} \rangle$, there exists a set $E_{n-1}$ such that $W_{a'}$ has
an erasure if and only if $y_1^{2^{n-1}} \in E_{n-1}$. If $\alpha_n = 0$, then $W_a$ is the result
of a `$-$'-transform of two BEC channels $W_{a'}$, so it has an erasure if and only if at least one
of them erases. In other words, $W_a$ has an erasure if and only if $y_1^{2^n} \in E_n$, $E_n
= \{y_1^{2^n}| y_1^{2^{n-1}} \in E_{n-1} \text{ or } y_{2^{n-1}+1}^{2^{n}} \in E_{n-1}\}$. If,
however, $\alpha_n = 1$, then $W_a$ is the result of a `$+$'-transform of two BEC channels $W_{a'}$,
so it has an erasure if and only if both of them erase. In other words, $W_a$ has an erasure if and only if $y_1^{2^n} \in E_n$, $E_n = \{y_1^{2^n}| y_1^{2^{n-1}} \in E_{n-1} \text{ and } y_{2^{n-1}+1}^{2^{n}} \in E_{n-1}\}$. Thus, the claim is true for $n$ as well, completing the proof. 
\end{IEEEproof}

\begin{IEEEproof}[Proof of \Cref{thm_for BEC P(E1UE2) is identical to P(EML1UEML2)}]
	By \Cref{lem_optimal phi2 is ML decoder}, a decoder $\phi_b$ that minimizes $\Prob{\mathcal{E}_a \cup \mathcal{E}_b}$ is an ML decoder. It remains to show that a minimizing $\phi_a$ is also an ML decoder. 
	Marginalizing the joint channel~\eqref{eq_Wab and its relationship to Wb} yields $W_a$:  
	\[ W_a(y_a | u_a) = \sum_{\substack{u_b,y_b}} W_b(y_b|u_b) \kindi{y_b = (y_a,u_a,y_{\ba})}.\]
	The ML decoder for channel $W_a$ maximizes $W_a(y_a|u_a)$ with respect to $u_a$; decoder $\phi_a$, on the other hand, maximizes $T(y_a|u_a)$, defined in~\eqref{eq_def of T(y_a|x_a)}. Using~\eqref{eq_Wab  and its relationship to Wb} we recast the expression for $T$ in the same form as the expression for $W_a$,
\iftoggle{twocol}{
\begin{align*} &T(y_a|u_a) \\&\quad = \sum_{u_b,y_b} W_b(y_b|u_b)\kindi{y_b = (y_a,u_a,y_{\ba})}\cdot\Prob{\phi_b(y_b) = u_b}.\end{align*}}{
\[ T(y_a|u_a) = \sum_{u_b,y_b} W_b(y_b|u_b)\kindi{y_b = (y_a,u_a,y_{\ba})}\cdot\Prob{\phi_b(y_b) = u_b}.\]}

	By \Cref{lem_for BEC erasures are based on channel outputs only}, whether $W_b$ has an erasure depends solely on the received channel symbols, which are wholly contained in $y_a$, and not on previous bit decisions. In particular, in computing $W_a$ or $T$, we either sum over only erasure symbols or over only non-erasure symbols.  
	Since $\phi_b$ is an ML decoder for $W_b$, if $y_b$ is an erasure of $W_b$ then $W_a(y_a|u_a) = 2 T(y_a|u_a)$; if $y_b$ is not an erasure of $W_b$ then $W_a(y_a|u_a) = T(y_a|u_a)$. In either case, it is clear that 
	the decision based on~\eqref{eq_phi1 as argmax} is identical to the ML decision. Therefore, $\phi_a$ is an ML decoder as well, implying that the IMJP decoder is an IML decoder.
\end{IEEEproof}

\section{Introduction to $D$-values}\label{ap_Definition of D values}
The decision of an ML decoder for a memoryless binary-input channel $W_{Y|U}$ may be based on any sufficient statistic of the channel output. One well-known sufficient statistic is the log-likelihood ratio (LLR), $l(y) = \log \left( \frac{W_{Y|U}(y|0)}{W_{Y|U}(y|1)} \right)$. When $l(y)$ is positive, the decoder declares that $0$ was transmitted; when $l(y)$ is negative, the decoder declares that $1$ was transmitted;   $l(y) = 0$ constitutes an erasure, at which the decoder makes some random choice. Another sufficient statistic is the $D$-value. 

The $D$-value of output $y$, $d(y)$, is given by 
\begin{equation}\label{eq_d} d(y) \triangleq W_{U|Y}(0|y)-W_{U|Y}(1|y).
\end{equation} Clearly, $-1 \leq d(y) \leq 1$. A maximum likelihood decoder makes its decision based on the sign of the $D$-value.
Assuming a symmetric channel input, $U=0,1$ with probability $1/2$, using Bayes' law on~\eqref{eq_d} yields
\begin{equation} \label{eq_d_alt} 
d(y) = \frac{W_{Y|U}(y|0)-W_{Y|U}(y|1)}{W_{Y|U}(y|0)+W_{Y|U}(y|1)}
\end{equation}

The input is binary, hence $W_{U|Y}(0|y)+W_{U|Y}(1|y)=1$. Consequently~\eqref{eq_d_alt} yields
\begin{align*}
 \frac{1+ d(y)}{2} &= \frac{W_{Y|U}(y|0)}{W_{Y|U}(y|0)+W_{Y|U}(y|1)} =  W_{U|Y}(0|y),\\  
 \frac{1 - d(y)}{2} &=  \frac{W_{Y|U}(y|1)}{W_{Y|U}(y|0)+W_{Y|U}(y|1)} = W_{U|Y}(1|y). 
\end{align*}
There is a one-to-one correspondence between $d(y)$ and $l(y)$,
$l(y) = \log \frac{1+d(y)}{1-d(y)},$ or, equivalently,
$ d(y) = \tanh(l(y)/2).$

If channel $W_{Y|U}$ is symmetric, for each output $y$ there is a conjugate output $\bar{y}$; their LLRs and $D$-values are related: 
$l(\bar{y}) = \frac{1}{l(y)}, d(\bar{y}) = -d(y).$

Since the $D$-value is a sufficient statistic of a BMS channel, we may replace the channel output with its $D$-value. Thus, we may assume that the output $y$ of channel $W_{Y|U}$ is a $D$-value, i.e., $y = W_{U|Y}(0|y)-W_{U|Y}(1|y)$. In this case, we say that $W$ is in $D$-value representation.  

Recall that every BMS channel can be decomposed into BSCs~\cite[Theorem 2.1]{Land2006}. We can think of the output of a BMS as consisting of the ``reliability'' of the BSC and its output. The absolute value of the $D$-value corresponds to the BSC's reliability and its sign to the BSC output ($0$ or $1$). 

A comprehensive treatment of $D$-values and LLRs in relation to BMS channels appears in~\cite[Chapter 4]{mct}.     

\section{Upgrades of a BMS Channel}\label{ap_BMS channel upgrades}

We state here in our notation the two upgrades of a BMS channel from~\cite{Tal_2013}.

Let $W$ be a discrete BMS whose outputs are $D$-values $\pm d_1, \pm d_2, \ldots, \pm d_m$, and let the probability of symbol $d_{\ell}$ be $\achanpi{d_{\ell}} \triangleq W(d_{\ell}|u) +  W(-d_{\ell}|u) = W(d_{\ell}|0) + W(d_{\ell}|1)$, $\ell = 1,\ldots,m$. Without loss of generality,  
 $0\leq d_1 \leq d_2 \leq \cdots \leq d_m \leq 1$. Clearly, $\achanpi{d_{\ell}} \geq 0$ for all $\ell$, and $\sum_{\ell=1}^m \achanpi{d_{\ell}} = 1$. Moreover, $\achanpi{d_{\ell}} = \achanpi{-d_{\ell}}$.
Namely, this is a BMS that decomposes to $m$ different BSCs, with crossover probabilities $(1-d_{\ell})/2$, $\ell=1,\ldots,m$. BSC channel $\ell$ is selected with probability $\achanpi{d_{\ell}}$. We have $W(d_{\ell}|u) = (\achanpi{d_{\ell}}/2) \cdot (1+(-1)^u d_{\ell})$ and $W(-d_{\ell}|u) = W(d_{\ell}|\bar{u})$. 

\subsection{The Upgrade-merge-2 Procedure}\label{ap_upgrade merge 2} The first upgrade-merge of~\cite{Tal_2013} takes two $D$-values $d_j \leq d_k$ and merges them by transferring the probability of $d_j$ to $d_k$. We call it  \emph{upgrade-merge-2}.  
Channel $W:\mathcal{U} \to \mathcal{Y}$ is upgraded to channel $Q^{(2)}:\mathcal{U} \to \mathcal{Z}$; the output alphabet of $Q^{(2)}$ is 
$ \mathcal{Z}  = (\mathcal{Y} \setminus \{d_j,d_k,-d_j,-d_k\}) \cup \{z_k, -z_k\},$
and  
\begin{equation} Q^{(2)}(z|u) = \begin{dcases} 
 \achanpi{z_{k}}\left(\frac{1 + (-1)^u d_k}{2}\right), & z = z_k \\[0.1cm]
 \achanpi{z_{k}}\left(\frac{1 - (-1)^u d_k}{2}\right), & z = -z_k\\[0.1cm] 
W(z|u), & \text{otherwise,}
\end{dcases}\label{eq_Formula for simple upgrade merge 2}
\end{equation}
where
\[ \achanpi{z_{\ell}} = \begin{cases} 0, & {\ell} = j \\ 
 						   \achanpi{d_j} + \achanpi{d_k}, & {\ell} = k \\ 
 						   \achanpi{d_{\ell}}, & \text{otherwise.}
 			 \end{cases} \]	

The degrading channel from $Q^{(2)}$ to $W$ is shown in  \Cref{fig_simple upgrading}. We show only the portion of interest, i.e., we do not show the symbols that this degrading channel does not change. The parameters of the degrading channel are 
\begin{align*}
p_1 &= \frac{\achanpi{d_j}}{\achanpi{d_j}+\achanpi{d_k}}\left(\frac{d_k+d_j}{2d_k}\right), \\
p_2 &= \frac{\achanpi{d_k}}{\achanpi{d_j}+\achanpi{d_k}},  \\
p_3	&= \frac{\achanpi{d_j}}{\achanpi{d_j}+\achanpi{d_k}}\left(\frac{d_k-d_j}{2d_k}\right). 
\end{align*}
Indeed, $p_1,p_2,p_3 \geq 0$ and $p_1+p_2+p_3 = 1$, so this constitutes a valid channel. Note that if $d_j = d_k$  then  $p_3 = 0$. 
\tikzset{->-/.style={decoration={markings,
			mark=at position #1 with {\arrow{>}}},postaction={decorate}}}
\begin{figure}[t]	\centering
\subfloat[Degrading channel from $Q^{(2)}$ to $W$ for upgrade-merge-2.]{
	\begin{tikzpicture}[>=latex]
	
	\node at (-1.5,-0.5) {$Q^{(2)}(z|u)$};
	\node at (4.5,-0.5) {$W(d|u)$};
	\fill (0, 0) node (Z2){} circle (2pt)  node[left=0.1]  {$z_k$}; 
	\fill (0,-1) node (Z2bar){} circle (2pt)  node[left=0.1]  {$-z_k$}; 
	\fill (3, 0) node (Y1){} circle (2pt)  node[right=0.3]  {$\,d_j$};
	\fill (3, 1) node (Y2){} circle (2pt)  node[right=0.3] {$\,d_k$}; 
	\fill (3,-1) node (Y1bar){} circle (2pt)  node[right=0.1]  {$-d_j$}; 
	\fill (3,-2) node (Y2bar){} circle (2pt)  node[right=0.1]  {$-d_k$};  
	
	\draw[->-=0.5] (Z2) -- node[above= -2pt] {$p_1$} (Y1); 
	\draw[->-=0.5] (Z2) -- node[above = -2pt] {$p_2$} (Y2);
	\draw[->-=0.3] (Z2) -- node[above = -3pt, near start] {$p_3$} (Y1bar); 
	\draw[->-=0.5] (Z2bar) -- node[below= -1pt] {$p_1$} (Y1bar); 
	\draw[->-=0.5] (Z2bar) -- node[below= -1pt] {$p_2$} (Y2bar);
	\draw[->-=0.3] (Z2bar) -- node[below= -3pt, near start] {$p_3$} (Y1); 
	
	\end{tikzpicture}
	\label{fig_simple upgrading}
	}\quad
\subfloat[Degrading channel from $Q^{(3)}$ to $W$ for upgrade-merge-3.]{
	\begin{tikzpicture}[>=latex]
	
	\node at (-1.5,-1) {$Q^{(3)}(z|u)$};
	\node at (4.5, -1) {$W(d|u)$};
	\fill (0, 0) node (Z3){} circle (2pt)  node[left=0.1]  {$z_k$}; 
	\fill (0,-0.5) node (Z1){} circle (2pt)  node[left=0.1]  {$z_i$}; 
	
	\fill (0, -1.5) node (Z1b){} circle (2pt)  node[left=0.1]  {$-z_i$}; 
	\fill (0,   -2) node (Z3b){} circle (2pt)  node[left=0.1]  {$-z_k$}; 
	
	\fill (3,  0.25) node (Y3){} circle (2pt)  node[right=0.3]  {$\,d_k$};
	\fill (3, -0.25) node (Y2){} circle (2pt)  node[right=0.3]  {$\,d_j$};
	\fill (3, -0.75) node (Y1){} circle (2pt)  node[right=0.3]  {$\,d_i$};

	\fill (3, -1.25) node (Y1b){} circle (2pt)  node[right =0.1]  {$-d_i$};
	\fill (3, -1.75) node (Y2b){} circle (2pt)  node[right =0.1]  {$-d_j$};
	\fill (3, -2.25) node (Y3b){} circle (2pt)  node[right =0.1]  {$-d_k$};

	\draw[->-=0.5] (Z3) -- node[above= -2pt] {\tiny $p_k$} (Y3); 
	\draw[->-=0.5] (Z3) -- node[above = -2pt] {\tiny $q_k$} (Y2);
	\draw[->-=0.5] (Z1) -- node[below= -2pt] {\tiny $q_i$} (Y2); 
	\draw[->-=0.5] (Z1) -- node[below = -2pt] {\tiny $p_i$} (Y1);
		
	\draw[->-=0.5] (Z3b) -- node[below= -2pt] {\tiny $p_k$} (Y3b); 
	\draw[->-=0.5] (Z3b) -- node[below = -2pt] {\tiny $q_k$} (Y2b);
	\draw[->-=0.5] (Z1b) -- node[above= -2pt] {\tiny $q_i$} (Y2b); 
	\draw[->-=0.5] (Z1b) -- node[above = -2pt] {\tiny $p_i$} (Y1b);

	\end{tikzpicture}
	\label{fig_complex upgrading}	
	}
	\caption{Degrading channels for the upgrade-merge-2 and upgrade-merge-3 procedures.}
	\label{fig_degrading channels for upgrade merge}	
	
\end{figure}
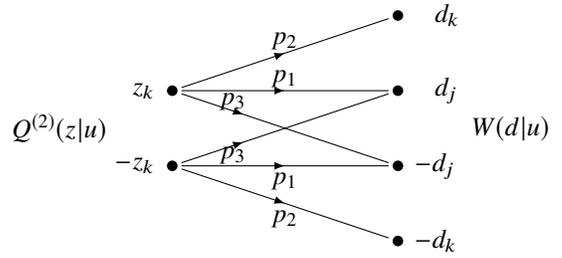
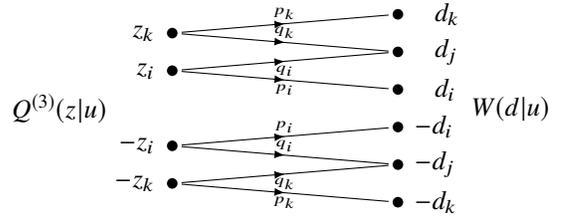

\subsection{The Upgrade-merge-3 Procedure}\label{ap_upgrade merge 3}
The second upgrade-merge of~\cite{Tal_2013} removes a $D$-value $d_j$ by splitting its probability between a preceding $D$-value $d_i \leq d_j$ and a succeeding $D$-value $d_k \geq d_j$. We call it \emph{upgrade-merge-3}. Unlike upgrade-merge-2, at least one of these inequalities must be strict (i.e., either $d_i < d_j$ or $d_j < d_k$). 
Channel $W:\mathcal{U}\to\mathcal{Y}$ is upgraded to channel $Q^{(3)}:\mathcal{U}\to \mathcal{Z}$ with output alphabet  
$\mathcal{Z} = (\mathcal{Y}\setminus\{d_{i},d_j,d_{k},-d_{i},-d_j,-d_{k}\})\cup\{z_{i},z_{k},-z_{i},-z_{k}\}, $ 
and \begin{equation} Q^{(3)}(z|u) = \begin{dcases}
 \achanpi{z_k}\left(\frac{1+ (-1)^ud_{k}}{2}\right), & z = z_{k} \\[0.1cm]
 \achanpi{z_i}\left(\frac{1+ (-1)^ud_{i}}{2}\right), & z = z_{i} \\[0.1cm]
 \achanpi{z_i}\left(\frac{1- (-1)^ud_{i}}{2}\right), & z = \bar{z}_{i} \\[0.1cm]
 \achanpi{z_k}\left(\frac{1- (-1)^ud_{k}}{2}\right), & z = \bar{z}_{k} \\[0.1cm]
W(z|u), & \text{otherwise,}
\end{dcases} \label{eq_formula for upgrade merge 3} \end{equation}
where
\[ \achanpi{z_{\ell}} = \begin{dcases} \achanpi{d_i} + \achanpi{d_j}\left(\frac{d_{k}-d_j}{d_{k}-d_{i}} \right), 	    & \ell = i \\[0.1cm]
  0, 	    & \ell = j \\[0.1cm]
 \achanpi{d_k} + \achanpi{d_j}\left(\frac{d_{j}-d_{i}}{d_{k}-d_{i}} \right), & \ell =k \\[0.1cm] 
 \pi_{\ell}, & \text{otherwise.}
 			 \end{dcases} \]	 
Note that \begin{equation}Q^{(3)}(z_k|u) + Q^{(3)}(z_i|u) = W(d_i|u)+W(d_j|u)+W(d_k|u).\label{eq_upgrade_merge_3}\end{equation}

The degrading channel from $Q^{(3)}(z|u)$ to $W(y|u)$ is shown in \Cref{fig_complex upgrading}, showing only the interesting portion of the channel. The parameters of the channel are
 $p_{\ell} = \achanpi{d_{\ell}}/\achanpi{z_{\ell}}$,  and $q_{\ell} = 1-p_{\ell}$, ${\ell}=i,k$. This is a valid channel as $\achanpi{z_{\ell}} \geq \achanpi{d_{\ell}}$. 
 
 It can be shown~\cite[Lemma 12]{Tal_2013} that $Q^{(2)} \succcurlyeq Q^{(3)} \succcurlyeq W$. That is, upgrade-merge-3 yields a better (closer) upgraded approximation of $W$ than does upgrade-merge-2. 


\end{document}